\documentclass[11pt,a4paper,english,twoside]{article}

\usepackage{a4wide}
\usepackage{amssymb, amsmath}
\usepackage{graphicx}
\usepackage{subcaption}
\usepackage[all]{xy}
\usepackage{enumerate}
\usepackage[pdftex,hyperref,svgnames]{xcolor}
\usepackage[pdftex,colorlinks=true,
pdfstartview=FitV,
pdfnewwindow=true,
linktoc = page,
linkcolor= Red,
citecolor= blue,
urlcolor= blue,
hyperindex=true,
hyperfigures=false]{hyperref}
\hypersetup{linktocpage}
\usepackage{dsfont}
\usepackage{empheq}
\usepackage{cite}
\usepackage{float}
\usepackage{cancel}
\usepackage{relsize}
\usepackage{soul}
\usepackage{enumitem}
\usepackage{hhline}

\usepackage{graphicx}

\newcommand{\beq}{\begin{equation}}
\newcommand{\eeq}{\end{equation}}
\def\bea#1\eea{\begin{align}#1\end{align}}
\def\beal#1\eeal{\begin{subequations}\begin{align}#1\end{align}\end{subequations}}
\newcommand{\nn}{\nonumber}
\newcommand{\w}{\wedge}
\newcommand{\R}{\mathcal{R}}

\newtheorem{thm}{Conjecture}

\def\d {{\rm d}}
\def\mmm {\mathcal{M}}

\begin{document}
\numberwithin{equation}{section}

\begin{titlepage}

\begin{center}

\phantom{DRAFT}

\vspace{2.0cm}

{\LARGE \bf{Charting the landscape of (anti-) de Sitter\vspace{0.3cm}\\ and Minkowski solutions of 10d supergravities}}\\

\vspace{2.2 cm} {\Large David Andriot$^{1}$, Ludwig Horer$^{2}$, Paul Marconnet$^{3}$}\\
\vspace{0.9 cm} {\small\slshape $^1$ Laboratoire d’Annecy-le-Vieux de Physique Th\'eorique (LAPTh),\\
UMR 5108, CNRS, Universit\'e Savoie Mont Blanc (USMB),\\
9 Chemin de Bellevue, 74940 Annecy, France}\\
\vspace{0.2 cm} {\small\slshape $^2$ Institute for Theoretical Physics, TU Wien\\
Wiedner Hauptstrasse 8-10/136, A-1040 Vienna, Austria}\\
\vspace{0.2 cm} {\small\slshape $^3$ Institut de Physique des 2 Infinis de Lyon\\
Universit\'{e} de Lyon, UCBL, UMR 5822, CNRS/IN2P3\\
4 rue Enrico Fermi, 69622 Villeurbanne Cedex, France}\\
\vspace{0.5cm} {\upshape\ttfamily andriot@lapth.cnrs.fr; ludwig.horer@tuwien.ac.at;\\
marconnet@ipnl.in2p3.fr}\\

\vspace{2.8cm}

{\bf Abstract}
\vspace{0.1cm}
\end{center}

\begin{quotation}
\noindent We classify solutions of 10d type IIA/B supergravities with orientifolds, on a 4d maximally symmetric spacetime times a 6d group manifold. We then look for new solutions in previously unexplored solution classes, and find some: (anti-) de Sitter solutions with intersecting $O_4$, $O_6$ and $D_6$, or Minkowski solutions with 3 intersecting $O_5$, among others. We provide the numerical code that we developed for this purpose. We also prove new no-go theorems against (anti-) de Sitter solutions. We finally conjecture the absence of de Sitter solution for 2 or less intersecting source sets, implying that a 4d effective theory with de Sitter is at most ${\cal N}=1$ supersymmetric.
\end{quotation}

\end{titlepage}

\newpage

\tableofcontents

\section{Introduction and results summary}

String theory backgrounds with maximally symmetric spacetimes have always been of prime importance. To start with, a 4-dimensional (4d) de Sitter spacetime is relevant for the connection to cosmology, since it can describe in good approximation our universe in an early inflation phase, or in the far future. A 4d Minkowski spacetime is relevant for particle physics models that may be derived from string theory, or to describe our nearby universe. Last but not least, 4d anti-de Sitter spacetimes are typically considered in a holographic context, but they also appear in various phenomenology related topics, for instance in certain constructions of de Sitter solutions \cite{Kachru:2003aw, Balasubramanian:2005zx} or in the matter of scale separation, revived recently in the swampland program \cite{Gautason:2018gln, Lust:2019zwm}. In addition, the general belief that non-supersymmetric solutions are unstable is shared in various forms for all maximally symmetric spacetimes: see for instance \cite{Andriot:2018wzk, Garg:2018reu, Ooguri:2018wrx, Andriot:2018mav, Andriot:2021rdy} for de Sitter, \cite{Acharya:2019mcu, Acharya:2020hsc} for Minkowski and \cite{Ooguri:2016pdq} for anti-de Sitter. The importance of such solutions, and the common properties they may share, motivates us in this paper to classify them, within a certain (standard) ansatz. In turn, the classification helps us finding new types of solutions, as well as new existence no-go theorems, and noticing few general properties. Further aspects of the new solutions found, such as their perturbative stability and mass spectrum, as well as properties of their extra dimensions, will be studied in a companion paper \cite{Andriot:2022yyj}.

We focus here on solutions of 10d type IIA/B supergravities with $D_p$-branes and orientifold $O_p$-planes, as candidates for classical string backgrounds. Checking whether the supergravity solutions actually meet the string effective theory requirements, allowing them to be in the classical string regime, is not always trivial: see e.g.~\cite{Andriot:2020vlg} for de Sitter and \cite{Cribiori:2021djm} for anti-de Sitter. We leave that question aside in this work, the classification provided remaining sufficient for classical string backgrounds. We further restrict ourselves to a historically standard ansatz,\footnote{Minkowski and de Sitter solutions of \cite{Macpherson:2016xwk, Apruzzi:2018cvq} and \cite{Cordova:2018dbb, Cribiori:2019clo, Cordova:2019cvf} typically do not obey this ansatz.} namely one where the extra dimensions are gathered as a group manifold. In addition, fluxes living there are constant, as well as the $O_p/D_p$ contributions to the equations: this last point corresponds to having $O_p/D_p$ sources smeared, or more precisely, considering an integrated version of the solution rather than a localized one. More details on this ansatz is provided in Section \ref{sec:ansatz} and in \cite{Andriot:2019wrs}. One interest of this ansatz is that it allows, through a consistent truncation, for an equivalent description as a 4d gauged supergravity, also sometimes used to find those solutions.\footnote{Among the vast literature studying or making use of this relation to 4d (see e.g.~Section 2.3 of \cite{Andriot:2019wrs}), one example with a nice title is \cite{Dibitetto:2011gm}.} A further reason to restrict to classical solutions with such an ansatz is the relative simplicity of the setting, of potential interest to further phenomenological applications, while still providing a variety of interesting examples: de Sitter solutions, (non-) supersymmetric Minkowski ones with fluxes, (non-) supersymmetric (non-) scale separated anti-de Sitter ones, etc.; references for all such solutions are given below. We will provide a classification for solutions with a 4d maximally symmetric spacetime and obeying this ansatz.

A crucial element in our ansatz and classification is the presence of orientifolds. $O_p$-planes are typically introduced to circumvent the Maldacena-Nu\~nez no-go theorem \cite{Maldacena:2000mw}, allowing to find Minkowski or de Sitter solutions. Anti-de Sitter solutions can however be found without. We will nevertheless consider $O_p$ there as well, to fit in our solution classification, but also because scale separation in anti-de Sitter solutions, one of the motivations here, is thought to be possible only with orientifold. Each $O_p$ imposes a projection, which, together with our ansatz, projects out certain flux components and other variables otherwise allowed in our solutions. Each solution class is thus defined by its $O_p$ (and labeled accordingly), which in turn selects a specific list of allowed variables (field content) in the solutions. We proceed with a systematic approach, described in Section \ref{sec:ansatz}, that uses the RR sourced Bianchi identities and the sources appearing there. As part of our ansatz, we do not consider the cases where $O_p$ and $D_p$ contributions exactly cancel each other, i.e.~the so-called ``tadpole cancelation'' always requires here fluxes. This allows us to identify all possible $O_p/D_p$ source configurations, the resulting field content, therefore all possible solution classes within our ansatz. This procedure is carried out in Section \ref{sec:singledim} and \ref{sec:multidim}, with a summary of the solution classes and their properties in Section \ref{sec:sumsingle} and  \ref{sec:summult}, as well as in Table \ref{tab:intro} here.
\begin{table}[H]   
  \begin{center}
    \begin{tabular}{|c|c|c||c|c|c|}
    \hline
Solution & Source & Field & dS sol. & Mink. sol. & AdS sol. \\
class & directions & content &  &  &  \\
    \hhline{===::===}
$s_3$ & \eqref{p=3sources1O3} & \eqref{variables1O3} & $\times$ & \cite{Giddings:2001yu} &  \\
     \hhline{---||---}
$s_4$ & \eqref{p=4sources1O4} & \eqref{variables1O4} & ? & \cite{Andriot:2016ufg} &  \\
     \hhline{---||---}
$s_5$ & \eqref{p=5sources1O5} & \eqref{variables1O5} & ? & \cite{Andriot:2016ufg} &  \\
\hhline{---||---}
 $s_{55}$ & \eqref{p=5sources2O5} & \eqref{variables2O5} & \cite{Andriot:2020wpp, Andriot:2021rdy}, $\checkmark$ & \cite{Grana:2006kf} & $\checkmark$ \\
     \hhline{---||---}
$s_{555}$ & \eqref{p=5sources3O5} & \eqref{variables3O5} & $\times$ & $\checkmark$ & $\times$ \\
\hhline{---||---}
$s_{6}$ & \eqref{p=6sources1O6} & \eqref{variables1O6} & ? & \cite{Andriot:2016ufg} &  \\
\hhline{---||---}
$s_{66}$ & \eqref{p=6sources2O6} & \eqref{variables2O6} & $\checkmark$ & \cite{Grana:2006kf} &  \\
\hhline{---||---}
$s_{6666}$ & \eqref{p=6sources3O6} & \eqref{variables3O6} & \cite{Danielsson:2011au}, $\checkmark$ & \cite{Camara:2005dc} & \cite{DeWolfe:2005uu, Camara:2005dc, Caviezel:2008ik} \\
\hhline{---||---}
$s_{7}$ & \eqref{p=7sources1O7} & \eqref{variables1O7} & $\times$ & \cite{Andriot:2016ufg} &  \\
\hhline{---||---}
$s_{77}$ & \eqref{p=7sources2O7} & \eqref{variables2O7} & $\times$ &  &  \\
    \hhline{===::===}
$m_4$ & \eqref{p=46sources1O4} & \eqref{variables1O4} & ? &  &  \\
\hhline{---||---}
$m_{46}$ & \eqref{p=64sources1O61O4} & \eqref{variables1O61O4} & $\checkmark$ & $\checkmark$ & $\checkmark$ \\
\hhline{---||---}
$m_{466}$ & \eqref{p=64sources2O61O4} & \eqref{variables2O61O4} & $\times$ & $\checkmark$ & $\times$ \\
\hhline{---||---}
$m_6$ & \eqref{p=64sources1O6} & \eqref{variables1O6} & ? &  &  \\
\hhline{---||---}
$m_{66}$ & \eqref{p=64sources2O6} & \eqref{variables2O6} & ? &  &  \\
    \hhline{===::===}
$m_5$ & \eqref{p=57sources1O5} & \eqref{variables1O5} & ? &  &  \\
\hhline{---||---}
$m_{55}$ & \eqref{p=57sources2O5} & \eqref{variables2O5} & $\checkmark$ &  &  \\
\hhline{---||---}
$m_{57}$ & \eqref{p=57sourcesO5O7} & \eqref{variablesO5O7} & ? &  &  \\
\hhline{---||---}
$m_{5577}$ & \eqref{p=57sources1O52O7} & \eqref{variables2O51O7} & \cite{Caviezel:2009tu}, $\checkmark$ &  & \cite{Caviezel:2008ik, Petrini:2013ika} \\
\hhline{---||---}
$m_7$ & \eqref{p=75sourcesO7} & \eqref{variables1O7} & ? &  &  \\
\hhline{---||---}
$m_{77}$ & \eqref{p=75sources2O7} & \eqref{variables2O7} & ? &  &  \\
    \hline
    \end{tabular}
     \caption{Classification of solutions on 4d maximally symmetric spacetimes with orientifold, within our ansatz. The solution classes are labeled with $s$ for $O_p/D_p$ sources of single dimensionality, and $m$ for multiple ones. The subscript corresponds to the dimensionalities of the $O_p$: for instance $s_{6666}$ means there are only $O_6$ and $D_6$, and 4 intersecting source sets with $O_6$. The details of the source configurations and the allowed field variables under the projections are given in corresponding equations. We further list for each solution class few references where explicit solutions were found; more references and comments are given in Section \ref{sec:resultssol} and Table \ref{tab:sol}. We indicate with a $\checkmark$ when we found new solutions in the present paper. On the contrary, a $\times$ indicates an existence no-go theorem against a solution, while a ? indicates that we have looked for solutions but have not found any. Finally, an empty box means that we have not looked for solutions, and do not know of solutions in the literature for that class.}\label{tab:intro}
  \end{center}
\end{table}
\noindent T-duality relations between the classes, as well as the supersymmetry preserved by the source configurations, are discussed in Section \ref{sec:Tdsusy}.

Given the complete classification of possible solutions on 4d maximally symmetric spacetimes with orientifold, within our ansatz, one may start identifying the classes for which solutions have been found, and look for new ones. We do so following a procedure described in Section \ref{sec:proc} and using the numerical code {\tt MaxSymSolSearch} ({\tt MSSS}) presented in Section \ref{sec:code}. Some searches in a given solution class do not provide any solution: this can either be a sign of computational complexity that we cannot overcome with our tools, or a hint at an existence no-go theorem. There are many no-go theorems against de Sitter solutions (see \cite{Andriot:2019wrs, Andriot:2020lea}); we discuss and find new ones here in Section \ref{sec:nogo} both for de Sitter and anti-de Sitter, the latter being less common. We finally present the new solutions found, together with those already known, in Section \ref{sec:resultssol}, as well as here in Table \ref{tab:intro}. The new solutions are also listed explicitly in Appendix \ref{ap:sol}. The study of their properties, namely their mass spectrum and stability, the identification of their group manifold and its compactness, are delayed to the companion paper \cite{Andriot:2022yyj}.

Several of the new solutions found in this paper deserve to be highlighted. We found de Sitter solutions in $m_{46}$, in particular some with, in terms of source sets, 1 $O_4$, 1 $O_6$ and 1 $D_6$, a source configuration that preserves ${\cal N}=1$ supersymmetry in 4d. These de Sitter solutions are the first of their kind. We also found Minkowski solutions in $s_{555}$ and $m_{466}$, i.e.~with the 3 corresponding $O_p$, which are the first of their kind. These classes are very special because as we prove here, there are no-go theorems against de Sitter and anti-de Sitter solutions. Finally, we find anti-de Sitter solutions in $s_{55}$ and $m_{46}$, also the first of their kind. More on our solutions can be found in Section \ref{sec:resultssol}. An important motivation in finding these completely new supergravity solutions is that they may provide new physics. This is particularly relevant for de Sitter solutions, for which the known examples are plagued with strong perturbative instabilities and a failure to be classical string backgrounds, as far as this could be tested. The search for new solutions remains here fairly generic, i.e.~not necessarily tailored to reveal new physics, and is meant as a proof of existence in these previously unexplored solution classes. It calls for more dedicated searches, e.g.~as in \cite{Andriot:2021rdy}, but exhibiting already these new examples offers the hope of new physics.

The classification and the extensive search for new solutions finally led us to notice general properties. We argue in Section \ref{sec:conjdS} in favor of conjecture \ref{conj4}, which can be thought of as extending conjecture 1 of \cite{Andriot:2019wrs}. The latter states the absence of classical de Sitter solutions with parallel sources (i.e.~1 source set). We repeat here for convenience conjecture \ref{conj4}, together with an implication of those results.
\bea
& \mbox{{\bf Conjecture 4 } {\it There is no classical de Sitter solution with 2 intersecting source sets.}} \nn\\
& \nn\\
& \mbox{\phantom{{\bf Conjecture 4} } {\it This implies that a 4d effective theory of a classical string compactification,}}\nn\\
& \mbox{\phantom{{\bf Conjecture 4} } {\it admitting a de Sitter critical point, can at most be ${\cal N} = 1$ supersymmetric.}}\nn
\eea
Several arguments supporting the conjecture, and a comparison to the literature of 4d gauged supergravities, are provided in Section \ref{sec:conjdS}. Interestingly, if true, these claims have surprising consequences for string phenomenology. Indeed, while looking for classical de Sitter solutions, one automatically obtains the stringy ingredients to build 4d ${\cal N} = 1$ (or ${\cal N} = 0$) supersymmetric particle physics models, namely the intersecting branes. It is remarkable that these two crucial phenomenological elements (cosmology and particle physics) {\it naturally} appear here together, hand-in-hand, from string theory.

\section{Classification of possible solutions}\label{sec:classif}

\subsection{Ansatz, systematic approach and solution classes}\label{sec:ansatz}

We are interested in solutions of 10d type II supergravities on a 4d maximally symmetric spacetime, i.e.~de Sitter, Minkowski or anti-de Sitter, times a 6d compact group manifold $\mmm$. Regarding supergravity, $D_p$-branes and orientifolds, we follow conventions of \cite{Andriot:2016xvq, Andriot:2017jhf}, and we restrict ourselves to the solution ansatz reviewed in \cite{Andriot:2019wrs}. Let us summarize now this ansatz and fix the notations. Along the 6 internal dimensions, we work in the 6d orthonormal coframe (sometimes referred to as the flat basis), in such a way that the metric is $\d s_6^2 = e^a \delta_{ab} e^b$, with the $1$-form $e^a = e^a{}_m \d y^m$. The spin connection can then be expressed in terms of $f^a{}_{bc}$ quantities. By restricting to a 6d group manifold (including a flat torus), the $f^a{}_{bc}$ are constant and correspond to structure constants of an underlying Lie algebra. This algebra allows to identify which group manifold we are facing. This information however does not fully characterise the global properties of $\mmm$, and in particular whether it is compact or not, a question we will come back to in great details in the companion paper \cite{Andriot:2022yyj}. We restrict ourselves to a basis where $f^a{}_{ac}=0$ {\it without sum} on $a$. The summed version is necessary for compactness, while the present unsummed version and corresponding choice of basis are preferred to prove the existence of lattices and thus compactness \cite{Andriot:2010ju}. We allow for $D_p$-branes and $O_p$ orientifold planes, collectively named sources; $p$ is their dimensionality, sometimes also called size. For a maximally symmetric spacetime, the sources must be along the 3 space extended dimensions, therefore restricting the dimensionality to $p \geq 3$. Again, for a maximally symmetric spacetime, supergravity fluxes can be purely internal (along the 6d) or spanning the whole 4d, leaving us to use only the 6d $k$-forms $F_{k=0, ..., 6}$ for RR fluxes and the 6d $3$-form $H$ for the NSNS one. As part of the ansatz, motivated in \cite{Andriot:2019wrs}, the fluxes components in the 6d orthonormal coframe are constant, and the sources are smeared. The latter means that there is no warp factor, the dilaton is constant and given by $e^{\phi} = g_s$, and the source contributions of each set $I$ (to be defined) are captured by the constant $T_{10}^I$. The latter can be understood as the sum of integrals of the $\delta$-functions localizing parallel sources. We will also consider the trace of the sources energy-momentum tensor $T_{10}= \sum_I T_{10}^I$. For a given source, we recall the notations of $a_{||}$ and $a_{\bot}$ for parallel and transverse internal dimensions. For fluxes, we also recall the notation $H^{(n)}$ for the form $H$ with only components along $n$ parallel dimensions. Overall, this ansatz has several advantages: it simplifies the equations and allows an alternative description using 4d gauged supergravities and their scalar potential \cite{Andriot:2019wrs}. In particular, let us emphasize that the only variables entering the equations will be the constants
\beq
f^a{}_{bc},\ H_{abc},\ g_s F_{q\, a_1 ...a_q},\ g_s T_{10}^I\ . \label{variables}
\eeq

An important distinction to make is whether sources $O_p/D_p$ (with single $p$) are parallel or intersecting. All sources that are along the same directions are said to be parallel and are part of one set $I$. We denote by $N$ the number of sets: if $N=1$, sources are said to be parallel, if $N>1$ they are intersecting. To stress the importance of this distinction, let us mention a few de Sitter examples. It turns out that de Sitter solutions with parallel sources are expected not to exist (conjecture 1 in \cite{Andriot:2019wrs}), while having intersecting ones was shown to help in \cite{Andriot:2017jhf} to get a positive cosmological constant. De Sitter solutions satisfying the above ansatz with $N>1$ were obtained in IIA for $p=6$, in particular in \cite{Danielsson:2011au}, and in IIB for $p=5$ in \cite{Andriot:2020wpp}. More examples will be given in Section \ref{sec:resultssol}.

We will also consider the case of sources with multiple dimensionalities $p$. For example, one such de Sitter solution is known, with $p=5, 7$ in \cite{Caviezel:2009tu}. This situation requires a few more notations, and we follow the formalism introduced in Section 6 of \cite{Andriot:2017jhf}. The sources energy momentum tensor and its trace is where changes occur, with respect to the case of single dimensionality. The formal definition of $T_{MN}$, as well as $T_{10} = g^{MN} T_{MN} = \eta^{AB} T_{AB}$, are still valid, but the sum over sources now has to be split into a further sum over the different values of $p$. The total contribution of sources for each dimensionality $p$ is now denoted $T_{10}^{(p)}$ (it was denoted $T_{10}^p$ in \cite{Andriot:2017jhf}), and one has
\beq
T_{10}= \sum_p T_{10}^{(p)} = \sum_p \sum_{I} T_{10}^{(p)I} \ ,
\eeq
with a split into all sets $I$ for a given $p$, introducing a $T_{10}^{(p)I}$. The case of single dimensionality $p$ is recovered by dropping the unnecessary upper ${}^{(p)}$; we will do so in the following. In addition, we now have the 4d trace $g^{MN}T_{MN=\mu\nu} = 4 \sum_p T_{10}^{(p)}/(p+1)$.

In IIA, we can a priori have $p=4,6,8$ sources, and in IIB, $p=3,5,7,9$. Our ansatz is however more restrictive. As we will see, sources are here visible only if their transverse volume appears in the right-hand side of the sourced RR Bianchi identity. This first implies that we cannot have $p=9$ sources, since we do not consider hypothetical $F_{-1}$ fluxes. In addition, since we consider constant fluxes, we get that $\d F_0 = 0$, so we cannot admit $p=8$ source. This leaves us in IIA with $p=4,6$, and $p=3,5,7$ in IIB; we will also see that $p=3$ sources cannot really contribute here.

In \cite{Andriot:2017jhf}, the following equation was established (with even/odd RR fluxes in IIA/IIB)
\beq
{\cal R}_4= g_s \sum_p \frac{T_{10}^{(p)}}{p+1} - g_s^2 \sum_{q=0}^6 |F_q|^2  \ , \label{R4T_{10}F}
\eeq
reminiscent of the Maldacena-Nu\~nez no-go theorem \cite{Maldacena:2000mw}, here in the case of multiple dimensionalities. It implies that de Sitter solutions, or Minkowski solutions with RR flux, require $\sum_p T_{10}^{(p)}/(p+1) > 0$. The latter implies that one needs an $O_p$ for some $p$, but interestingly, not necessarily for all $p$. This result motivates us to systematically include $O_p$.\\

Given the above ansatz for solutions on a 4d maximally symmetric spacetime with orientifolds, we now present a systematic approach that will allow us to determine all possible source configurations, as well as the allowed fields or variables among \eqref{variables}.
\begin{enumerate}
  \item We first consider an orientifold $O_p$-plane, and place it along the first $p-3$ internal dimensions, in the set $I=1$.
  \item An $O_p$ imposes a projection. Because of our ansatz where variables are constant, the projection sets to zero many flux components and structure constants \cite{Andriot:2019wrs}. We then give the explicit list of remaining variables.
  \item We finally look at each sourced RR Bianchi identity
  \beq
  \d F_{8-p} - H \w F_{6-p} = \varepsilon_p \sum_{I} \frac{T_{10}^{(p)I}}{p+1} \, {\rm vol}_{\bot_{(p)I}} \ ,\quad \varepsilon_p= (-1)^{p+1} (-1)^{[\frac{9-p}{2}]} \ ,
  \eeq
  where the right-hand side indicates the various $p$-sources which are present, through the total contribution $T_{10}^{(p)I}$ in each set $I$, together with the transverse volume form ${\rm vol}_{\bot_{(p)I}}$ to this set. Using the Maurer-Cartan equation $\d e^a= -\tfrac{1}{2} f^{a}{}_{bc} e^b\w e^c$, we compute the various components on the left-hand side using the list of variables remaining after the projection. Each potentially non-zero component can be interpreted as giving rise to a ${\rm vol}_{\bot_{(p)I}}$ in the right-hand side, with a non-zero $T_{10}^{(p)I}$. On the contrary, there cannot be any source (in our ansatz) whose transverse directions do not appear in the left-hand side.\footnote{Proceeding this way, we neglect the possibility of having sources in a set $I$ such that $T_{10}^{(p)I}=0$, i.e.~where $O_p$ and $D_p$ contributions perfectly cancel each other. We view this here as going beyond our ansatz. See in particular \cite{Andriot:2015sia} for a discussion of such Minkowski solutions.} We identify this way the possible placements of source sets, i.e.~the allowed source configurations, and they will turn out to be very constrained.
  \item We then start over by adding another $O_p$ in a different set, studying the resulting projection, the allowed variables and remaining sources. In case this leads to a contradiction, we take it back and conclude that other sets $J$ can at best contain $D_p$-branes, implying in our conventions $T_{10}^{(p)J} \leq 0$.
\end{enumerate}

We will proceed in the following with this systematic approach, first considering a single dimensionality $p$ and then allowing for multiple ones. This will result in identifying all possible source configurations and the associated sets of variables. This provides a natural classification of the possible solutions, and we will distinguish the various possibilities into so-called solution classes.\\

Anticipating on our results, we now present these classes and the symbols to denote them. Given our ansatz, a solution class is defined by the number and dimensionalities of $O_p$. This defines the allowed variables under the corresponding projections. The symbol to be used is $s$ for single dimensionality of $O_p$ and $D_p$, and $m$ for multiple dimensionalities. To this, we add a subscript carrying the $p$'s of the $O_p$: for instance $m_{5577}$ stands for 2 $O_5$ and 2 $O_7$. In the case where different choices of $O_p$ lead to the same set of allowed variables and sources (up to the nature of the latter, i.e.~$D_p$ or $O_p$), then the class is defined by the maximum number of $O_p$. This will become clear in the following, but for example, it will be the case for $s_{6666}$ instead of $s_{666}$, or $m_{5577}$ instead of $m_{577}$. Finally, note that a solution is sometimes searched within a certain class, but once found, it ends up having many variables and sources set to zero, in such a way that it can belong to another class with more $O_p$: in that case, the convention is to place it in the latter.

The classes defined as above amount to consider some sets of source directions as ``equivalent''. This means that directions are equivalent up to a relabeling, not considering the orientation. For instance, given an $O_p$ along directions 123, choosing another set along 145 or 245 is equivalent since one can consider $1 \leftrightarrow 2$, which does not change the placement of the first $O_p$, ignoring orientation. However, when it comes to a concrete solution, doing a relabeling (or more generally a change of basis) that does not preserve source volume forms, in particular their orientation, typically does not lead to a solution. Indeed, changing the orientation of a source can be compensated by changing the sign of the corresponding $T_{10}^{(p)I}$, which however does not solve the equations anymore: see appendix \ref{ap:chgbasis}. Because of this, solution classes could be split into subclasses: the variables and source directions of each subclasses can be mapped into each other by some transformation, e.g.~a relabeling, but actual solutions do not survive this transformation. We will have an example of this for $m_{5577}$, and will then introduce the subclass $m_{5577}^*$.

We now turn to the systematic determination of the solution classes, and will provide summaries of those in sections \ref{sec:sumsingle} and \ref{sec:summult}.

\subsection{Source configurations and fields for a single dimensionality $p$}\label{sec:singledim}

\subsubsection{$O_3$-plane}

The case of $p=3$ sources is special, because they are transverse to all 6d dimensions, and are thus only points in $\mmm$. A first consequence is that no structure constant $f^a{}_{bc}$ survives the orientifold projection. In addition, given our ansatz, only the following flux components remain after an $O_3$ projection
\beq
O_3: \quad F_3^{(0)},\ H^{(0)} \ ,
\eeq
or in other words
\beq
s_3: \quad F_{3\ abc}\,, \quad H_{abc},\quad a,b,c=1,...,6 \ . \label{variables1O3}
\eeq
Considering an $O_3$, the left-hand side of the Bianchi identity for $F_5$ boils down to $-H\w F_3$, which can be non-zero with the above components, and proportional to the 6d volume form. We can then have $p=3$ sources. In short we get
\beq
s_3: \quad 1\ O_3\ \mbox{(at a point)} \Rightarrow\ \mbox{$p=3$ sources at points}\ . \label{p=3sources1O3}\\
\eeq

\subsubsection{$O_4$-plane}

The orientifold projection of an $O_4$ restricts the structure constants and fluxes to be \cite{Andriot:2018ept}
\beq
O_4: \quad f^{a_{||}}{}_{b_{\bot}c_{\bot}},\ f^{a_{\bot}}{}_{b_{\bot}c_{||}}, \quad F_2^{(0)},\ F_4^{(1)},\ H^{(0)} \ .
\eeq
The other possible type of structure constants, $f^{a_{||}}{}_{b_{||}c_{||}}$, is vanishing due to the antisymmetry of $b,c$, since there is only one direction parallel to the $O_4$. We also recall that we restrict ourselves for simplicity to a basis where $f^{a}{}_{ac} = 0$ without sum. We now place an $O_4$ in the set $I=1$ along the internal direction 1. We deduce the following list of remaining variables after one $O_4$ projection
\bea
s_4:\quad F_2:&\quad F_{2\ 23} \,, \quad F_{2\ 24} \,, \quad F_{2\ 25} \,, \quad F_{2\ 26} \,, \quad F_{2\ 34} \,, \quad F_{2\ 35} \,, \quad F_{2\ 36} \,, \quad F_{2\ 45} \,, \nn\\
&\quad F_{2\ 46} \,, \quad F_{2\ 56} \,, \nn\\
F_4:&\quad F_{4 \ 1234} \,, \quad F_{4 \ 1235} \,, \quad F_{4 \ 1236} \,, \quad F_{4 \ 1245} \,, \quad F_{4 \ 1246} \,, \quad F_{4 \ 1256} \,, \quad F_{4 \ 1345} \,, \nn\\
&\quad F_{4 \ 1346} \,, \quad F_{4 \ 1356} \,, \quad F_{4 \ 1456} \,, \nn\\
H:&\quad H_{234} \,, \quad H_{235} \,, \quad H_{236} \,, \quad H_{245} \,, \quad H_{246} \,, \quad H_{256} \,, \quad H_{345} \,, \quad H_{346} \,, \nn\\
&\quad H_{356} \,, \quad H_{456} \,, \label{variables1O4}\\
f^{a_{||_1}}{}_{b_{\bot_1} c_{\bot_1}}:&\quad f^1{}_{23}\,, \quad f^1{}_{24}\,, \quad f^1{}_{25}\,, \quad f^1{}_{26}\,, \quad f^1{}_{34}\,, \quad f^1{}_{35}\,, \quad f^1{}_{36}\,, \quad f^1{}_{45}\,, \quad f^1{}_{46}\,, \quad f^1{}_{56}\,, \nn\\
f^{a_{\bot_1}}{}_{b_{\bot_1} c_{||_1}}:&\quad f^2{}_{31} \,, \quad f^2{}_{41} \,, \quad f^2{}_{51}\,, \quad f^2{}_{61}\,, \quad f^3{}_{21}\,, \quad f^3{}_{41}\,, \quad f^3{}_{51}\,, \quad f^3{}_{61}\,, \quad f^4{}_{21} \,, \quad f^4{}_{31} \,, \nn\\
&\quad f^4{}_{51}\,, \quad f^4{}_{61}\,, \quad f^5{}_{21}\,, \quad f^5{}_{31}\,, \quad f^5{}_{41}\,, \quad f^5{}_{61}\,, \quad f^6{}_{21} \,, \quad f^6{}_{31} \,, \quad f^6{}_{41}\,, \quad f^6{}_{51}\,. \nn
\eea

There are 30 fluxes and 30 structure constants. From this list, it is straightforward to compute the components of $\d F_4$ and $H\w F_2$. It is easy to verify that $e^1$ never appears in these 5-forms. This means the components are purely along directions 23456, i.e.~${\rm vol}_{\bot_1}$, in other words
\beq
s_4:\quad 1\ O_4\ \mbox{(along direction 1)}\ \Rightarrow\ \mbox{$p=4$ sources along direction}\ 1 \ . \label{p=4sources1O4}
\eeq
We conclude that imposing one $O_4$ projection with our ansatz allows to have only $N=1$ set of sources: the one with the $O_4$. In other words, with an $O_4$, $p=4$ sources can only be parallel!

\subsubsection{$O_5$-plane}

We turn to $p=5$ and proceed as above. We recall the choice of working in a basis where $f^{a}{}_{ac} = 0$ without sum. One $O_5$ projection then leaves us with
\beq
O_5: \quad f^{a_{||}}{}_{b_{\bot}c_{\bot}},\ f^{a_{\bot}}{}_{b_{\bot}c_{||}}, \quad F_1^{(0)},\ F_3^{(1)},\ F_5^{(2)},\ H^{(0)},\ H^{(2)}  \ .
\eeq
Having the $O_5$ in a set $I=1$ along directions 12, one is left with the following variables
\bea
s_5: \quad F_1:&\quad F_{1 \ 3} \,, \quad F_{1 \ 4} \,, \quad F_{1 \ 5} \,, \quad F_{1 \ 6} \,,\nn\\
F_3:&\quad F_{3\ 134} \,, \quad F_{3\ 135} \,, \quad F_{3\ 136} \,, \quad F_{3\ 145} \,, \quad F_{3\ 146} \,, \quad F_{3\ 156} \,, \quad F_{3\ 234} \,, \quad F_{3\ 235} \,,\nn\\
& \quad F_{3\ 236} \,, \quad F_{3\ 245} \,, \quad F_{3\ 246} \,, \quad F_{3\ 256} \,,\nn\\
F_5:&\quad F_{5 \ 12345} \,, \quad F_{5 \ 12346} \,, \quad F_{5 \ 12356} \,, \quad F_{5 \ 12456} \,, \nn\\
H:&\quad H_{123} \,, \quad H_{124} \,, \quad H_{125} \,, \quad H_{126} \,, \quad H_{345} \,, \quad H_{346} \,, \quad H_{356} \,, \quad H_{456} \,, \label{variables1O5}\\
f^{a_{||_1}}{}_{b_{\bot_1} c_{\bot_1}}:&\quad f^{1}{}_{34} \,, \quad f^{1}{}_{35} \,, \quad f^{1}{}_{36} \,, \quad f^{1}{}_{45} \,, \quad f^{1}{}_{46} \,, \quad f^{1}{}_{56} \,, \quad f^{2}{}_{34} \,, \quad f^{2}{}_{35} \,, \quad f^{2}{}_{36} \,,\nn\\
& \quad f^{2}{}_{45} \,, \quad f^{2}{}_{46} \,, \quad f^{2}{}_{56} \,,\nn\\
f^{a_{\bot_1}}{}_{b_{\bot_1}c_{||_1}}:&\quad f^{3}{}_{14} \,, \quad f^{3}{}_{15} \,, \quad f^{3}{}_{16} \,, \quad f^{3}{}_{24} \,, \quad f^{3}{}_{25} \,, \quad f^{3}{}_{26} \,, \quad f^{4}{}_{13} \,, \quad f^{4}{}_{15} \,, \quad f^{4}{}_{16} \,,\nn\\
& \quad f^{4}{}_{23} \,, \quad f^{4}{}_{25} \,, \quad f^{4}{}_{26} \,, \quad f^{5}{}_{13} \,, \quad f^{5}{}_{14} \,, \quad f^{5}{}_{16} \,, \quad f^{5}{}_{23} \,, \quad f^{5}{}_{24} \,, \quad f^{5}{}_{26} \,, \nn\\
& \quad f^{6}{}_{13} \,, \quad f^{6}{}_{14} \,, \quad f^{6}{}_{15} \,, \quad f^{6}{}_{23} \,, \quad f^{6}{}_{24} \,, \quad f^{6}{}_{25} \,,\nn
\eea
namely 28 flux components and 36 structure constants. From there one computes $\d F_3$, $H\w F_1$, and deduces the possible non-zero components. It is straightforward to deduce that source sets can be along the following directions
\bea
s_5: \quad 1\ O_5\ \mbox{(along directions 12)}\ \Rightarrow\ \mbox{$p=5$ sources along directions}\ & 12, 34, 35, 36, \label{p=5sources1O5}\\
& 45, 46, 56 \ . \nn
\eea
So contrary to $p=4$, we can have here intersecting sources. The source directions in \eqref{p=5sources1O5} are, apart from 12, all equivalent. We then place a second $O_5$ in a second set $I=2$ along 34, and determine the remaining variables to be
\bea
s_{55}:\quad F_1:&\quad F_{1 \ 5} \,, \quad F_{1 \ 6} \,,\nn\\
F_3:&\quad F_{3\ 315} \,, \quad F_{3\ 316} \,, \quad F_{3\ 325} \,, \quad F_{3\ 326} \,, \quad F_{3\ 415} \,, \quad F_{3\ 416} \,, \quad F_{3\ 425} \,, \quad F_{3\ 426} \,,\nn\\
F_5:&\quad F_{5 \ 34125} \,, \quad F_{5 \ 34126} \,, \nn\\
H:&\quad H_{125} \,, \quad H_{126} \,, \quad H_{345} \,, \quad H_{346} \,,\label{variables2O5}\\
f^{a_{||_2}}{}_{b_{\bot_2} c_{\bot_2}}:&\quad f^{3}{}_{15} \,, \quad f^{3}{}_{16} \,, \quad f^{3}{}_{25} \,, \quad f^{3}{}_{26} \,, \quad f^{4}{}_{15} \,, \quad f^{4}{}_{16} \,, \quad f^{4}{}_{25} \,, \quad f^{4}{}_{26} \,,\nn\\
f^{a_{\bot_2}}{}_{b_{\bot_2}c_{||_2}}:&\quad f^{1}{}_{53} \,, \quad f^{1}{}_{63} \,, \quad f^{1}{}_{54} \,, \quad f^{1}{}_{64} \,, \quad f^{2}{}_{53} \,, \quad f^{2}{}_{63} \,, \quad f^{2}{}_{54} \,, \quad f^{2}{}_{64} \,, \nn\\
&\quad f^{5}{}_{13} \,, \quad f^{5}{}_{23} \,, \quad f^{5}{}_{14} \,, \quad f^{5}{}_{24} \,, \quad f^{6}{}_{13} \,, \quad f^{6}{}_{23} \,, \quad f^{6}{}_{14} \,, \quad f^{6}{}_{24} \,,\nn
\eea
namely 16 fluxes and 24 structure constants. We then compute the Bianchi identity components, and verify that sources can be along
\beq
s_{55}:\quad 2\ O_5\ \mbox{(along directions 12, 34)}\ \Rightarrow\ \mbox{$p=5$ sources along directions}\ 12, 34, 56 \ . \label{p=5sources2O5}
\eeq
We finally consider a third $O_5$ in $I=3$ along 56. In this case, the structure constants are not constrained further, but only the $F_3$ flux remains, i.e.~the variables are
\bea
s_{555}: \quad F_3:&\quad F_{3\ 135} \,, \quad F_{3\ 136} \,, \quad F_{3\ 145} \,, \quad F_{3\ 146} \,,  \quad F_{3\ 235} \,, \quad F_{3\ 236} \,, \quad F_{3\ 245} \,, \quad F_{3\ 246} \,,\nn\\
f^{a_{||_2}}{}_{b_{\bot_2} c_{\bot_2}}:&\quad f^{3}{}_{15} \,, \quad f^{3}{}_{16} \,, \quad f^{3}{}_{25} \,, \quad f^{3}{}_{26} \,, \quad f^{4}{}_{15} \,, \quad f^{4}{}_{16} \,, \quad f^{4}{}_{25} \,, \quad f^{4}{}_{26} \,,\label{variables3O5}\\
f^{a_{\bot_2}}{}_{b_{\bot_2}c_{||_2}}:&\quad f^{1}{}_{53} \,, \quad f^{1}{}_{63} \,, \quad f^{1}{}_{54} \,, \quad f^{1}{}_{64} \,, \quad f^{2}{}_{53} \,, \quad f^{2}{}_{63} \,, \quad f^{2}{}_{54} \,, \quad f^{2}{}_{64} \,, \nn\\
&\quad f^{5}{}_{13} \,, \quad f^{5}{}_{23} \,, \quad f^{5}{}_{14} \,, \quad f^{5}{}_{24} \,, \quad f^{6}{}_{13} \,, \quad f^{6}{}_{23} \,, \quad f^{6}{}_{14} \,, \quad f^{6}{}_{24} \,,\nn
\eea
for a total of 8 flux components and 24 structure constants. We compute the Bianchi identity components, and those remain as above
\beq
\hspace{-0.1in} s_{555}:\quad 3\ O_5\ \mbox{(along directions 12, 34, 56)}\ \Rightarrow\ \mbox{$p=5$ sources along directions}\ 12, 34, 56 \ . \label{p=5sources3O5}
\eeq

\subsubsection{$O_6$-plane}

We proceed as above. One $O_6$ projection allows for the following variables
\beq
O_6: \quad f^{a_{||}}{}_{b_{\bot}c_{\bot}},\ f^{a_{\bot}}{}_{b_{\bot}c_{||}},\ f^{a_{||}}{}_{b_{||}c_{||}}, \quad F_0^{(0)},\ F_2^{(1)},\ F_4^{(2)},\ F_6^{(3)},\ H^{(0)},\ H^{(2)} \ .
\eeq
Placing the $O_6$ in the first set $I=1$ along 123, we are left with the following variables
\bea
s_6:\quad F_0:&\quad F_0 \,,\nn\\
F_2:&\quad F_{2\ 14} \,, \quad F_{2\ 15} \,, \quad F_{2\ 16} \,, \quad F_{2\ 24} \,, \quad F_{2\ 25} \,, \quad F_{2\ 26} \,, \quad F_{2\ 34} \,, \quad F_{2\ 35} \,, \quad F_{2\ 36} \,, \nn\\
F_4:&\quad F_{4 \ 1245} \,, \quad F_{4 \ 1246} \,, \quad F_{4 \ 1256} \,, \quad F_{4 \ 1345} \,, \quad F_{4 \ 1346} \,, \quad F_{4 \ 1356} \,, \quad F_{4 \ 2345} \,, \nn\\
&\quad F_{4 \ 2346} \,, \quad F_{4 \ 2356} \,, \nn\\
F_6:&\quad F_{6 \ 123456} \,, \nn\\
H:&\quad H_{124} \,, \quad H_{125} \,, \quad H_{126} \,, \quad H_{134} \,, \quad H_{135} \,, \quad H_{136} \,, \quad H_{234} \,, \quad H_{235} \,, \nn\\
&\quad H_{236} \,, \quad H_{456} \,, \label{variables1O6}\\
f^{a_{||_1}}{}_{b_{\bot_1} c_{\bot_1}}:&\quad f^1{}_{45}\,, \quad f^1{}_{46}\,, \quad f^1{}_{56}\,, \quad f^2{}_{45}\,, \quad f^2{}_{46}\,, \quad f^2{}_{56}\,, \quad f^3{}_{45}\,, \quad f^3{}_{46}\,, \quad f^3{}_{56}\,, \nn\\
f^{a_{\bot_1}}{}_{b_{\bot_1} c_{||_1}}:&\quad f^4{}_{51} \,, \quad f^4{}_{61} \,, \quad f^5{}_{41}\,, \quad f^5{}_{61}\,, \quad f^6{}_{41}\,, \quad f^6{}_{51}\,, \quad f^4{}_{52} \,, \quad f^4{}_{62} \,, \quad f^5{}_{42}\,, \quad f^5{}_{62}\,, \nn\\
&\quad f^6{}_{42}\,, \quad f^6{}_{52}\,, \quad f^4{}_{53} \,, \quad f^4{}_{63} \,, \quad f^5{}_{43}\,, \quad f^5{}_{63}\,, \quad f^6{}_{43}\,, \quad f^6{}_{53}\,, \nn\\
f^{a_{||_1}}{}_{b_{||_1} c_{||_1}}:&\quad f^1{}_{23}\,, \quad f^2{}_{31}\,, \quad f^3{}_{12}\,, \nn
\eea
namely 30 fluxes and 30 structure constants (as for $p=4$). From those we compute $\d F_2$, $H \w F_0$, and deduce that the directions of possible sources
\bea
s_6:\quad 1\ O_6\ \mbox{(along directions 123)}\ \Rightarrow\ \mbox{$p=6$ sources along directions}\ 123, 145, 146 & ,  \label{p=6sources1O6} \\
156, 245, 246, 256, 345, 346, 356 & \ . \nn
\eea
This means that we can have intersecting sources. Given directions 123, the other possible sets in \eqref{p=6sources1O6} are all equivalent. We place a second $O_6$ in set $I=2$ along 145, and we are left with the following variables
\bea
s_{66}:\quad F_0:&\quad F_0 \,,\nn\\
F_2:&\quad F_{2\ 16} \,, \quad F_{2\ 24} \,, \quad F_{2\ 25} \,, \quad F_{2\ 34} \,, \quad F_{2\ 35} \,, \nn\\
F_4:&\quad F_{4 \ 1246} \,, \quad F_{4 \ 1256} \,, \quad F_{4 \ 1346} \,, \quad F_{4 \ 1356} \,, \quad F_{4 \ 2345} \,, \nn\\
F_6:&\quad F_{6 \ 123456} \,, \nn\\
H:&\quad H_{124} \,, \quad H_{125} \,, \quad H_{134} \,, \quad H_{135} \,, \quad H_{236} \,, \quad H_{456} \,, \label{variables2O6}\\
f^{a_{||_1}}{}_{b_{\bot_1} c_{\bot_1}}:&\quad f^1{}_{45}\,, \quad f^2{}_{46}\,, \quad f^2{}_{56}\,, \quad f^3{}_{46}\,, \quad f^3{}_{56}\,, \nn\\
f^{a_{\bot_1}}{}_{b_{\bot_1} c_{||_1}}:&\quad f^4{}_{51} \,, \quad f^5{}_{41}\,, \quad f^4{}_{62} \,, \quad f^5{}_{62}\,, \quad f^6{}_{42}\,, \quad f^6{}_{52}\,, \quad f^4{}_{63} \,, \quad f^5{}_{63}\,, \quad f^6{}_{43}\,, \quad f^6{}_{53}\,, \nn\\
f^{a_{||_1}}{}_{b_{||_1} c_{||_1}}:&\quad f^1{}_{23}\,, \quad f^2{}_{31}\,, \quad f^3{}_{12}\,, \nn
\eea
namely 18 fluxes and 18 structure constants. We proceed as above and obtain the following sources
\bea
s_{66}:\quad 2\ O_6\ \mbox{(along directions 123, 145)}\ \Rightarrow\ \mbox{$p=6$ sources along directions}\ 123, 145 & , \label{p=6sources2O6}\\
246, 256, 346, 356 &\ . \nn
\eea
Given 123 and 145, the other directions in \eqref{p=6sources2O6} are equivalent. We place a third $O_6$ in set $I=3$ along 256, and we are left with
\bea
s_{6666}:\quad F_q:&\quad F_0 \,, \quad F_{2\ 16} \,, \quad F_{2\ 24} \,, \quad F_{2\ 35} \,, \quad F_{4 \ 1246} \,, \quad F_{4 \ 1356} \,, \quad F_{4 \ 2345} \,, \quad F_{6 \ 123456} \,, \nn\\
H:&\quad H_{125} \,, \quad H_{134} \,, \quad H_{236} \,, \quad H_{456} \,, \nn\\
f^{a_{||_1}}{}_{bc}:&\quad f^1{}_{45}\,, \quad f^2{}_{56}\,, \quad f^3{}_{46}\,, \quad f^1{}_{23}\,, \quad f^2{}_{31}\,, \quad f^3{}_{12}\,, \label{variables3O6}\\
f^{a_{\bot_1}}{}_{b_{\bot_1} c_{||_1}}:&\quad f^4{}_{51}\,, \quad f^4{}_{63}\,, \quad f^5{}_{41}\,, \quad f^5{}_{62}\,, \quad f^6{}_{52}\,, \quad f^6{}_{43}\,, \nn
\eea
namely 12 fluxes and 12 structure constants. Proceeding as above we obtain
\bea
s_{6666}:\quad 3\ O_6\ \mbox{(along directions 123, 145, 256)}\ \Rightarrow\ \mbox{$p=6$ sources along directions} &  \label{p=6sources3O6}\\
123, 145, 256, 346 & \ .\nn
\eea
Placing an $O_6$ along the last set of directions 346 preserves exactly the same variables and sources. This was already noticed in \cite{Danielsson:2011au}: the fourth orientifold involution comes for free. As explained in Section \ref{sec:ansatz}, this is why we name the above class $s_{6666}$.

\subsubsection{$O_7$-plane}

We finally consider $p=7$. The components allowed under an $O_7$ projection are
\beq
O_7: \quad f^{a_{||}}{}_{b_{\bot}c_{\bot}},\ f^{a_{\bot}}{}_{b_{\bot}c_{||}},\ f^{a_{||}}{}_{b_{||}c_{||}}, \quad F_1^{(1)},\ F_3^{(2)},\ F_5^{(3)},\ H^{(2)} \ .
\eeq
We start with one $O_7$ along 1234 and get after projection the following list of variables
\begin{equation}
	\label{variables1O7}
	\begin{aligned}
		s_{7}:\quad F_1:&\quad F_{1 \ 1} \,, \quad F_{1 \ 2} \,, \quad F_{1 \ 3} \,,\quad F_{1 \ 4} \,, \\
		F_3:&\quad F_{3\ 125} \,, \quad F_{3\ 126} \,, \quad F_{3\ 135} \,, \quad F_{3\ 136} \,, \quad F_{3\ 145} \,,\quad F_{3\ 146} \,, \\
		&\quad F_{3\ 235} \,, \quad F_{3\ 236} \,, \quad F_{3\ 245} \,, \quad F_{3\ 246} \,, \quad F_{3\ 345} \,,\quad F_{3\ 346} \,, \\
		F_5:&\quad F_{5 \ 12356} \,, \quad F_{5 \ 12456} \,, \quad F_{5 \ 13456} \,, \quad F_{5 \ 23456} \,, \\
		H:& \quad H_{125} \,, \quad H_{126} \,, \quad H_{135} \,, \quad H_{136} \,, \quad H_{145} \,, \quad H_{146} \,, \\
		& \quad H_{235} \,, \quad H_{236} \,, \quad H_{245} \,, \quad H_{246} \,, \quad H_{345} \,, \quad H_{346} \,, \\
		f^{a_{||_1}}{}_{b_{\bot_1} c_{\bot_1}}:&\quad f^{1}{}_{56} \,, \quad f^{2}{}_{56} \,, \quad f^{3}{}_{56} \,, \quad f^{4}{}_{56} \,, \\
		f^{a_{\bot_1}}{}_{b_{\bot_1}c_{||_1}}:&\quad f^{5}{}_{16} \,, \quad f^{5}{}_{26} \,, \quad f^{5}{}_{36} \,, \quad f^{5}{}_{46} \,,  \quad f^{6}{}_{15} \,, \quad f^{6}{}_{25} \quad f^{6}{}_{35} \,,\quad f^{6}{}_{45}\,, \\	
		f^{a_{||_1}}{}_{b_{||_1}c_{||_1}}:&\quad f^{1}{}_{23} \,, \quad f^{1}{}_{24} \,,\quad f^{1}{}_{34} \,, \quad f^{2}{}_{13} \,,  \quad f^{2}{}_{14} \,,
		\quad f^{2}{}_{34} \,, \\
		&\quad f^{3}{}_{12} \,, \quad f^{3}{}_{14} \,, \quad f^{3}{}_{24} \,, \quad f^{4}{}_{12} \,, \quad f^{4}{}_{13} \,, \quad f^{4}{}_{23} \,.
	\end{aligned}
\end{equation}
There are here 32 flux components and 24 structure constants. We obtain the following possible sources
\bea
s_7:\quad 1\ O_7\ \mbox{(along 1234)}\ \Rightarrow\  \mbox{$p=7$ sources along}\ & 1234, 3456, 2456, 2356,\label{p=7sources1O7} \\
& 1456, 1356, 1256 \ .\nn
\eea
From \eqref{p=7sources1O7}, we add a second $O_7$ in the set $I=2$ along 1256 and obtain the following variables
\begin{equation}
\label{variables2O7}
\begin{aligned}
s_{77}:\quad F_1:&\quad F_{1 \ 1} \,, \quad F_{1 \ 2} \,, \\
F_3:&\quad F_{3\ 135} \,, \quad F_{3\ 136} \,, \quad F_{3\ 145} \,,\quad F_{3\ 146} \,, \\
&\quad F_{3\ 235} \,, \quad F_{3\ 236} \,, \quad F_{3\ 245} \,, \quad F_{3\ 246} \,, \\
F_5:&\quad F_{5 \ 13456} \,, \quad F_{5 \ 23456} \,, \\
H:& \quad H_{135} \,, \quad H_{136} \,, \quad H_{145} \,, \quad H_{146} \,,
\quad H_{235} \,, \quad H_{236} \,, \quad H_{245} \,, \quad H_{246} \,, \\
f^{a_{||_1}}{}_{b_{\bot_1} c_{\bot_1}}:&\quad f^{1}{}_{56} \,, \quad f^{2}{}_{56} \,, \\
f^{a_{\bot_1}}{}_{b_{\bot_1}c_{||_1}}:&\quad f^{5}{}_{16} \,, \quad f^{5}{}_{26} \,,  \quad f^{6}{}_{15} \,, \quad f^{6}{}_{25} \,, \\	
f^{a_{||_1}}{}_{b_{||_1}c_{||_1}}:&\quad f^{1}{}_{34}  \,,
\quad f^{2}{}_{34} \,,\quad f^{3}{}_{14} \,, \quad f^{3}{}_{24} \,, \quad f^{4}{}_{13} \,, \quad f^{4}{}_{23} \,.
\end{aligned}
\end{equation}
The Bianchi identity allows for the following source configurations
\beq
s_{77}:\quad 2\ O_7\ \mbox{(along 1234 and 1256)}\ \Rightarrow\ \mbox{$p=7$ sources along}\ 1234, 1256 \ . \label{p=7sources2O7}
\eeq

\subsubsection{Summary}\label{sec:sumsingle}

Our ansatz with at least one $O_p$ and a single dimensionality $p$ allows for only few possible source configurations and associated field content (variables). As explained in Section \ref{sec:ansatz}, this information allows to classify possible solutions into classes. We summarize these results in Table \ref{tab:sourcessinglesize}.
\begin{table}[H]
  \begin{center}
    \begin{tabular}{|c|c||c|c|c|}
    \hline
Solution & Number of sets & $O_p$ sets & Possible $D_p$ sets & Field \\
class & with $O_p$ & directions & directions & content \\
    \hhline{==::===}
$s_3$ & 1 $O_3$ & pt &  & \eqref{variables1O3} \\
     \hhline{--||---}
$s_4$ & 1 $O_4$ & 1 &  & \eqref{variables1O4} \\
     \hhline{--||---}
$s_5$ & 1 $O_5$ & 12 & 34, 35, 36, 45, 46, 56 & \eqref{variables1O5} \\
      \hhline{--||---}
$s_{55}$ & 2 $O_5$ & 12, 34 & 56 & \eqref{variables2O5} \\
     \hhline{--||---}
$s_{555}$ & 3 $O_5$ & 12, 34, 56 &  & \eqref{variables3O5} \\
     \hhline{--||---}
$s_{6}$ & 1 $O_6$ & 123 & 145, 146, 156, 245, 246, & \eqref{variables1O6} \\
 & & & 256, 345, 346, 356 &  \\
     \hhline{--||---}
$s_{66}$ & 2 $O_6$ & 123, 145 & 246, 256, 346, 356 & \eqref{variables2O6} \\
     \hhline{--||---}
$s_{6666}$ & 3 or 4 $O_6$ & 123, 145, 256, (346) & (346) & \eqref{variables3O6} \\
\hhline{--||---}
$s_7$ & 1 $O_7$ & 1234 & 3456, 2456, 2356, & \eqref{variables1O7} \\
 & & & 1456, 1356, 1256 &  \\
     \hhline{--||---}
$s_{77}$ & 2 $O_7$ & 1234, 1256 &  & \eqref{variables2O7} \\
    \hline
    \end{tabular}
     \caption{Source configurations allowed in our ansatz for a single dimensionality $p$, with at least one $O_p$. To each of those corresponds a list of variables or field content (structure constants, fluxes) allowed by the orientifold projection. Together, these pieces of information define a solution class.}\label{tab:sourcessinglesize}
  \end{center}
\end{table}

\subsubsection{Comments on de Sitter solutions}\label{sec:singlepdS}

The results obtained offer a new light on various observations and constraints regarding de Sitter solutions, and we pause here to comment on those. The question of finding de Sitter solutions with our ansatz and a single dimensionality $p$ has been studied in great details and is highly constrained \cite{Andriot:2018ept, Andriot:2019wrs, Andriot:2020lea}. A first result is that only $p=4$, $5$ or $6$ would allow for solutions, the other $p$'s leading to no-go theorems. In addition, we mentioned already that de Sitter solutions with parallel sources are conjectured not to exist \cite{Andriot:2019wrs}, while solutions have been found with intersecting sources for $p=5$ or 6. The details of the intersection was shown in \cite{Andriot:2017jhf} to play a role: let us add a word on this point. Of particular interest was the case of ``homogeneous overlap'', where each single set $I$ overlaps the other sets in the same manner, i.e.~the number of common directions is the same with each other set and denoted $N_o$. Different situations are named ``inhomogeneous overlap''. In general, one has $0 \leq N_o \leq p-3$, where $N_o=p-3$ means that sources are parallel. It was noticed in \cite{Andriot:2017jhf} that the choice of homogeneous overlap with $N_o = p-5$ for $p=5,6$ would simplify the equations of motion and make these two cases very analogous. As it turns out, the de Sitter solutions obtained for $p=5,6$ both verified $N_o=p-5$. Contrary to the other $p$'s, the case of $p=4$ could only be loosely constrained in \cite{Andriot:2017jhf}.\footnote{We note that our ansatz, in particular the specification to group manifolds, was not implemented in \cite{Andriot:2017jhf}. One constraint put forward in that paper for $p=4$ is (4.3): this requirement turns out to be automatically satisfied when specializing to our ansatz, since then $F_6=0, {\cal R}_{||_I}=0, {\cal R}_{||_I}^{\bot_I}\leq 0$, as shown there in section 4.3. Another constraint, (4.5), remains less trivial.} An obvious difference with $p=5,6$ is that $O_4/D_4$ only wrap one internal dimension, so they cannot overlap;\footnote{We do not consider here sources placed at angles, but only orthogonal ones. It could be that the former is however contained in the latter by projection.} few more differences were pointed out in \cite{Andriot:2017jhf}.

As we now explain, the results obtained here clarify the previous observations:
\begin{itemize}
  \item $p=4$: we obtain that the only possible source configuration is a single set of parallel sources. This emphasizes the difference with $p=5,6$, and makes it very unlikely to find any de Sitter solution (with our ansatz) for $p=4$.
  \item $p=5,6$ and homogeneous overlap: when looking at all possible source configurations, we discover to our surprise that the only possible cases of homogeneous overlap correspond to those already studied and obey $N_o=p-5$.\footnote{Although of no interest for de Sitter, the case $p=7$ also exhibits a single possibility of homogeneous overlap, that is, the source configuration of $s_{77}$ (possibly extended with an extra source in $s_{7}$), which also obeys $N_o=p-5$.} What appeared previously as an interesting choice turns out to be the only possibility. This possibility also has a relation to supersymmetry, as we will see in section \ref{sec:susy}.

      In more details, the source configurations are those of classes $s_{6666}$, $s_{55}$ or $s_{555}$; de Sitter solutions, however, cannot be found in the latter \cite{Andriot:2020wpp}, but were found in the two former. The same homogeneous overlap can also be achieved in the other classes $s_5$, $s_6$ and $s_{66}$, by turning off some of the sets with $D_p$.

  \item $p=5,6$ and inhomogeneous overlap: such source configurations can appear in classes $s_5$, $s_6$ and $s_{66}$. Only those may then provide new and different examples of de Sitter solutions.
\end{itemize}

\subsection{Source configurations and fields for multiple dimensionalities}\label{sec:multidim}

We proceed systematically to determine the solution classes in the case of multiple dimensionalities, as done previously for a single dimensionality.

\subsubsection{IIA systematics}\label{sec:IIAsys}

Given our ansatz, one can have in type IIA supergravity $p=4, 6$ sources. We first consider one $O_6$ along directions 123: the variables allowed by the projection are given in \eqref{variables1O6}. From those we compute the components of the Bianchi identity sourcing the $p=4$ sources, namely the terms $\d F_4$ and $H\w F_2$, while the analysis for $p=6$ sources is already made in \eqref{p=6sources1O6}. We obtain the following possible sources
\bea
m_{6}:\quad 1\ O_6\ \mbox{(along directions 123)}\ \Rightarrow\ & \mbox{$p=4$ sources along directions}\  4, 5, 6, \label{p=64sources1O6}\\
& \mbox{$p=6$ sources along directions}\ 123, 145, 146, 156, 245, \nn \\
& \phantom{\mbox{$p=6$ sources along directions}\ } 246, 256, 345, 346, 356 \ . \nn
\eea
From this point we have two options: adding an $O_6$ or an $O_4$. If we add an $O_6$, the list of remaining variables is given in \eqref{variables2O6}, and now the Bianchi identities give
\bea
m_{66}:\quad 2\ O_6\ \mbox{(along directions 123, 145)}\ \Rightarrow\ &\mbox{$p=4$ sources along direction}\ 6, \label{p=64sources2O6}\\
& \mbox{$p=6$ sources along directions}\ 123, 145, 246, \nn\\
& \phantom{\mbox{$p=6$ sources along directions}\ } 256, 346, 356 \ . \nn
\eea
From \eqref{p=64sources2O6}, we can add an $O_4$ along $6$, which up to relabeling, is considered below in \eqref{p=64sources2O61O4}. The other option is to add a third $O_6$. However, in that case the remaining variables are given in \eqref{variables3O6}, and the $F_4$ Bianchi identity does not allow for any $p=4$ source; this is thus not a multiple dimensionalities configuration.

If from \eqref{p=64sources1O6} we rather add an $O_4$ along 4 (all directions being equivalent), the list of remaining variables is
\bea
m_{46}:\quad F_2:&\quad F_{2\ 15} \,, \quad F_{2\ 16} \,, \quad F_{2\ 25} \,, \quad F_{2\ 26} \,, \quad F_{2\ 35} \,, \quad F_{2\ 36} \,, \nn\\
F_4:&\quad F_{4 \ 1245} \,, \quad F_{4 \ 1246} \,, \quad F_{4 \ 1345} \,, \quad F_{4 \ 1346} \,, \quad F_{4 \ 2345} \,, \quad F_{4 \ 2346} \,, \nn\\
H:&\quad H_{125} \,, \quad H_{126} \,, \quad H_{135} \,, \quad H_{136} \,, \quad H_{235} \,, \quad H_{236}\,, \label{variables1O61O4}\\
f^{a_{||_1}}{}_{b_{\bot_1} c_{\bot_1}}:&\quad f^1{}_{45}\,, \quad f^1{}_{46}\,, \quad f^2{}_{45}\,, \quad f^2{}_{46}\,, \quad f^3{}_{45}\,, \quad f^3{}_{46}\,, \nn\\
f^{a_{\bot_1}}{}_{b_{\bot_1} c_{||_1}}:&\quad f^4{}_{51} \,, \quad f^4{}_{61} \,, \quad f^5{}_{41}\,, \quad f^6{}_{41}\,, \quad f^4{}_{52} \,, \quad f^4{}_{62} \,, \quad f^5{}_{42}\,, \quad f^6{}_{42}\,, \quad f^4{}_{53} \,,  \nn\\
&\quad f^4{}_{63} \,, \quad f^5{}_{43}\,, \quad f^6{}_{43}\,, \nn
\eea
namely 18 flux components and 18 structure constants, and the Bianchi identities for $p=4$ and $p=6$ give
\bea
m_{46}:\quad 1\ O_6\ \mbox{(along 123) and}\ 1\ O_4\ \mbox{(along 4)}\ \Rightarrow\ & \mbox{$p=4$ sources along}\ 4, \label{p=64sources1O61O4}\\
& \mbox{$p=6$ sources along}\ 123, 156, 256, 356 \ .\nn
\eea
Directions being equivalent, we can add an $O_6$ along 156. The remaining variables are now
\bea
m_{466}:\quad F_2:&\quad F_{2\ 25} \,, \quad F_{2\ 26} \,, \quad F_{2\ 35} \,, \quad F_{2\ 36} \,, \nn\\
F_4:&\quad F_{4 \ 1245} \,, \quad F_{4 \ 1246} \,, \quad F_{4 \ 1345} \,, \quad F_{4 \ 1346}  \,, \nn\\
H:&\quad H_{125} \,, \quad H_{126} \,, \quad H_{135} \,, \quad H_{136} \,, \label{variables2O61O4}\\
f^{a_{||_1}}{}_{b_{\bot_1} c_{\bot_1}}:&\quad f^2{}_{45}\,, \quad f^2{}_{46}\,, \quad f^3{}_{45}\,, \quad f^3{}_{46}\,, \nn\\
f^{a_{\bot_1}}{}_{b_{\bot_1} c_{||_1}}:&\quad f^4{}_{52} \,, \quad f^4{}_{62} \,, \quad f^5{}_{42}\,, \quad f^6{}_{42}\,, \quad f^4{}_{53} \,, \quad f^4{}_{63} \,, \quad f^5{}_{43}\,, \quad f^6{}_{43}\,, \nn
\eea
namely 12 flux components and 12 structure constants, and the Bianchi identities give
\bea
m_{466}:\quad 2\ O_6\ \mbox{(along 123, 156) and}\ 1\ O_4\ \mbox{(along 4)}\ \Rightarrow\ & \mbox{$p=4$ sources along}\ 4, \label{p=64sources2O61O4}\\
& \mbox{$p=6$ sources along}\ 123, 156 \ .\nn
\eea
We cannot add any more source.

Finally, if we rather start by considering one $O_4$ along 4, the list of allowed variables is given up to relabeling in \eqref{variables1O4}. The result of the Bianchi identity of $F_4$ is given in \eqref{p=4sources1O4}, while the $F_2$ leads us to the following sources
\bea
m_{4}:\quad 1\ O_4\ \mbox{(along directions 4)}\ \Rightarrow\ & \mbox{$p=4$ sources along direction}\ 4, \label{p=46sources1O4}\\
& \mbox{$p=6$ sources along directions}\  123, 125, 126, 135, 136, \nn \\
& \phantom{ \mbox{$p=6$ sources along directions}\ } 156, 235, 236, 256, 356 \ . \nn
\eea
From there, we can add an $O_6$ along 123, bringing us back to the case \eqref{p=64sources1O61O4}. Overall, this gives 5 possible solution classes.

\subsubsection{IIB systematics}\label{sec:IIBsys}

We turn to type IIB supergravity, where our ansatz allows for $p=3, 5, 7$ sources. We first consider one $O_5$ along directions 12. The list of allowed variables is given in \eqref{variables1O5}, and $p=5$ sources are given in \eqref{p=5sources1O5}. The $F_5$ Bianchi identity does not allow for any $p=3$ source. The Bianchi identity for $F_1$ leads us to the following
\bea
m_{5}:\quad 1\ O_5\ \mbox{(along directions 12)}\ \Rightarrow\ & \mbox{$p=5$ sources along directions}\ 12, 34, 35, 36, \label{p=57sources1O5}\\
& \phantom{ \mbox{$p=5$ sources along directions}\ } 45, 46, 56, \nn\\
& \mbox{$p=7$ sources along directions}\ 2456, 2356, 2346, 2345, \nn\\
& \phantom{\mbox{$p=7$ sources along directions}\ } 1456, 1356, 1346, 1345 \ .\nn
\eea
From there, we can either add one $O_5$ or one $O_7$. Another $O_5$ can be placed without loss of generality along 34. The remaining variables are given in \eqref{variables2O5} and from the Bianchi identities, we deduce
\bea
m_{55}:\quad 2\ O_5\ \mbox{(along directions 12, 34)}\ \Rightarrow\ & \mbox{$p=5$ sources along directions}\ 12, 34, 56, \label{p=57sources2O5}\\
& \mbox{$p=7$ sources along directions}\ 2456, 2356,  1456, 1356 \ . \nn
\eea
From \eqref{p=57sources2O5}, we can add an $O_7$, which will be considered below in \eqref{p=57sources2O51O7}. We rather add a third $O_5$ along 56. The remaining variables are given in \eqref{variables3O5}, and the allowed $p=5$ sources are the same. However, the $F_1$ Bianchi identity does not allow for any $p=7$ source, which brings us back to a single dimensionality case.

If from \eqref{p=57sources1O5} we rather add an $O_7$ along 2456 (all directions being equivalent) the list of remaining variables is
\begin{equation}
\begin{aligned}
m_{57}:\quad F_1:&\quad F_{1 \ 4} \,, \quad F_{1 \ 5} \,, \quad F_{1 \ 6} \,, \\
F_3:&\quad F_{3\ 145} \,, \quad F_{3\ 146} \,, \quad F_{3\ 156} \,, \quad F_{3\ 234} \,, \quad F_{3\ 235} \,,\quad F_{3\ 236} \,, \\
F_5:&\quad F_{5 \ 12345} \,, \quad F_{5 \ 12346} \,, \quad F_{5 \ 12356} \,, \\
H:& \quad H_{124} \,, \quad H_{125} \,, \quad H_{126} \,, \quad H_{345} \,, \quad H_{346} \,, \quad H_{356} \,, \label{variablesO5O7}\\
f^{a_{||_1}}{}_{b_{\bot_1} c_{\bot_1}}:&\quad f^{1}{}_{34} \,, \quad f^{1}{}_{35} \,, \quad f^{1}{}_{36} \,, \\
f^{a_{\bot_1}}{}_{b_{\bot_1}c_{||_1}}:&\quad f^{3}{}_{14} \,, \quad f^{3}{}_{15} \,, \quad f^{3}{}_{16} \,, \quad f^{4}{}_{13} \,,  \quad f^{5}{}_{13} \,, \quad f^{6}{}_{13} \,, \\
f^{a_{||_2}}{}_{b_{||_2}c_{||_2}}:&\quad f^{2}{}_{45} \,, \quad f^{2}{}_{46} \,, \quad f^{2}{}_{56} \,,\quad f^{4}{}_{25} \,, \quad f^{4}{}_{26} \,,  \quad f^{5}{}_{24} \,, \quad f^{5}{}_{26}
\quad f^{6}{}_{24} \,,\quad f^{6}{}_{25} \,.
\end{aligned}
\end{equation}
There are 18 flux components and 18 structure constants, and the Bianchi identities allow the following sources
\bea
m_{57}:\quad 1\ O_5\ \mbox{(along 12) and}\ 1\ O_7\ \mbox{(along 2456)}\ \Rightarrow\ & \mbox{$p=5$ sources along}\ 12, 34, 35, 36, \label{p=57sourcesO5O7}\\
& \mbox{$p=7$ sources along}\ 2456, 1356, 1346, 1345 \ .\nn
\eea
From there we add an $O_5$ along 34 (all directions being equivalent), and obtain the following list of variables
\begin{equation}
\label{variables2O51O7}
\begin{aligned}
m_{5577}:\quad F_1:& \quad F_{1 \ 5} \,, \quad F_{1 \ 6} \,, \\
F_3:&\quad F_{3\ 145} \,, \quad F_{3\ 146} \,, \quad F_{3\ 235} \,,\quad F_{3\ 236} \,, \\
F_5:&\quad F_{5 \ 12345} \,, \quad F_{5 \ 12346} \,, \\
H:& \quad H_{125} \,, \quad H_{126} \,, \quad H_{345} \,, \quad H_{346} \,, \\
f^{a_{||_1}}{}_{b_{\bot_1} c_{\bot_1}}:& \quad f^{1}{}_{35} \,, \quad f^{1}{}_{36} \,, \\
f^{a_{\bot_1}}{}_{b_{\bot_1}c_{||_1}}:& \quad f^{3}{}_{15} \,, \quad f^{3}{}_{16} \,, \quad f^{5}{}_{13} \,, \quad f^{6}{}_{13} \,, \\
f^{a_{||_2}}{}_{b_{||_2}c_{||_2}}:&\quad f^{2}{}_{45} \,, \quad f^{2}{}_{46} \,,\quad f^{4}{}_{25} \,, \quad f^{4}{}_{26} \,,  \quad f^{5}{}_{24} \,,
\quad f^{6}{}_{24} \,.
\end{aligned}
\end{equation}
This gives 12 flux components and 12 structure constants, and Bianchi identities give
\bea
m_{5577}:\quad 2\ O_5\ \mbox{(along 12, 34) and}\ 1\ O_7\ \mbox{(along 2456)}\ \Rightarrow\ & \mbox{$p=5$ sources along}\ 12, 34, \label{p=57sources2O51O7}\\
& \mbox{$p=7$ sources along}\ 2456, 1356 \ .\nn
\eea
The last thing we can add from here is an $O_7$ along 1356. Doing so, one obtains the exact same list of variables as in \eqref{variables2O51O7}, so in that sense, the last $O_7$ ``comes for free''. This explains why the solution class was already called $m_{5577}$. The Bianchi identities give the same sources as above, and one cannot add any more source. From \eqref{p=57sourcesO5O7}, we could also have added an $O_7$ along 1356. In this case, one ends up again with the same variables as in \eqref{variables2O51O7}. The Bianchi identities therefore give the same sources, namely
\bea
\hspace{-0.1in} m_{5577}:\quad 1\ O_5\ \mbox{(along 12) and}\ 2\ O_7\ \mbox{(along 2456, 1356)}\ \Rightarrow\ & \mbox{$p=5$ sources along}\ 12, 34, \label{p=57sources1O52O7}\\
& \mbox{$p=7$ sources along}\ 2456, 1356 \ .\nn
\eea
From there, we can only add an $O_5$ along 34 to reach the case discussed above. It is obvious that this extra $O_5$ also ``comes for free''.

Starting now with one $O_7$ along 1234, the list of variables is given in \eqref{variables1O7} and the corresponding $p=7$ sources in \eqref{p=7sources1O7}. Possible sources are as follows
\bea
m_{7}:\quad 1\ O_7\ \mbox{(along 1234)} \Rightarrow\ & \mbox{$p=3$ sources at points},\nn\\
& \mbox{$p=5$ sources along}\ 15, 16, 25, 26, 35, 36, 45, 46, \label{p=75sourcesO7}\\
& \mbox{$p=7$ sources along}\ 1234, 3456, 2456, 2356, 1456, 1356, 1256 \ .\nn
\eea
From there, we can add an $O_5$, bringing us back to \eqref{p=57sourcesO5O7}. We rather add a second $O_7$ along 1256 and obtain the variables given in \eqref{variables2O7}. The Bianchi identities allow for the following sources
\bea
m_{77}:\quad 2\ O_7\ \mbox{(along 1234 and 1256)} \Rightarrow\ & \mbox{$p=3$ sources at points},\nn\\
& \mbox{$p=5$ sources along}\ 46, 45, 36, 35, \label{p=75sources2O7}\\
& \mbox{$p=7$ sources along}\ 1234, 1256 \ .\nn
\eea
From there, we can add an $O_5$, bringing us back to \eqref{p=57sources1O52O7}. We will also see below that the addition of an $O_3$ from \eqref{p=75sourcesO7} and \eqref{p=75sources2O7} is not possible with multiple dimensionalities.

We finally consider from the start an $O_3$. The allowed variables are given in \eqref{variables1O3}. Having $p=3$ sources is then possible as indicated in \eqref{p=3sources1O3}. However, the Bianchi identities of $F_1$ and $F_3$ do not allow for any $p=7$ nor $p=5$ source, bringing us to a single dimensionality case. Overall, we find 6 distinct solution classes (not counting the ``free'' $O_7$ or $O_5$).

\subsubsection{Summary}\label{sec:summult}

We summarize in Table \ref{tab:sourcesmulsizeIIA} and \ref{tab:sourcesmulsizeIIB} the source configurations with multiple dimensionalities and associated field content (allowed variables) obtained in Section \ref{sec:IIAsys} and \ref{sec:IIBsys}. This information defines solution classes.

\begin{table}[H]
  \begin{center}
    \begin{tabular}{|c|c||c|c|c|}
    \hline
Solution & Number of sets & $O_p$ sets & Possible $D_p$ sets & Field \\
class & with $O_p$ & directions & directions & content \\
    \hhline{==::===}
$m_4$ & 1 $O_4$ & 4 & 123, 125, 126, 135, 136, & \eqref{variables1O4} \\
 & & & 156, 235, 236, 256, 356 &  \\
     \hhline{--||---}
$m_{46}$ & 1 $O_6$, 1 $O_4$ & 123, 4 & 156, 256, 356  & \eqref{variables1O61O4} \\
     \hhline{--||---}
$m_{466}$ & 2 $O_6$, 1 $O_4$ & 123, 156, 4 &  & \eqref{variables2O61O4} \\
    \hhline{--||---}
$m_6$ & 1 $O_6$ & 123 & 4,5,6 &  \\
 & & & 145, 146, 156, 245, 246, & \eqref{variables1O6} \\
 & & & 256, 345, 346, 356 &  \\
     \hhline{--||---}
$m_{66}$ & 2 $O_6$ & 123, 145 & 6, 246, 256, 346, 356 & \eqref{variables2O6} \\
    \hline
    \end{tabular}
     \caption{Source configurations in type IIA supergravity allowed by our ansatz for multiple dimensionalities with at least one $O_p$, together with the associated list of allowed variables or field content. These two pieces of information form together a solution class.}\label{tab:sourcesmulsizeIIA}
  \end{center}
\end{table}

\begin{table}[H]
  \begin{center}
    \begin{tabular}{|c|c||c|c|c|}
    \hline
Solution & Number of sets & $O_p$ sets & Possible $D_p$ sets & Field \\
class & with $O_p$ & directions & directions & content \\
    \hhline{==::===}
$m_5$ & 1 $O_5$ & 12 & 34, 35, 36, 45, 46, 56 &  \\
 & & & 2456, 2356, 2346, 2345, & \eqref{variables1O5} \\
 & & & 1456, 1356, 1346, 1345 &  \\
    \hhline{--||---}
$m_{55}$ & 2 $O_5$ & 12, 34 & 56 & \eqref{variables2O5} \\
 & & & 2456, 2356, 1456, 1356 &  \\
    \hhline{--||---}
$m_{57}$ & 1 $O_5$, 1 $O_7$ & 12, 2456 & 34, 35, 36 & \eqref{variablesO5O7} \\
 & & & 1356, 1346, 1345 &  \\
    \hhline{--||---}
$m_{5577}$ & 1 (or 2) $O_5$, & 12, (34), & (34), (1356) &  \eqref{variables2O51O7} \\
 & 1 (or 2) $O_7$ & 2456, (1356) &  &  \\
    \hhline{--||---}
$m_7$ & 1 $O_7$ & 1234 & 15, 16, 25, 26, 35, 36, 45, 46 &  \\
 & & & {\rm pt}, 3456, 2456, 2356,  & \eqref{variables1O7} \\
 & & & 1456, 1356, 1256 &  \\
    \hhline{--||---}
$m_{77}$ & 2 $O_7$ & 1234, 1256 & {\rm pt}, 35, 36, 45, 46 & \eqref{variables2O7} \\
    \hline
    \end{tabular}
     \caption{Source configurations in type IIB supergravity allowed by our ansatz for multiple dimensionalities with at least one $O_p$, together with the associated list of allowed variables or field content. These two pieces of information form together a solution class. The $p=3$ sources are located at a point, denoted ${\rm pt}$. The cases of 1 $O_5$, 2 $O_7$, or 2 $O_5$, 1 $O_7$, or 2 $O_5$, 2 $O_7$ form just one class: they all give the same allowed variables and the same four sources (up to their $O_p/D_p$ nature).}\label{tab:sourcesmulsizeIIB}
  \end{center}
\end{table}

\subsection{T-duality and supersymmetry}\label{sec:Tdsusy}

Having identified all solution classes within our ansatz with at least one $O_p$, a few comments are in order regarding T-duality relations and the unbroken supersymmetries.

\subsubsection{T-duality}\label{sec:Td}

Some source configurations are T-dual to others. For example, the configuration of $m_{46}$ with 3 sets, i.e. 1 $O_4$ along 4, 1 $O_6$ along 123 and 1 $D_6$ along 156 of \eqref{p=64sources1O61O4} is T-dual to the configuration of $s_{55}$ of \eqref{p=5sources2O5}. This can be seen by performing a T-duality along direction 1 and relabeling 2 $\leftrightarrow$ 4. Turning the $D_p$-branes into $O_p$-planes, we get that the sources of $m_{466}$ in \eqref{p=64sources2O61O4} are T-dual to those of $s_{555}$ in \eqref{p=5sources3O5}. Similarly, the configuration of $m_{5577}$, i.e. 2 $O_5$ along 12, 34 and 2 $O_7$ along 1356, 2456, mentioned below \eqref{p=57sources2O51O7}, is T-dual to the configuration of $s_{6666}$ with 4 $O_6$ given below \eqref{p=6sources3O6}. One should perform the T-duality along 6, followed by a relabeling $1\leftrightarrow 2, \; 3\leftrightarrow 6$. Finally, T-dual source configurations with 2 sets are given in \eqref{Td2Op} and \eqref{Td1Op1Dp}.

One may then wonder whether beyond the source configurations, the solution classes as a whole are actually T-dual. This would mean that the allowed variables or fields are transformed into each other by T-duality. This is however not the case, as one can verify using the standard T-duality rules on flux indices \cite{Shelton:2005cf}. This point was made in \cite{Caviezel:2009tu} regarding the de Sitter solution with $O_5 \& O_7$ (belonging to $m_{5577}$): T-duality on the fields leads to non-geometric $Q$-fluxes, which are not allowed in our settings. So the solutions cannot be concluded to be T-dual. Anticipating on our solutions, the same will be true here. In type IIB, we also find de Sitter solutions in $m_{5577}$. All of them have either a $f^2{}_{46}$ or a $f^4{}_{26}$ non-zero. T-duality then gives $f^a{}_{b6} \rightarrow  Q_{b}{}^{a6}$, i.e.~generates a non-geometric $Q$-flux, so our solutions are not T-dual to geometric ones and are truly new. Similarly in IIA, our de Sitter solutions in $m_{46}$ admit $f^4{}_{15}, f^4{}_{16}$ non-zero. The T-duality rule is $f^a{}_{1c} \rightarrow  - Q_{c}{}^{a1}$, so again, a non-geometric $Q$-flux would be generated from our solutions, so they are not T-dual to known geometric ones and are truly new.\footnote{It is conceivable that the problematic structure constants $f^a{}_{1c}$ disappear via a rotation among e.g.~directions 1 and 2, applied to the Maurer-Cartan equations. However, such a rotation may in turn transform a single set of $D_6$ along 156 into two sets along 156 and 256; see appendix \ref{ap:chgbasis}. As a consequence, the rewritten solution, assuming it is still a solution, would have a source configuration which is not T-dual anymore to the one in $s_{55}$, when performing T-duality along the new, rotated, directions. This is another way to view the obstruction.} Even though the complete solution classes are not T-dual, the source configurations still are, and this will be useful for the supersymmetry analysis that we now turn to.

\subsubsection{Unbroken supersymmetries}\label{sec:susy}

We have considered various configurations of static branes and orientifolds. It would be interesting to determine whether those preserve some supersymmetry (i.e.~are ``mutually BPS''), first to have a chance to obtain a 4d supersymmetric effective theory, but also to avoid possible (open string) instabilities. One set of parallel branes breaks half of type II supersymmetries. Two orthogonal sets break a further half, i.e.~a quarter, if and only if their total number of Neumann-Dirichlet (ND) boundary conditions, ${\cal N}_{ND}$, is a multiple of 4: ${\cal N}_{ND} = 4i >0$ for some positive integer $i$. In other words, the total number of directions belonging to one set and not to the other, which is always an even number, should be a multiple of 4. If this does not hold, supersymmetry is broken.

Given we work in the orthonormal (co)frame, our sets are orthogonal, it is then straightforward to perform the supersymmetry check. Our configurations are along the 3 external space dimensions, and have for each set $p-3$ internal dimensions; only the latter can be ND. Let us first consider configurations of single dimensionality $p\geq4$. For $p=4$, two different sets only have ${\cal N}_{ND} = 1+1 =2$ so intersecting $p=4$ sources break supersymmetry. For $p\geq 5$, let us denote $N_o\geq 0$ the number of common internal directions of two orthogonal sets. Two conditions need to be obeyed:
\beq
4i = 2 (p-3- N_o) \ ,\qquad 2 (p-3- N_o) + N_o \leq 6 \ .
\eeq
The first one is the condition for unbroken supersymmetry; the second one is the requirement that the total of internal dimensions of the two sets is smaller than 6. We deduce that $1 \leq i \leq \frac{3}{2}$, and conclude that we must have $i=1$, i.e.~$N_o=p-5$, to preserve supersymmetry. Since there is no other possible value for $N_o$, this means that this value of $N_o$ must hold for all pairs of sets, in other words there is an homogenous overlap. As discussed in Section \ref{sec:singlepdS}, the value $N_o=p-5$ is precisely the one for which solutions with homogeneous overlap can be found here, as initially advocated from equations of motion for de Sitter solutions in \cite{Andriot:2017jhf}. This result also implies that configurations with inhomogeneous overlap break supersymmetry.

We extend the analysis to source configurations with multiple dimensionalities. We apply the rule ${\cal N}_{ND} = 4i >0$ to pairs of sources in such configurations, allowing us to determine those that leave some unbroken supersymmetry. A remaining question is the amount of supersymmetry that is left unbroken. Each pair of orthogonal sets verifying ${\cal N}_{ND} = 4i >0$ preserves a quarter of the initial amount. It is well-known that adding a third source as we do with $D_5/O_5$ (along 12, 34, 56) or $D_6/O_6$ (along 123, 145, 256) breaks supersymmetry by a further half. For $D_6/O_6$ one can add the fourth source along 346 for free. In those two cases, one then preserves ${\cal N}=1$ in 4d. Thanks to the T-duality relation of these source configurations to those with $O_4/O_6$ and $O_5/O_7$ (see Section \ref{sec:Td}), we deduce the number of preserved supersymmetries for configurations with multiple dimensionalities. The results are summarized in Table \ref{tab:sourcessusy}.

\begin{table}[H]
  \begin{center}
    \begin{tabular}{|c||c|c|}
    \hline
Solution & Sources directions & 4d ${\cal N}$ of \\
class & allowing for unbroken SUSY & preserved SUSY \\
    \hhline{=::==}
$s_3$ & pt & 4 \\
     \hhline{-||--}
$s_4$ & 1 & 4 \\
     \hhline{-||--}
$s_5$ & 12, (34, 56) & 4,(2,1) \\
\hhline{-||--}
 $s_{55}$ & 12, 34, (56) & 2,(1) \\
     \hhline{-||--}
$s_{555}$ & 12, 34, 56 & 1 \\
\hhline{-||--}
$s_{6}$ & 123, (145, 256, 346) & 4,(2,1)\\
\hhline{-||--}
$s_{66}$ & 123, 145, (256, 346) & 2,(1)\\
\hhline{-||--}
$s_{6666}$ & 123, 145, 256, (346) & 1\\
\hhline{-||--}
$s_{7}$ & 1234, (1256, 3456) & 4,(2,1) \\
\hhline{-||--}
$s_{77}$ & 1234, 1256 & 2 \\
    \hhline{=::==}
$m_4$ & 4, 123, (156) & 2,(1) \\
\hhline{-||--}
$m_{46}$ & 4, 123, (156) & 2,(1)\\
\hhline{-||--}
$m_{466}$ & 4, 123, 156 & 1\\
\hhline{-||--}
$m_6$ & 4, 123, (156) & 2,(1)\\
\hhline{-||--}
$m_{66}$ & 6, 123, 145 & 1\\
    \hhline{=::==}
$m_5$ & 12, 2456, (34, 1356) & 2,(1)\\
\hhline{-||--}
$m_{55}$ & 12, 34, 2456, (1356) & 1\\
\hhline{-||--}
$m_{57}$ & 12, 2456, (34, 1356) & 2,(1)\\
\hhline{-||--}
$m_{5577}$ & 12, 34, 2456, (1356) & 1\\
 & 12, 2456, 1356, (34) & 1\\
\hhline{-||--}
$m_7$ & pt, 1234, (1256, 3456) & 4,(2,1)\\
 & 15, 1234, (26, 3456) & 2,(1)\\
\hhline{-||--}
$m_{77}$ & pt, 1234, 1256 & 2\\
 & 35, 1234, 1256, (46) & 1\\
    \hline
    \end{tabular}
     \caption{Internal directions of the source sets that allow for unbroken supersymmetry, for each solution class previously identified. The corresponding number ${\cal N}$ of preserved supersymmetries in 4d is given. The sets in parentheses are optional in the class, to which the amount ${\cal N}$ in parentheses corresponds. Some classes allow for different supersymmetry-preserving configurations, then specified on distinct rows.}\label{tab:sourcessusy}
  \end{center}
\end{table}
We see through Table \ref{tab:sourcessusy} that among all possible solution classes, it is eventually only few redundant source configurations that appear and preserve supersymmetry.

\section{Solutions}

We classified in Section \ref{sec:classif} the possible solutions with our ansatz into a list of solution classes, summarized in Section \ref{sec:sumsingle} and \ref{sec:summult}. In this section, we look for new solutions in those classes, most of our efforts being dedicated to de Sitter ones.

\subsection{Procedure to find solutions}\label{sec:proc}

To find a solution with a 4d maximally symmetric spacetime within our ansatz, one has to solve the equations listed in Appendix \ref{ap:eq}, together with the constraints
\begin{equation}
\label{constraints}
\R_4 \ {\rm sign} \,, \qquad T_{10}^{J\, =\, D_p \, {\rm only}} \leq 0 \ ,
\end{equation}
where the sign constraint on the 4d curvature is e.g.~$\R_4 >0$ for de Sitter, and the second constraint is about possible source sets without $O_p$. This is done numerically as further detailed in Section \ref{sec:code}.

Prior to this, one should start by choosing a solution class among those listed in Table \ref{tab:sourcessinglesize}, \ref{tab:sourcesmulsizeIIA} and \ref{tab:sourcesmulsizeIIB}. This determines the source configuration as well as the allowed variables to be considered non-zero in the equations; considering that list of variables amounts to ensure that the orientifold projection is satisfied. Given the source configuration, the labeling of the sets can be fixed, and the transverse volume forms as well as the internal energy momentum tensor can be determined; those quantities are needed in the equations to solve. We provide in the following two examples of single dimensionality (and a priori inhomogeneous overlap), but the procedure extends to cases of multiple dimensionalities.

\begin{itemize}
  \item {\bf Category $s_5$}:

  We label the source sets as in Table \ref{tab:sourcesIO51}.
  \begin{table}[H]
  \begin{center}
    \begin{tabular}{|c||c||c|c|c|c|c|c|c|c|c|}
    \hline
Set $I$ & Sources & \multicolumn{9}{c|}{Space dimensions} \\
\hhline{~||~||---------}
 &  & \multicolumn{3}{c|}{4d} & 1 & 2 & 3 & 4 & 5 & 6 \\
    \hhline{=::=::=========}
1 & $O_5, (D_5)$ & $\otimes$ & $\otimes$ & $\otimes$ & $\otimes$ & $\otimes$ & & & & \\
     \hhline{-||-||---------}
2 & $(D_5)$ & $\otimes$ & $\otimes$ & $\otimes$ & & & $\otimes$ & $\otimes$ & & \\
      \hhline{-||-||---------}
3 & $(D_5)$ & $\otimes$ & $\otimes$ & $\otimes$ & & & $\otimes$ & & $\otimes$ & \\
       \hhline{-||-||---------}
4 & $(D_5)$ & $\otimes$ & $\otimes$ & $\otimes$ & & & $\otimes$ & & & $\otimes$ \\
 	   \hhline{-||-||---------}
5 & $(D_5)$ & $\otimes$ & $\otimes$ & $\otimes$ & & & & $\otimes$ & $\otimes$ & \\
       \hhline{-||-||---------}
6 & $(D_5)$ & $\otimes$ & $\otimes$ & $\otimes$ & & & & $\otimes$ & & $\otimes$ \\
      \hhline{-||-||---------}
7 & $(D_5)$ & $\otimes$ & $\otimes$ & $\otimes$ & & & & & $\otimes$ & $\otimes$ \\
    \hline
    \end{tabular}
     \caption{$O_5/D_5$ source configuration in the class $s_5$. Sources in parentheses are not mandatory.}\label{tab:sourcesIO51}
  \end{center}
\end{table}
This leads us to the following volume forms
\beq
\begin{aligned}
	\label{setsIO51}
& I=1:\quad {\rm vol}_{||_1} = e^1 \w e^2 \ ,\ {\rm vol}_{\bot_1} = e^3 \w e^4 \w e^5 \w e^6 \ ,\\
& I=2:\quad {\rm vol}_{||_2} = e^3 \w e^4 \ ,\ {\rm vol}_{\bot_2} = e^1 \w e^2 \w e^5 \w e^6 \ ,\\
& I=3:\quad {\rm vol}_{||_3} = e^3 \w e^5 \ ,\ {\rm vol}_{\bot_3} = -\, e^1 \w e^2 \w e^4 \w e^6 \ ,\\
& I=4:\quad {\rm vol}_{||_4} = e^3 \w e^6 \ ,\ {\rm vol}_{\bot_4} = e^1 \w e^2 \w e^4 \w e^5 \ ,\\
& I=5:\quad {\rm vol}_{||_5} = e^4 \w e^5 \ ,\ {\rm vol}_{\bot_5} = e^1 \w e^2 \w e^3 \w e^6 \ ,\\
& I=6:\quad {\rm vol}_{||_6} = e^4 \w e^6 \ ,\ {\rm vol}_{\bot_6} = -\, e^1 \w e^2 \w e^3 \w e^5 \ ,\\
& I=7:\quad {\rm vol}_{||_7} = e^5 \w e^6 \ ,\ {\rm vol}_{\bot_7} = e^1 \w e^2 \w e^3 \w e^4 \ ,
\end{aligned}
\eeq
where the $-1$ in ${\rm vol}_{\bot_I}$ are due to the 6d orientation. The source energy momentum tensor is given by
\bea
\quad T_{ab} = \rm{diag} \bigg(&\frac{T_{10}^1}{6},\ \frac{T_{10}^1}{6},\ \frac{T_{10}^2 +T_{10}^3 +T_{10}^4}{6},\ \frac{T_{10}^2 + T_{10}^5 + T_{10}^6}{6},  \\
& \frac{T_{10}^3 + T_{10}^5 + T_{10}^7}{6},\ \frac{T_{10}^4 + T_{10}^6 + T_{10}^7}{6} \bigg) \nn\ .
\eea
The list of variables is given in \eqref{variables1O5}.  The equations to solve are given in appendix \ref{ap:eq}, setting $T_{10}^{(3)_I} = T_{10}^{(7)_I} = 0$, thus giving $T_{10} =T_{10}^{(5)}$.\footnote{These equations are formally the same as in \cite{Andriot:2020wpp}, up to the volume forms and the $T_{ab}$. One should pay attention to the typo in that paper: the missing sign $\varepsilon_5$. We corrected this here.} As in \cite{Andriot:2020wpp}, simplifications occur in the equations. Because of the $O_5$ projection and the fluxes being constant, the e.o.m. for $F_1$ and the BI for $F_5$ are trivially satisfied: these six-forms vanish identically. In addition, following the reasoning of Section 3.2 of \cite{Andriot:2019wrs}, the off-diagonal ${}_{a_{||_I}b_{\bot_I}}$ Einstein equations for a set $I$ with an $O_5$ projection are trivially satisfied. With such a projection for $I=1$, we are left with the Einstein equations along the blocks $12$ and $3456$, i.e.~13 equations.

  \item {\bf Category $s_{66}$}:

  Similarly, we label the source sets as in Table \ref{tab:sourcesIO62}.
\begin{table}[H]
  \begin{center}
    \begin{tabular}{|c||c||c|c|c|c|c|c|c|c|c|}
    \hline
Set $I$ & Sources & \multicolumn{9}{c|}{Space dimensions} \\
\hhline{~||~||---------}
 &  & \multicolumn{3}{c|}{4d} & 1 & 2 & 3 & 4 & 5 & 6 \\
    \hhline{=::=::=========}
1 & $O_6, (D_6)$ & $\otimes$ & $\otimes$ & $\otimes$ & $\otimes$ & $\otimes$ & $\otimes$ & & & \\
     \hhline{-||-||---------}
2 & $O_6, (D_6)$ & $\otimes$ & $\otimes$ & $\otimes$ & $\otimes$ & & & $\otimes$ & $\otimes$ & \\
      \hhline{-||-||---------}
3 & $(D_6)$ & $\otimes$ & $\otimes$ & $\otimes$ & & $\otimes$ & & $\otimes$ & & $\otimes$ \\
      \hhline{-||-||---------}
4 & $(D_6)$ & $\otimes$ & $\otimes$ & $\otimes$ & & $\otimes$ & & & $\otimes$ & $\otimes$ \\
 	   \hhline{-||-||---------}
5 & $(D_6)$ & $\otimes$ & $\otimes$ & $\otimes$ & & & $\otimes$ & $\otimes$ & & $\otimes$ \\
      \hhline{-||-||---------}
6 & $(D_6)$ & $\otimes$ & $\otimes$ & $\otimes$ & & & $\otimes$ & & $\otimes$ & $\otimes$ \\
    \hline
    \end{tabular}
     \caption{$O_6/D_6$ source configuration in the class $s_{66}$. Sources in parentheses are not mandatory.}\label{tab:sourcesIO62}
  \end{center}
\end{table}
This leads us to the following volume forms
\beq
\begin{aligned}
	\label{setsIO62}
& I=1:\quad {\rm vol}_{||_1} = e^1 \w e^2 \w e^3 \ ,\ {\rm vol}_{\bot_1} = e^4 \w e^5 \w e^6 \ ,\\
& I=2:\quad {\rm vol}_{||_2} = e^1 \w e^4 \w e^5 \ ,\ {\rm vol}_{\bot_2} = e^2 \w e^3 \w e^6 \ ,\\
& I=3:\quad {\rm vol}_{||_3} = e^2 \w e^4 \w e^6 \ ,\ {\rm vol}_{\bot_3} = e^1 \w e^3 \w e^5 \ ,\\
& I=4:\quad {\rm vol}_{||_4} = e^2 \w e^5 \w e^6 \ ,\ {\rm vol}_{\bot_4} = -\, e^1 \w e^3 \w e^4 \ ,\\
& I=5:\quad {\rm vol}_{||_5} = e^3 \w e^4 \w e^6 \ ,\ {\rm vol}_{\bot_5} = -\, e^1 \w e^2 \w e^5 \ ,\\
& I=6:\quad {\rm vol}_{||_6} = e^3 \w e^5 \w e^6 \ ,\ {\rm vol}_{\bot_6} = e^1 \w e^2 \w e^4 \ .
\end{aligned}
\eeq
The source energy momentum tensor is given by
\bea
\quad T_{ab} = \rm{diag} \bigg(&\frac{T_{10}^1 +T_{10}^2}{7},\ \frac{T_{10}^1 +T_{10}^3 +T_{10}^4}{7},\ \frac{T_{10}^1 +T_{10}^5 +T_{10}^6}{7},\ \nn\\
&\frac{T_{10}^2 + T_{10}^3 + T_{10}^5}{7}, \ \frac{T_{10}^2 + T_{10}^4 + T_{10}^6}{7},\ \frac{T_{10}^3 + T_{10}^4 + T_{10}^5 + T_{10}^6}{7} \bigg) \nn\ .
\eea
The list of variables is given in \eqref{variables2O6}. The equations to solve are those of appendix \ref{ap:eq}, setting $T_{10}^{(4)_I} = T_{10}^{(8)_I} = 0$, thus giving $T_{10} =T_{10}^{(6)}$. Regarding simplifications in our setting, let us mention that the $F_0$ BI is automatically satisfied given our ansatz. As for $p=5$, all off-diagonal ${}_{a_{||_I}b_{\bot_I}}$ Einstein equations for a set $I$ with an $O_6$ projection are trivially satisfied. Here the projection is for $I=1,2$, leaving the Einstein equations along $11$, $66$, and the blocks $23$ and $45$, giving 8 equations.

\end{itemize}

Last but not least, once a solution to the equations of Appendix \ref{ap:eq} and the constraints \eqref{constraints} is found, a remaining task is to verify that the group manifold $\mmm$, for which structure constants have been found, is compact. We will come back in detail in the companion paper \cite{Andriot:2022yyj} to this matter of the compactness.

Let us add that the equations to solve enjoy an overall scaling symmetry, associated to a parameter $\lambda$ in \cite[Sec. 4.2]{Andriot:2020wpp}. To help in the search for solutions, this freedom is sometimes fixed at first, by giving a value to one of the variables, e.g.~$g_s\, T_{10}^1=10$. One should keep in mind that all variable values can later be rescaled at will, in the absence of possible further constraints, such as requiring the supergravity solution to be a classical string background. This is true in particular for the values of the curvatures $|{\cal R}_4|$ and $|{\cal R}_6|$, that can for instance be lowered.

\subsection{Numerical code}\label{sec:code}

The numerical code {\tt MaxSymSolSearch} ({\tt MSSS}) used to find new solutions follows the one used and described in \cite{Andriot:2020wpp}, with several extensions and improvements; we refer to that paper for more technical details. As explained in Section \ref{sec:proc}, the first step is to choose a solution class. This amounts to indicate to the code the sets of sources, their directions, and the sets in which there are $O_p$. From this information, the code deduces the theory to use (IIA or IIB supergravity), and the list of allowed variables under the orientifold projection. As shown in Section \ref{sec:proc}, the code also works out the volume forms and the energy momentum tensor components. From there, the code obtains all equations to solve, listed in Appendix \ref{ap:eq}: the equations of motion (e.o.m.) for the fluxes, their Bianchi identities, the Einstein equations, the dilaton e.o.m., and the Jacobi identities on the structure constants. These equations are derived in components, i.e~as scalar, quadratic and algebraic equations. As indicated in Section \ref{sec:proc}, simplifications can occur depending on the class, because some equations are trivially satisfied. The remaining number $N_{{\rm eq}}$ of non-trivial equations varies according to the class,\footnote{We note a typo in \cite{Andriot:2020wpp} on this matter: the number of non-trivial equations for $s_{55}$ is $N_{{\rm eq}}=56$.} as well as the number of variables.

At this stage, the code is ready to solve the equations, i.e.~look for a solution. Prior to this, one can indicate at this point further specifications on the solution Ansatz. This typically amounts to set to zero, or to some other value, some of the variables, either to help the code or to look for more specific solutions within a class. Two kinds of Ans\"atzen are often used. The first one consists in leaving almost all allowed variables free, and letting the program look for solutions. This has the potential drawback to leave too much freedom, potentially diluting too much solutions, if any, in the parameter space, such that the program may not find them. The second Ansatz consists in starting close to another class for which we know there exist solutions. This means setting many of the extra (compared to the other class) variables to zero, leaving free only few of them. This may have the drawback of being too restrictive, and then missing possible solutions in the starting class. Finally, a more refined approach can be taken to choose an Ansatz, for instance setting progressively one by one some variables to zero, to simplify step by step a solution.

Once the Ansatz is further specified, the code tries to solve the set of $N_{{\rm eq}}$ equations. To that end, we make use of minimisation techniques and proceed as follows. Every equation is written in the form $E_i = 0$, $i=1,...,N_{{\rm eq}}$, and we consider the following quantity
\begin{equation}
S = \sum_{i=1}^{N_{{\rm eq}}} (E_i)^2 \,, \qquad \text{such that} \qquad (S = 0) \; \Longleftrightarrow \; (E_i = 0 \quad \forall i = 1,..., N_{{\rm eq}}) \,.
\end{equation}
We then use algorithms implemented in {\tt Wolfram Mathematica} to minimise $S$ to zero. As indicated in Section \ref{sec:proc}, we add at this point the constraints \eqref{constraints}. In particular, we ask for the condition ${\cal R}_4 > \epsilon$, $-\tilde{\epsilon} < {\cal R}_4 < \tilde{\epsilon}$ or ${\cal R}_4 < -\epsilon$ (for instance $\epsilon = 10^{-3},\, \tilde{\epsilon} = 10^{-7}$) depending on whether we look for de Sitter, Minkowski or anti-de Sitter solutions. A solution is considered valid if one reaches the standard precision $S \sim 10^{-30} \ll \epsilon, \tilde{\epsilon}$ \cite{Andriot:2020wpp}, with in addition for a Minkowski solution ${\cal R}_4 \sim 10^{-15} \sim \sqrt{S}$. This usually requires to do a two-step minimisation: we first run a \texttt{NMinimize} that yields a first solution $s_1$. One then refines it by running a \texttt{FindMinimum}, where one uses $s_1$ as a starting point. This last step can sometimes be iterated. Further final validity checks of the solution include the evaluation of the maximum $|E_i|$, which should be of order $10^{-15}$.

For a given Ansatz (solution class and further specifications of variables), if we obtain the desired precision after the previous steps, we conclude that we have found a solution, otherwise we only claim that we were not able to find any solution in this set-up (this corresponds to the question mark in Table \ref{tab:sol}). This last situation does not necessarily mean that such solutions do not exist, as e.g.~in the case of a no-go theorem. It can also simply underline the computational complexity.

\subsection{(Anti-) de Sitter no-go theorems}\label{sec:nogo}

Prior to looking for new solutions, it is useful to be aware of existence no-go theorems for some solution classes. In turn, a failure in finding new solutions may signal a previously unnoticed no-go theorem. Let us say a few words on these no-go theorems, and find new ones for both de Sitter or anti-de Sitter solutions.

Most of the de Sitter no-go theorems were established for sources of single dimensionality $p$ (see e.g.~ \cite{Andriot:2018ept, Andriot:2019wrs, Andriot:2020lea}). The case of multiple dimensionalities was poorly studied. For the latter, a few no-go theorems were still obtained, beyond group manifolds, in Section 6 of \cite{Andriot:2017jhf}. In particular, $p=3 \& 7$ was excluded for de Sitter solutions. Some constraints were also established in IIA, but we note from \eqref{p=64sources1O6} that the overlap of $p=4$ and $p=6$ considered in \cite{Andriot:2017jhf} cannot happen in our ansatz. We will still make use of some of the equations of \cite{Andriot:2017jhf} in the following.

It has been noticed with single dimensionality that having intersecting sources instead of only parallel ones helps to get de Sitter solutions. One may then expect that having multiple dimensionalities can also help. We illustrate this idea in Section \ref{sec:warmup46}, before finding a new no-go theorem when adding too many orientifolds in Section \ref{sec:newnogo}. Finally, this last situation will lead as well to existence no-go theorems for anti-de Sitter, discussed in Section \ref{sec:nogoAdS}.

\subsubsection{Warm-up: $p = 4 \& 6$}\label{sec:warmup46}

De Sitter no-go theorems for single dimensionality sources do not generalize easily to the case of multiple dimensionalities. We know that for $p=6$ alone, one needs for de Sitter $F_0 \neq 0$ and ${\cal R}_6 <0$. How do those statements get modified with $p=4$ on top? This is a priori an important question when looking for solutions with $p = 4 \& 6$, because an $O_4$ imposes $F_0=0$ in our ansatz. Let us reproduce here these no-go theorems while extending to multiple dimensionalities with $p=4 \& 6$. We follow the reasoning of \cite{Andriot:2016xvq} to derive the initial no-go theorems.

We first need preliminary equations. In Section 6 of \cite{Andriot:2017jhf} are given the dilaton e.o.m., the ten-dimensional Einstein trace and the four-dimensional one. Combining those, one reaches there (6.7) given by
\beq
\mathcal{R}_{4} + 2 \mathcal{R}_{10} - |H|^2 + g_s^2 \sum_{q=0}^6 |F_q|^2 = 0  \ .  \label{comb1}
\eeq
Introducing a parameter $p_0 \geq 3$, another combination leads to (6.10) given by
\beq
(p_0-3) \mathcal{R}_{4} = -2 |H|^2 + g_s^2 \sum_{q=0}^6 (8-p_0-q) |F_q|^2 + g_s \left(\sum_p \frac{p_0+1}{p+1} T_{10}^{(p)} -T_{10} \right)  \ .  \label{comb2}
\eeq
Now, equating (\ref{comb1}) and (\ref{comb2}) for $p_0=6$ yields
\bea
\frac{9}{2} \mathcal{R}_{4} &= -3 |H|^2 + 3 g_s^2 \left(|F_0|^2-|F_4|^2-2|F_6|^2 \right) + \frac{3}{5} g_s T_{10}^{(4)} \label{R4T104} \\
& = -2 \mathcal{R}_{6} - g_s^2 \left(|F_2|^2+2|F_4|^2+3|F_6|^2 \right) + \frac{g_s}{5}  T_{10}^{(4)} \nn \ .
\eea
We recover for $T_{10}^{(4)}=0$ the requirements $F_0 \neq 0$ and ${\cal R}_6 <0$. However, with $T_{10}^{(4)}>0$ as can be the case with $O_4$, these requirements do not hold anymore.

Similarly, equating (\ref{comb1}) and three times (\ref{comb2}) for $p_0=4$, we get
\bea
\frac{7}{2} \mathcal{R}_{4} &= -7 |H|^2 + 7 g_s^2 \left(2 |F_0|^2+|F_2|^2-|F_6|^2 \right) - g_s T_{10}^{(6)}  \\
& = -2 \mathcal{R}_{6} + g_s^2 \left(|F_0|^2-|F_4|^2-2|F_6|^2 \right) - \frac{g_s}{7}  T_{10}^{(6)} \nn \ .
\eea
With an $O_4$ setting $F_0=0$ and with $T_{10}^{(6)}=0$, we recover the known de Sitter requirements $F_2 \neq 0$ and ${\cal R}_6 <0$ for $p=4$. Now with $T_{10}^{(6)}<0$, i.e.~with $D_6$ and possibly a few $O_6$, we can again wave these requirements. We conclude that allowing for multiple dimensionalities gives more freedom, i.e.~relaxes some of the de Sitter no-go theorems for single dimensionality.

\subsubsection{A new de Sitter no-go theorem for $m_{466}$}\label{sec:newnogo}

Adding too many orientifolds in different sets may however lead to new no-go theorems, because one projects out many fields. It was already noticed to be the case for 3 $O_5$, i.e.~the class $s_{555}$, that project out $F_1$ thus forbidding de Sitter solutions \cite{Andriot:2020wpp}. As mentioned in Section \ref{sec:Td}, the source configuration in class $m_{466}$ is T-dual to the previous one, so one would as well expect a no-go theorem against de Sitter solutions in $m_{466}$. We found one and prove it in the following.

Inserting the 4d Einstein equation \eqref{Einstein4d} in the 6d one \eqref{Einstein6d}, one obtains
\bea
{\cal R}_{ab} & = \frac{g_s^2}{2}\left(F_{2\ ac}F_{2\ b}^{\ \ \ c} +\frac{1}{3!} F_{4\ acde}F_{4\ b}^{\ \ \ cde} \right)+ \frac{1}{4} H_{acd}H_b^{\ \ cd}  \nn\\
& \ + \frac{g_s}{2} \left( T_{ab} - \delta_{ab} \sum_p \frac{T_{10}^{(p)}}{p+1} \right)  + \frac{\delta_{ab}}{4} \left( {\cal R}_4 + 2 g_s^2 |F_6|^2  \right)  \ ,\label{Einstein6dsimp}
\eea
which was used in \cite{Andriot:2019wrs} to get constraints on the Ricci tensor components. We are now interested in the case of 2 $O_6$ along 123 and 156, and an $O_4$ along 4, i.e.~class $m_{466}$ with variables \eqref{variables2O61O4} and sources \eqref{p=64sources2O61O4}. In that case the sources along 1 are all the $p=6$ ones, so one has
\bea
T_{11} = \sum_I \frac{T_{10}^{(6)I}}{7} = \frac{T_{10}^{(6)}}{7} \ .
\eea
Most interestingly, these three orientifolds impose so much projection that no $f^a{}_{bc}$ with an index along 1 is left. This implies that ${\cal R}_{11} =0$. From \eqref{Einstein6dsimp}, we deduce, using further the remaining flux components in \eqref{variables2O61O4}
\beq
g_s \frac{T_{10}^{(4)}}{5}  = \frac{1}{2} {\cal R}_4 + |H|^2 + g_s^2 |F_4|^2  + g_s^2 |F_6|^2  \ . \label{eqnogo}
\eeq
The fact there is no $F_2$ contribution is due to $F_2$ having no component along $1$: this, together with the absence of $F_0$, is the T-dual statement of having no $F_1$ in IIB with 3 $O_5$, and this will allow us to conclude on the no-go. We now compare \eqref{eqnogo} to \eqref{R4T104} (where we recall that $F_0=0$ with $O_4$). It is expected that the coefficients differ given the former is one equation and the latter is a trace. We deduce
\beq
{\cal R}_4  + g_s^2 |F_6|^2  = 0\ \Rightarrow\ {\cal R}_4 \leq 0 \ . \label{nogo}
\eeq
This is a no-go theorem against de Sitter solutions. It would be interesting to derive a 4d version of it, for which one would need the tools developed for no-go theorem 9 of \cite{Andriot:2020lea}. This would allow us to compute the constant $c$ that characterises the obstruction and compare it to the swampland de Sitter conjectures \cite{Obied:2018sgi, Bedroya:2019snp, Andriot:2020lea, Rudelius:2021oaz}.

\subsubsection{Anti-de Sitter no-go theorems for $s_{555}$ and $m_{466}$}\label{sec:nogoAdS}

Anti-de Sitter solutions are rarely constrained by no-go theorems, since they typically are favored by the equations of motion (see e.g.~\cite{Andriot:2016ufg}). When having however too many orientifolds, we will see as above that anti-de Sitter solutions are not possible. We start with $s_{555}$, the class with 3 $O_5$. Combining equations of motion in the case of intersecting sources with single dimensionality $p$ (with our ansatz on the warp factor and the dilaton), we obtained (3.5) in \cite{Andriot:2017jhf} that we repeat here
\beq
(p-3) {\cal R}_4  =  - 2 |H|^2 + g_s^2 \left( (7-p) |F_1|^2 + (5-p) |F_3|^2 + (3-p) |F_5|^2 \right) \ . \label{R4flux}
\eeq
It is easy to deduce the no-go theorem for de Sitter mentioned above, for $p=5$ and $F_1=0$ as in $s_{555}$. Looking closer at the allowed variables \eqref{variables3O5} in that class, we actually realise that $H=F_1=F_5=0$. This implies
\beq
{\cal R}_4 = 0 \ , \label{R4=0}
\eeq
which in particular gives a no-go theorem for anti-de Sitter solutions as well.

As motivated in Section \ref{sec:newnogo}, it is expected to get a similar no-go theorem for the T-dual source configuration, i.e.~2 $O_6$ and 1 $O_4$ as in the class $m_{466}$. The equations derived in Section \ref{sec:newnogo} are independent of the sign of ${\cal R}_4$, so we consider the last equation, namely \eqref{nogo}. Again, a closer look at the variables \eqref{variables2O61O4} allowed in $m_{466}$ makes us realise that $F_6=0$. This implies once again \eqref{R4=0} for $m_{466}$, and a no-go theorem for anti-de Sitter. These no-go theorems make Minkowski solutions in those 2 classes very special.

\subsection{Results: known and found solutions}\label{sec:resultssol}

We present in Table \ref{tab:sol} the known and found solutions in our ansatz, classified in solution classes. To find new ones, we have proceeded as indicated in Section \ref{sec:proc}. We looked for de Sitter solutions in each solution class, while we only searched for Minkowski and anti-de Sitter solutions in some of them. The new solutions were not tailored in any specific way in this paper. They rather serve as a proof of concept for their existence in certain classes, whenever no solution was previously known in the given class.

We label the solutions with the class name, the sign of the cosmological constant, and the number of the solution in its class: for instance $m_{46}^+ 1$ is the de Sitter solution number 1 in the class $m_{46}$. For $s_{55}$, we start counting the de Sitter solutions at 28, since 1-27 refer to the solutions found in \cite{Andriot:2020wpp, Andriot:2021rdy}; for all other classes we start at 1.

\begin{table}[H]
  \begin{center}
    \begin{tabular}{|c||c|c|c|}
    \hline
Solution & dS sol. & Mink. sol. & AdS sol. \\
class &  &  &  \\
    \hhline{=::===}
$s_3$ & $\times$ & \cite{Giddings:2001yu} &  \\
     \hhline{-||---}
$s_4$ & ? & \cite{Andriot:2016ufg} &  \\
     \hhline{-||---}
$s_5$ & ? & \cite{Andriot:2016ufg} &  \\
\hhline{-||---}
 $s_{55}$ & \cite{Andriot:2020wpp, Andriot:2021rdy}, {\bf 28} & \cite{Grana:2006kf, Andriot:2020wpp} & {\bf 1-4} \\
     \hhline{-||---}
$s_{555}$ & $\times$ & {\bf 1-4} & $\times$ \\
\hhline{-||---}
$s_{6}$ & ? & \cite{Andriot:2016ufg} &  \\
\hhline{-||---}
$s_{66}$ & {\bf 1} & \cite{Grana:2006kf} &  \\
\hhline{-||---}
$s_{6666}$ & \cite{Danielsson:2011au}, {\bf 1-4} & \cite{Camara:2005dc, Marchesano:2019hfb} & \cite{DeWolfe:2005uu, Camara:2005dc, Caviezel:2008ik} \\
\hhline{-||---}
$s_{7}$ & $\times$ & \cite{Andriot:2016ufg} &  \\
\hhline{-||---}
$s_{77}$ & $\times$ &  &  \\
    \hhline{=::===}
$m_4$ & ? &  &  \\
\hhline{-||---}
$m_{46}$ & {\bf 1-10} & {\bf 1-2} & {\bf 1-5} \\
\hhline{-||---}
$m_{466}$ & $\times$ & {\bf 1-6} & $\times$ \\
\hhline{-||---}
$m_6$ & ? &  &  \\
\hhline{-||---}
$m_{66}$ & ? &  &  \\
    \hhline{=::===}
$m_5$ & ? &  &  \\
\hhline{-||---}
$m_{55}$ & {\bf 1-4} &  &  \\
\hhline{-||---}
$m_{57}$ & ? &  &  \\
\hhline{-||---}
$m_{5577}$ & \cite{Caviezel:2009tu}, {\bf 1-12}, {\bf 1${}^*$} &  & \cite{Caviezel:2008ik, Petrini:2013ika} \\
\hhline{-||---}
$m_7$ & ? &  &  \\
\hhline{-||---}
$m_{77}$ & ? &  &  \\
    \hline
    \end{tabular}
     \caption{Solutions on a 4d maximally symmetric spacetime with orientifold that accommodate our ansatz. A reference or a number indicates that solutions have been found. A reference means that an example can be found in the corresponding paper; more references are provided in the main text. A number labels a new solution that has been found in this paper; without a reference, it means the solution is the first in its class up to our knowledge. We recall that when a solution is found in a class, it does not mean that all possible sources in the class are turned on. A cross ($\times$) indicates a no-go theorem against finding a solution in a given class. A question mark means that we have searched for solutions without finding any, potentially hinting at a no-go theorem, or at computational complexity. An empty box means that we have not searched for a solution, and are not aware of corresponding solutions in the literature.}\label{tab:sol}
  \end{center}
\end{table}

Let us first comment on de Sitter solutions. To start with, we found new solutions in classes where solutions were already known: $s_{55}$ \cite{Andriot:2020wpp, Andriot:2021rdy}, $s_{6666}$ \cite{Caviezel:2008tf, Flauger:2008ad, Danielsson:2010bc, Danielsson:2011au, Roupec:2018mbn} and $m_{5577}$ \cite{Caviezel:2009tu}. Only one solution was known in $m_{5577}$ and we found here several. In addition, those fall in the two subclasses, $m_{5577}$ and $m_{5577}^*$, as explained in Section \ref{sec:ansatz}. We recall that these solutions have source configurations which are T-dual to those of $s_{6666}$, but the solutions are not, because of the field content (see Section \ref{sec:Td}). Secondly, we found new solutions in the new class $m_{46}$ where no solution was known before. Some of these solutions have, in terms of sets, 1 $O_4$ along 4, 1 $O_6$ along 123 and 1 $D_6$ along 156. This source configuration is T-dual to that of $s_{55}$, so finding solutions there may not be surprising; however because of the field content, the solutions are again not T-dual to geometric ones and are thus truly new (see Section \ref{sec:Td}). These seemingly T-dual solutions in $m_{46}$, $s_{55}$, $s_{6666}$ and $m_{5577}$, all have source configurations that preserve supersymmetry, as discussed in Section \ref{sec:susy}. It is also the case for the new solution found in the new class $s_{66}$, which essentially differs from those in $s_{6666}$ because of the field content. Finally, we find new solutions, which have source configurations that break supersymmetry. It is the case of the other solutions found in $m_{46}$, with 2 additional $D_6$ along 256 and 356 (discussed in Appendix \ref{ap:chgbasis}), as well as the new solutions found in the new class $m_{55}$, which have at least three sets of $D_p$-branes.

We turn to Minkowski solutions, for which a few references already appear in Table \ref{tab:sol}. A list of known Minkowski solutions on group manifolds (found using supersymmetry conditions) with parallel sources (i.e.~only one set) is given in Section 2.4 of \cite{Andriot:2015sia}; all but one of them fit the ansatz of \cite{Andriot:2016ufg}. That ansatz includes some group manifolds, but also goes beyond them. Regarding intersecting sources, a list of known Minkowski solutions on group manifolds is given in Section 5 of \cite{Andriot:2017jhf}; a new one, $s_{55}^0$1, was found in \cite{Andriot:2020wpp}. All those had only 2 sets of sources. To those one should add the solutions in the class $s_{6666}$ indicated in Table \ref{tab:sol}. Here, we found new Minkowski solutions in new classes with 3 sets of sources: 4 in $s_{555}$ and 6 in $m_{466}$. These two classes are very special because of the no-go theorems against de Sitter and anti-de Sitter solutions, discussed in Section \ref{sec:newnogo} and \ref{sec:nogoAdS}. We also find 2 new Minkowski solutions in the new class $m_{46}$.

We finally consider anti-de Sitter solutions. Beyond references given in Table \ref{tab:sol}, we can mention the recent works \cite{Marchesano:2019hfb, Cribiori:2021djm} that have solutions in $s_{6666}$. We find here new solutions in new classes: 5 in $m_{46}$ (with 3 sets of sources), and 4 in $s_{55}$ (with 2 sets of sources). For the latter, we do not find solutions with 3 sets, i.e.~with a last set having $D_5$. This may hint at a (surprising) no-go theorem or at computational complexity.

For Minkowski and anti-de Sitter solutions, similar points on T-duality relations and on the supersymmetry preserved by the source configurations can be made, as for de Sitter solutions. It would also be interesting to know whether these new Minkowski and anti-de Sitter solutions preserve (bulk) supersymmetry, given their fluxes and geometry. We do not investigate this question here, but it could in principle be done following the material reviewed in \cite{Koerber:2010bx}.

For all solutions, a remaining important point is to identify the 6d group manifold and verify its compactness. This requires tools that will be developed and presented in the companion paper \cite{Andriot:2022yyj}. Identifying the manifold, in particular the underlying algebra, is not straightforward because in each solution, the structure constants are obtained in an arbitrary basis, suited to the placement of the sources, with metric $\delta_{ab}$. Through an appropriate change of basis, the structure constants can be brought to a form where the algebra can be recognised. From there one can discuss whether the group manifold can be compact.

We provide in Appendix \ref{ap:sol} the complete list of new solutions found in this work. Let us recall, as indicated at the end of Section \ref{sec:proc}, that an overall rescaling is available to modify together the value of all variables in these solutions; $|{\cal R}_4|$ and $|{\cal R}_6|$ can in particular be lowered.

\section{Conjecture 4: de Sitter, intersecting sources and 4d ${\cal N}=1$ supersymmetry}\label{sec:conjdS}

\setcounter{thm}{3}

In \cite{Andriot:2019wrs}, 3 conjectures on classical de Sitter solutions have been formulated. Since that paper, no proof nor counterexample to these conjectures have been found, in spite of progress in this field. These conjectures are believed to hold true at least for the ansatz presented in \cite{Andriot:2019wrs} and used in the present paper, if not beyond it. Of particular interest here is conjecture 1, claiming that there exists no classical de Sitter solution with parallel sources, i.e.~a single set of sources. We propose here a fourth conjecture, that somehow extends conjecture 1.
\begin{thm}
There is no classical de Sitter solution with 2 intersecting source sets.\label{conj4}
\end{thm}
We believe again that this is true within our solution ansatz, and maybe beyond it, hence the general formulation of the conjecture.

A first argument in favor of conjecture \ref{conj4} is that among all de Sitter solutions known and found listed in Table \ref{tab:sol}, none of them has less than 3 source sets ($I=1,2,3$). We also ran specific searches for de Sitter solutions with only 2 sets in several classes, but did not find any. Let us emphasize that the claim of conjecture \ref{conj4} is made possible here for the first time thanks to the classification of solutions, that provides an overview of the possibilities. The T-duality relations \eqref{Td2Op} and \eqref{Td1Op1Dp} described below also make manifest the role played by our classification to support this conjecture.

A second argument comes from T-duality. As discussed in Section \ref{sec:Td}, some source configurations are T-dual to each other. If we restrict ourselves to exactly 2 source sets, we have the following T-duality chains between source configurations of classes
\bea
\mbox{T-dual source config.~with 2 $O_p$:}& \quad \xymatrix{s_{77}\, \ar@{<->}[r] &\, s_{66}\, \ar@{<->}[r] &\, s_{55}\, \ar@{<->}[r] &\, m_{46}\\
 & & \ar@{<->}[ul] \, m_{57}\, \ar@{<->}[ur] & }  \label{Td2Op}\\
\mbox{T-dual source config.~with 1 $O_p$, 1 $D_p$:}& \quad\xymatrix{s_{7}\, \ar@{<->}[r] &\, s_{6}\, \ar@{<->}[r] &\, s_{5}\, \ar@{<->}[r] &\, m_{4}\\
 & & \ar@{<->}[ul] \, m_{5}\, \ar@{<->}[ur] &\ar@{<->}[ul] \, m_{6} \\
 & & \ar@{<->}[uul] \, m_{7}\, \ar@{<->}[ur] & } \label{Td1Op1Dp}
\eea
where by definition, $m_4$ with 2 source sets means 1 $O_4$ and 1 $D_6$, etc. The different arrows on one class correspond to different possible choices of directions along which to perform the T-duality. The placement of source sets and T-duality directions is not detailed here, but should be clear when looking at the information of each class. Interestingly, we cover with these two T-duality chains (which only differ by the nature of the sources) all classes that can admit exactly 2 sets; in particular, no class is left isolated. The argument in favor of conjecture \ref{conj4} is now the following: we know of no-go theorems on de Sitter solutions in $s_7$ and $s_{77}$ with (at least) 2 sets \cite{Andriot:2017jhf}. The T-duality relation to the configurations of the other classes suggests no-go theorems there as well. As discussed in Section \ref{sec:Td}, a complete T-duality relation of the classes is not strictly speaking realised, because few fields in one class can prevent a solution to be T-dual to another one, even if the source configurations are. This remains a strong hint in favor of an obstruction on de Sitter for all classes in these T-duality chains, which are all classes allowing for exactly 2 source sets. This supports conjecture \ref{conj4}. We tried to prove such a no-go theorem for $s_{66}$, inspired by the T-duality relation to $s_{77}$, but failed. The proof may again be difficult to achieve, as for the case of a single set in conjecture 1. Having it would still be very interesting, and further deriving through a 4d version the corresponding value of $c$ in de Sitter swampland conjectures \cite{Obied:2018sgi, Bedroya:2019snp, Andriot:2020lea, Rudelius:2021oaz}.\\

\setcounter{thm}{3}

An interesting consequence of conjecture \ref{conj4} is the result in a 4d effective theory. The ``effectiveness'' refers here to a certain truncation, and for the following implication to hold, we need a truncation that preserves the contribution of sources (source sets should not be erased through the truncation). This includes at least consistent truncations as we will see. If classical de Sitter solutions require at least 3 intersecting source sets, this implies, as discussed in Section \ref{sec:susy}, that supersymmetry is reduced at least by $2^3$, and since the truncation considered preserves the sources, we conclude
\begin{thm}
{\bf implies} \quad  A 4d effective theory of a classical string compactification,\linebreak \phantom{{\bf Conjecture 4 implies} \quad} admitting a de Sitter critical point, can at most be ${\cal N} = 1$ \linebreak \phantom{{\bf Conjecture 4 implies} \quad} supersymmetric. \label{conj4impl}
\end{thm}

As mentioned in Section \ref{sec:resultssol}, we know examples of de Sitter solutions admitting source configurations that break completely supersymmetry, so a non-supersymmetric 4d theory is a priori also possible (coming then probably with instabilities).

Our solution ansatz actually allows for a consistent truncation giving a 4d gauged supergravity (see Section 2.3 of \cite{Andriot:2019wrs}). This means that any solution in the latter will be one in our 10d theory. But there exists many 4d gauged supergravities, defined in particular by their turned-on gaugings, most of them having no higher dimensional origin as a classical compactification. It has been already noticed that finding de Sitter solutions in 4d gauged supergravities with extended supersymmetries (${\cal N} > 1$) seems difficult precisely when requiring a classical compactification origin. This observation is exactly in agreement with conjecture \ref{conj4} and its implication. (Meta-) stable de Sitter solutions in 4d gauged supergravities with \textsl{extended} supersymmetries typically require Fayet-Iliopoulos terms (whose higher dimensional origin is in general unclear), non-compact gaugings (which would correspond to non-compact extra dimensions) \cite{Fre:2002pd, deRoo:2002jf, Ogetbil:2008tk, Roest:2009tt, DallAgata:2012plb, Cribiori:2020use} or non-geometric fluxes \cite{Dibitetto:2011gm} (whose stringy origin is not in a standard classical compactification); some of the latter examples are even disputed in \cite{Plauschinn:2020ram}. More examples and arguments were provided recently in \cite{Cribiori:2020use, DallAgata:2021nnr}, beyond the (meta-) stable case. Further references can be found in Footnote 7 of \cite{Andriot:2021rdy}. These observations on 4d gauged supergravities are consistent with our claims, thanks to the consistent truncation allowed by our solution ansatz.

The constraint phrased in conjecture \ref{conj4} on classical de Sitter solutions and resulting 4d effective theories is interesting for phenomenology. The configurations of intersecting sources considered here typically allow to build particle physics models (see \cite{Andriot:2017jhf} or the recent works \cite{Loges:2021hvn, He:2021gug}). Having at most ${\cal N} = 1$ in 4d {\it naturally} allows for chirality in those models, as wished for phenomenology. Looking for classical de Sitter solutions thus provides unexpectedly several required ingredients to build, {\it together with cosmology}, viable particle physics models. We hope to investigate more these possibilities in future work.

\vfill

\subsection*{Acknowledgements}

We warmly thank E.~Dudas and D.~Tsimpis for helpful exchanges during the completion of this work, as well as D.~Chicherin and S.~Zoia regarding Section \ref{sec:nogoAdS}. D.~A.~acknowledges support from the Austrian Science Fund (FWF): project number M2247-N27. D.~A.~and L.~H.~acknowledge support from the Austrian Science Fund (FWF): project number P34562-N.

\newpage

\begin{appendix}

\section{Changes of basis}\label{ap:chgbasis}

In this paper we aim at classifying solutions to the equations in Appendix \ref{ap:eq} into classes. It is therefore important to analyse the symmetries of these equations, namely transformations leaving them invariant or covariant, and transforming a solution into another one. Of prime importance to us will be what we call a change of basis, used in \cite{Andriot:2020lea, Andriot:2020wpp}: these are represented by constant matrices $M \in GL(6)$, which transform the one-forms $e^a$ as ${e^a}'= M^a{}_b e^b$. Each object with an orthonormal index then has to transform by a multiplication with $M^a{}_b$ or its inverse, as a tensor would: this applies to flux components but also to $f^a{}_{bc}$. In case $M \in$ SO(6), then the metric $\delta_{ab}$ is preserved, otherwise it gets transformed as well into $g=M^{-T} \delta M^{-1}$ of components $g_{ab}$. Scalar quantities such as ${\cal R}_6$, $|F_q|^2$, or even forms as $H$, are invariant under such transformations, thanks to the contraction of indices, while tensors such as ${\cal R}_{ab}$ transform covariantly.

Are the equations listed in Appendix \ref{ap:eq} invariant or covariant under a change of basis? This is easily verified to be true, except for the source terms. Let us first discuss properties of the source contributions with respect to diffeomorphisms. From the perspective of the action, it is clear that the 10d = 4d + 6d diffeomorphism invariance gets broken, due to the presence of extended objects, to the invariance of their world-volume. In particular, their internal volume forms ${\rm vol}_{||_{(p)I}}$ should remain invariant, and since ${\rm vol}_6$ is  by definition invariant, the transverse ones ${\rm vol}_{\bot_{(p)I}}$ are as well. This amounts to say that each $T_{10}^{(p)I}$ is a scalar under diffeomorphisms. Similarly we can break $GL(6)$ to consider $M$ that preserve ${\rm vol}_{\bot_{(p)I}}$ (without forgetting the volume factor in there). This means that the transformation does not mix directions of different sets, and $T_{10}^{(p)I}$ is invariant. The equations are then clearly invariant or covariant, and a solution is transformed into another one.\\

This restriction has the important implication that sources are fixed once and for all, i.e.~are not modified through such a change of basis. Could this restriction be relaxed? Is it possible to find a change of basis that does not preserve the sources, but such that the transformed fields still solve the equations? Let us discuss here such a possibility. What enters the dilaton equation of motion and the ${\cal R}_4$ definition is the sum $T_{10}^{(p)} = \sum_I T_{10}^{(p)I}$ for each $p$, rather than the $T_{10}^{(p)I}$ separately. We may then consider a change of basis that preserves the sum, rather than the individual $T_{10}^{(p)I}$. One example is a relabeling transformation that exchanges two sets of sources, in particular two ${\rm vol}_{\bot_{(p)I}}$, to which we can add an exchange of two $T_{10}^{(p)I}$. The Bianchi identities are then invariant, as well as the dilaton equation of motion and the ${\cal R}_4$ definition. The source terms in the Einstein equations are transformed accordingly, and are then covariant (as well as the Jacobi identities). Such a discrete transformation relaxes the previous restriction on the preservation of individual sets of sources.

We tried to apply such discrete transformations to solutions in $m_{5577}$. There, some solutions admit 2 $D_7$ along 2456 and 1356, while another one has 2 $D_7$ along 2356 and 1456. Despite the similarity, we did not find a change of basis that would exchange ${\rm vol}_{\bot_{(p)I}}$ as well as $T_{10}^{(p)I}$, to bring one solution in the form of the other. This is due to the orientation in these volume forms that leads to minus signs. Those signs could be transferred to the $T_{10}^{(p)I}$ but the $\sum_I T_{10}^{(p)I}$ is then not preserved. As mentioned in Section \ref{sec:ansatz}, we then split that class into two subclasses, one denoted as $m_{5577}^*$.\\

Another attempt to consider transformations that do not preserve each individual ${\rm vol}_{\bot_{(p)I}}$ is a rotation. Consider for example the general form combination $C_1\, e^1\w \omega + C_2\, e^2\w \omega$ where $C_{1,2}$ are non-zero coefficients and $\omega$ is a form that does not contain $e^1$ or $e^2$. The combination can be rewritten as follows, with $\varepsilon^2=1$
\bea
C_1\, e^1\w \omega + C_2\, e^2 \w \omega & = \varepsilon \sqrt{C_1^2 +C_2^2}\, \left(\varepsilon \frac{C_1}{\sqrt{C_1^2 +C_2^2}}\, e^1 +  \varepsilon \frac{C_2}{\sqrt{C_1^2 +C_2^2}} \, e^2 \right) \w \omega \\
& \equiv \varepsilon \sqrt{C_1^2 +C_2^2}\, \left( \cos \alpha\, e^1 + \sin \alpha\, e^2 \right) \w \omega = \varepsilon \sqrt{C_1^2 +C_2^2}\ {e^1}' \w \omega \ .\nn
\eea
In other words, we can perform a rotation
\beq
\left( \begin{array}{c} {e^1}' \\ {e^2}' \end{array} \right) = \left( \begin{array}{cc} \cos \alpha & \sin \alpha  \\ -\sin \alpha & \cos \alpha \end{array} \right) \left( \begin{array}{c} e^1 \\ e^2 \end{array} \right) \ ,
\eeq
that turns the initial combination of $e^1, e^2$ into just ${e^1}'$, adjusting the angle in terms of the coefficients. We introduce the sign $\varepsilon$ to show that there is a freedom in the final sign. The form $e^1 \w e^2$ is ``preserved'', in the sense that $e^1 \w e^2 = {e^1}' \w {e^2}'$.

We tried to use such rotations for solutions of class $m_{46}$, that have 1 $O_4$ along 4, 1 $O_6$ along 123 and 3 $D_6$ along 156, 256, 356. Indeed, considering the Bianchi identity of $F_2$
\bea
\d F_2  - H\w F_0 = \frac{1}{7} \Big(& T_{10}^{(6)_1}\, e^4 \w e^5 \w e^6 \\
 +& T_{10}^{(6)_2}\, e^2 \w e^3 \w e^4  - T_{10}^{(6)_3}\, e^1 \w e^3 \w e^4  + T_{10}^{(6)_4}\, e^1 \w e^2 \w e^4  \Big) \ ,\nn
\eea
one may wonder whether the solution can be transformed into a solution with 1 $O_4$ along 4, 1 $O_6$ along 123 and 1 $D_6$ along 156, thanks to rotations of the above type. While one can indeed reach the appropriate transverse volume form, the $T_{10}^{(6)_I}$ is however changed, in such a way that $\sum_I T_{10}^{(6)I}$ is not preserved. Indeed, one rather has to consider the quantity $\sqrt{( T_{10}^{(6)_2})^2 + ( T_{10}^{(6)_3})^2 + ( T_{10}^{(6)_4})^2 }$. So the dilaton equation of motion is not satisfied anymore. Even though such rotations can be useful, they do not provide us here with less restrictive changes of basis that still transform a solution to another one. In particular, solutions with 3 $D_6$ in $m_{46}$ seem definitely different from those with 1 $D_6$.

\section{Equations}\label{ap:eq}

Building on \cite{Andriot:2016xvq, Andriot:2017jhf, Andriot:2019wrs}, we give in this appendix the type II supergravities equations of motion (e.o.m.) and Bianchi identities (BI), in our framework with sources of multiple dimensionalities, $3\leq p \leq 8$. These equations encompass in particular the case of single dimensionality sources, obtained by setting to zero the appropriate source variables. Notations are introduced in Section \ref{sec:ansatz}; we recall in particular $T_{10}= \sum_p T_{10}^{(p)}=\sum_{p,I} T_{10}^{(p)I}$. By combining a few equations as in \cite{Andriot:2017jhf}, one obtains the useful expression \eqref{R4T_{10}F} of ${\cal R}_4$, that we repeat here for completeness
\beq
{\cal R}_4= g_s \sum_p \frac{T_{10}^{(p)}}{p+1} - g_s^2 \sum_{q=0}^6 |F_q|^2  \ .
\eeq

\subsection{IIA supergravity}

\begin{itemize}
\item the fluxes e.o.m.
\bea
& \d ( *_6 H) - g_s^2  \left(F_{0} \w *_6 F_{2}+F_{2} \w *_6 F_{4}+F_{4} \w *_6 F_{6} \right)   = 0 \ ,\\
& \d( *_6 F_2 ) + H \w *_6 F_{4} = 0\ , \label{eomF2}\\
&\d ( *_6 F_4 ) + H \w *_6 F_{6}= 0 \ ,
\eea
\item the fluxes BI
\bea
& \d H =0 \ ,\\
& \d F_0= - \sum_{I} \frac{T_{10}^{(8)I}}{9} \, {\rm vol}_{\bot_{(8)I}}  \ ,  \label{BI0}\\
& \d F_2 - H \w F_0 = \ \ \sum_{I} \frac{T_{10}^{(6)I}}{7} \, {\rm vol}_{\bot_{(6)I}}  \ ,  \label{BI2}\\
& \d F_4 - H \w F_2 = -\sum_{I} \frac{T_{10}^{(4)I}}{5} \, {\rm vol}_{\bot_{(4)I}} \ , \label{BI4}
\eea
\item the dilaton e.o.m.
\beq
2 {\cal R}_{4}+ 2{\cal R}_6 + g_s \sum_p \frac{T_{10}^{(p)}}{p+1} -|H|^2 = 0 \ ,
\eeq
\item the 4d Einstein equation (equivalent to its trace)
\beq
4 {\cal R}_4 = g_s \sum_p \frac{7-p}{p+1}\, T_{10}^{(p)} - 2|H|^2 + g_s^2 ( |F_0|^2 - |F_2|^2 - 3 |F_4|^2 - 5 |F_6|^2 ) \ , \label{Einstein4d}
\eeq
\item the 6d (trace-reversed) Einstein equation
\bea
{\cal R}_{ab} & = \frac{g_s^2}{2}\left(F_{2\ ac}F_{2\ b}^{\ \ \ c} +\frac{1}{3!} F_{4\ acde}F_{4\ b}^{\ \ \ cde} \right)+ \frac{1}{4} H_{acd}H_b^{\ \ cd}  \nn\\
& \ + \frac{g_s}{2}T_{ab} + \frac{\delta_{ab}}{16} \left( - g_s T_{10} - 2|H|^2 + g_s^2 ( |F_0|^2 - |F_2|^2 - 3 |F_4|^2 + 3 |F_6|^2 ) \right)  \ ,\label{Einstein6d}\\
{\rm with} &\ \ T_{ab} = \sum_p \sum_I \delta^{a_{||_I}}_{a} \delta^{b_{||_I}}_{b} \delta_{a_{||_I}b_{||_I}} \frac{T_{10}^{(p)I}}{p+1} \ ,
\eea
\item the Jacobi identity (or Riemann BI)
\beq
f^a{}_{e[b} f^e{}_{cd]}=0 \ .
\eeq
\end{itemize}

\subsection{IIB supergravity}

\begin{itemize}
\item the fluxes e.o.m.
\bea
& \d ( *_6 H) - g_s^2 ( F_{1} \w *_6 F_{3} + F_{3} \w *_6 F_{5} )  = 0 \ ,\\
& \d( *_6 F_1 ) + H \w *_6 F_{3} = 0\ , \label{F_1eom} \\
& \d( *_6 F_3 ) + H \w *_6 F_{5} = 0\ ,\\
&\d ( *_6 F_5 ) = 0 \ ,
\eea
\item the fluxes BI
\bea
& \d H =0 \ ,\\
& \d F_1= -\sum_{I} \frac{T_{10}^{(7)I}}{8} \, {\rm vol}_{\bot_{(7)I}} \ ,\\
& \d F_3 - H \w F_1 = \ \ \sum_{I} \frac{T_{10}^{(5)I}}{6} \, {\rm vol}_{\bot_{(5)I}}  \ , \label{BI3}\\
& \d F_5 - H \w F_3 = - \sum_{I} \frac{T_{10}^{(3)I}}{4} \, {\rm vol}_{\bot_{(3)I}} \ , \label{BI5}
\eea
\item the dilaton e.o.m.
\beq
2 {\cal R}_{4}+ 2{\cal R}_6 + g_s \sum_p \frac{T_{10}^{(p)}}{p+1} -|H|^2 = 0 \ ,
\eeq
\item the 4d Einstein equation (equivalent to its trace)
\beq
4 {\cal R}_4 = g_s \sum_p \frac{7-p}{p+1}\, T_{10}^{(p)} - 2|H|^2 - g_s^2 ( 2 |F_3|^2 + 4 |F_5|^2 ) \ ,
\eeq
\item the 6d (trace-reversed) Einstein equation
\bea
{\cal R}_{ab} & = \frac{g_s^2}{2}\left(F_{1\ a}F_{1\ b} +\frac{1}{2!} F_{3\ acd}F_{3\ b}^{\ \ \ cd} + \frac{1}{2 \cdot 4!} F_{5\ acdef}F_{5\ b}^{\ \ \ cdef} - \frac{1}{2} *_6 F_{5\ a} *_6 F_{5\ b} \right) \nn\\
& \ + \frac{1}{4} H_{acd}H_b^{\ \ cd} + \frac{g_s}{2}T_{ab} + \frac{\delta_{ab}}{16} \left( - g_s T_{10} - 2|H|^2 - 2 g_s^2 |F_3|^2 \right)  \ ,\\
{\rm with} &\ \ T_{ab} = \sum_p \sum_I \delta^{a_{||_I}}_{a} \delta^{b_{||_I}}_{b} \delta_{a_{||_I}b_{||_I}} \frac{T_{10}^{(p)I}}{p+1} \ ,
\eea
\item the Jacobi identity (or Riemann BI)
\beq
f^a{}_{e[b} f^e{}_{cd]}=0 \ .
\eeq
\end{itemize}

\section{List of solutions}\label{ap:sol}

We list the new solutions found in this work, presented in Section \ref{sec:resultssol}. They are first ordered according to their cosmological constant, in Appendix \ref{ap:dS}, \ref{ap:Mink} and \ref{ap:AdS}. In each of those, we follow the order of solution classes of Table \ref{tab:sol}. Solutions are labeled accordingly, as described there. While solutions have been found to a satisfactory precision level (see Section \ref{sec:code}), we round them here to 5 significant digits for readability. The variables are expressed with the following symbols: $T_{10}[I]$ for $g_s T_{10}^I$, $F_q[a_1,..., a_q]$ for $g_s F_{q\, a_1... a_q}$, $H[a,b,c]$ for $H_{abc}$, $f[a,b,c]$ for $f^a{}_{bc}$. Only the non-zero variables are given. We also provide ${\cal R}_4$ and ${\cal R}_6$. All values should be understood in units of $2\pi l_s$. Note though that as indicated at the end of Section \ref{sec:proc}, each solution can go through an overall rescaling of its values. Contrary to the bulk of the paper, the source sets are here labeled with a single index $I$, independently of the dimensionality $p$. We thus specify the internal directions wrapped by each set. The sets with $O_p$ are the first ones, and their number can be read from the class name.

For each solution, we provide stability data drawn from the companion paper \cite{Andriot:2022yyj}. We first give the mass spectrum, namely the mass matrix eigenvalues: $\text{masses}^2$. For de Sitter solutions, we give an eigenvector $\vec{v}$ in field space (in the $(\rho,\tau,\sigma_I)$ basis) corresponding to the tachyonic direction. For (anti-) de Sitter solutions, we also compute the parameter $\eta_V$. We refer to \cite{Andriot:2022yyj} for more details. Algebras corresponding to the $f^a{}_{bc}$ are identified in that paper for all solutions.

\subsection{De Sitter solutions}\label{ap:dS}

\subsection*{$\boldsymbol{s_{55}^+ 28}$}

\begin{equation*}
\begin{aligned}
& I=1\!: 12\, ,\ I=2\!: 34\, ,\ I=3\!: 56,\\[6pt]&
 T_{10}[1] = 2.2601, T_{10}[2] = 5.167, T_{10}[3] = -0.46062, F_1[5] = 1,
 F_3[1, 3, 5] = -0.16181,\\[6pt]&
F_3[1, 3, 6] = -0.0038892,
F_3[1, 4, 6] = 0.16022,
F_3[2, 3, 5] = -0.0001645,
F_3[2, 3, 6] = -0.1408,\\[6pt]&
F_3[2, 4, 5] = -0.13582,
F_3[2, 4, 6] = -0.0045142,
H[1, 2, 5] = 0.057659,
H[1, 2, 6] = 0.84321,\\[6pt]&
 H[3, 4, 5] = -0.38,
 H[3, 4, 6] = 0.26525,
 f[1, 4, 5] = -0.052068,
f[1, 4, 6] = -0.68811,\\[6pt]&
f[2, 3, 5] = -0.062695,
f[2, 4, 5] = 0.00011416,
f[2, 4, 6] = 0.00020752,
 f[3, 1, 5] = -0.00002051,\\[6pt]&
f[3, 2, 5] = -0.015941,
f[4, 1, 5] = -0.0099985,
f[4, 1, 6] = -0.13214,
f[6, 1, 4] = -0.54525,\\[6pt]&
 f[3, 1, 6] = -0.00020752,
\end{aligned}
\end{equation*}
\begin{equation*}
{\cal R}_4 = 0.070923 \,, \quad  {\cal R}_6 = -0.18693 \,, \quad \eta_V= -3.2374 \,,
\end{equation*}
\begin{equation*}
\text{masses}^2 = (2.6997, 0.83912, 0.17195, -0.057401) \,, \;  \vec{v} = (0.39746, 0.91322, 0.04005, 0.080317) \,.
\end{equation*}

\subsection*{$\boldsymbol{s_{66}^+ 1}$}

\begin{equation*}
\begin{aligned}
&I=1\!: 123\, ,\ I=2\!: 145\, ,\ I=3\!: 256\, ,\ I=4\!: 346,\\[6pt]&
T_{10}[1] = 10, T_{10}[2] = 8.6452, T_{10}[3] = -0.8438, T_{10}[4] = -0.8438, \\[6pt]&
F_{0} = 1.2685, F_{2}[1,6] = -0.053287, F_{2}[2,4] = 0.18526, F_{2}[2,5] = -0.44051, \\[6pt]&
F_{2}[3,4] = 0.44051, F_{2}[3,5] = 0.18526, F_{4}[1,2,4,6] = 0.20229, F_{4}[1,2,5,6] = 0.080552, \\[6pt]&
F_{4}[1,3,4,6] = -0.080552, F_{4}[1,3,5,6] = 0.20229, H[1,2,5] = -0.095031, H[1,3,4] = -0.095031, \\[6pt]&
H[2,3,6] = -0.69318, H[4,5,6] = -0.79175, f[1,2,3] = -0.48293, f[1,4,5] = 0.50315, \\[6pt]&
f[2,4,6] = -0.38643, f[2,5,6] = -0.15388, f[3,4,6] = 0.15388, f[3,5,6] = -0.38643, \\[6pt]&
f[4,2,6] = -0.37091, f[4,3,6] = 0.14769, f[5,2,6] = -0.14769, f[5,3,6] = -0.37091,
\end{aligned}
\end{equation*}
\begin{equation*}
{\cal R}_4 = 0.25915 \,, \quad  {\cal R}_6 = -0.90769 \,, \quad \eta_V= -3.617 \,, \end{equation*} \begin{equation*} \text{masses}^2 = (5.5537, 1.1135, 0.91621,
0.39899, -0.23434) \,, \;  \vec{v} = (0.21153, 0.95997, 0.13038, 0.1293, 0) \,. \end{equation*}

\subsection*{$\boldsymbol{s_{6666}^+ 1}$}

\begin{equation*}
\begin{aligned}
& I=1\!: 123\, ,\ I=2\!: 145\, ,\ I=3\!: 256\, ,\ I=4\!: 346,\\[6pt]&
T_{10}[1] = 10, T_{10}[2] = -0.64856, T_{10}[3] = -0.28674, T_{10}[4] = -0.62522, \\[6pt]&
F_{0} = -0.5871, F_{2}[1,6] = 0.56149, F_{2}[2,4] = 0.49009, F_{2}[3,5] = -0.551, \\[6pt]&
H[1,2,5] = 0.24103, H[1,3,4] = 0.098649, H[2,3,6] = -0.25101, H[4,5,6] = 0.45935, \\[6pt]&
f[1,2,3] = 0.11063, f[1,4,5] = -0.71643, f[2,1,3] = -0.028654, f[2,5,6] = -0.73726, \\[6pt]&
f[3,1,2] = 0.10166, f[3,4,6] = -0.71741, f[4,1,5] = 0.10447, f[4,3,6] = 0.11384, \\[6pt]&
f[5,1,4] = -0.027882, f[5,2,6] = 0.11078, f[6,2,5] = -0.10152, f[6,3,4] = -0.027844,
\end{aligned}
\end{equation*}
\begin{equation*}
{\cal R}_4 = 0.0019 \,, \quad  {\cal R}_6 = -0.4338 \,, \quad \eta_V= -18.445 \,,
\end{equation*}
\begin{equation*}
\text{masses}^2 = (2.6542, 0.51278, 0.11832, 0.025737, -0.0087616) \,,
\end{equation*}
\begin{equation*}
\vec{v} = (-0.21114, -0.97063, -0.10494, -0.0044338, 0.047676) \,.
\end{equation*}

\subsection*{$\boldsymbol{s_{6666}^+ 2}$}

\begin{equation*}
\begin{aligned}
& I=1\!: 123\, ,\ I=2\!: 145\, ,\ I=3\!: 256\, ,\ I=4\!: 346,\\[6pt]&
T_{10}[1] = 0.28747, T_{10}[2] = 10, T_{10}[3] = -0.27065, T_{10}[4] = -0.29230, \\[6pt]&
F_{0} = 0.59808, F_{2}[1,6] = 0.72514, F_{2}[2,4] = 0.48711, F_{2}[3,5] = 0.50209, \\[6pt]&
H[1,2,5] = -0.026084, H[1,3,4] = -0.12392, H[2,3,6] = -0.55827, H[4,5,6] = -0.074278, \\[6pt]&
f[1,2,3] = -0.79263, f[1,4,5] = 0.048506, f[2,1,3] = 0.032869, f[2,5,6] = -0.032187, \\[6pt]&
f[3,1,2] = -0.023594, f[3,4,6] = 0.032142, f[4,1,5] = 0.023626, f[4,3,6] = 0.52596, \\[6pt]&
f[5,1,4] = -0.032823, f[5,2,6] = -0.52523, f[6,2,5] = 0.035605, f[6,3,4] = -0.049535,
\end{aligned}
\end{equation*}
\begin{equation*}
{\cal R}_4 = 0.016318 \,, \quad  {\cal R}_6 = -0.54431 \,, \quad \eta_V= -2.6435 \,,
\end{equation*}
\begin{equation*}
\text{masses}^2 = (2.6325, 0.25768, 0.021208, 0.011834, -0.010784) \,,
\end{equation*}
\begin{equation*}
\vec{v} = (-0.11873, -0.95004, -0.14484, -0.23405, -0.087002) \,.
\end{equation*}

\subsection*{$\boldsymbol{s_{6666}^+ 3}$}

\begin{equation*}
\begin{aligned}
& I=1\!: 123\, ,\ I=2\!: 145\, ,\ I=3\!: 256\, ,\ I=4\!: 346,\\[6pt]&
T_{10}[1] = 10, T_{10}[2] = -0.022254, T_{10}[3] = -0.56181, T_{10}[4] = -0.022254, \\[6pt]&
F_{0} = 0.27366, F_{2}[1,6] = 0.15807, F_{2}[2,4] = -0.88473, F_{2}[3,5] = -0.66851, \\[6pt]&
F_{4}[1,2,4,6] = -0.040498, F_{4}[2,3,4,5] = 0.064472, H[1,2,5] = -0.011617, H[1,3,4] = -0.13984, \\[6pt]&
H[2,3,6] = 0.011617, H[4,5,6] = -0.19831, f[1,4,5] = -0.50423, f[2,1,3] = 0.25402, \\[6pt]&
f[2,5,6] = 0.85671, f[3,4,6] = -0.80274, f[5,1,4] = -0.23802, f[6,3,4] = -0.14951,
\end{aligned}
\end{equation*}
\begin{equation*}
{\cal R}_4 = 0.0066281 \,, \quad  {\cal R}_6 = -0.64803 \,, \quad \eta_V= -2.3772 \,,
\end{equation*}
\begin{equation*}
\text{masses}^2 = (2.1906, 0.2362, 0.036109, 0.003449, -0.0039391) \,,
\end{equation*}
\begin{equation*}
\vec{v} = (-0.10262, -0.93062, -0.25695, 0, -0.23957) \,.
\end{equation*}

\subsection*{$\boldsymbol{s_{6666}^+ 4}$}

\begin{equation*}
\begin{aligned}
& I=1\!: 123\, ,\ I=2\!: 145\, ,\ I=3\!: 256\, ,\ I=4\!: 346,\\[6pt]&
T_{10}[1] = 10, T_{10}[2] = -0.11433, T_{10}[3] = -0.68982, T_{10}[4] = -0.11433, \\[6pt]&
F_{0} = 0.68725, F_{2}[1,6] = -0.49234, F_{2}[2,4] = -0.48960, F_{2}[3,5] = 0.55498, \\[6pt]&
F_{4}[1,2,4,6] = 0.023125, F_{4}[2,3,4,5] = -0.024432, H[1,2,5] = -0.023765, H[1,3,4] = -0.32922, \\[6pt]&
H[2,3,6] = 0.023765, H[4,5,6] = -0.55765, f[1,4,5] = 0.73078, f[2,1,3] = 0.1262, \\[6pt]&
f[2,5,6] = 0.52498, f[3,4,6] = 0.77209, f[5,1,4] = 0.18561, f[6,3,4] = 0.17568,
\end{aligned}
\end{equation*}
\begin{equation*}
{\cal R}_4 = 0.033794 \,, \quad  {\cal R}_6 = -0.47223 \,, \quad \eta_V= -3.6231 \,, \end{equation*} \begin{equation*} \text{masses}^2 = (2.876, 0.62526, 0.047899, 0.017462, -0.03061) \,, \;  \vec{v} = (0.17002, 0.96611, 0.16803, 0, 0.097448) \,. \end{equation*}

\subsection*{$\boldsymbol{m_{46}^+ 1}$}

\begin{equation*}
\begin{aligned}
& I=1\!: 4\, ,\ I=2\!: 123\, ,\ I=3\!: 156,\\[6pt]&
T_{10}[1] = 0.055518, T_{10}[2] = 0.91544, T_{10}[3] = -0.012619, F_{2}[1,5] = 0.068521, \\[6pt]&
F_{2}[1,6] = -0.12113, F_{2}[2,5] = 0.029417, F_{2}[2,6] = -0.27374, F_{2}[3,5] = -0.024587, \\[6pt]&
F_{2}[3,6] = 0.20380, F_{4}[1,2,4,5] = -0.012313, F_{4}[1,2,4,6] = 0.0020369, F_{4}[1,3,4,5] = 0.0086934, \\[6pt]&
F_{4}[1,3,4,6] = 0.0013222, H[1,2,5] = 0.020813, H[1,2,6] = -0.040268, H[1,3,5] = 0.025725, \\[6pt]&
H[1,3,6] = -0.054176, H[2,3,5] = -0.030633, H[2,3,6] = 0.021713, f[1,4,5] = 0.12212, \\[6pt]&
f[1,4,6] = 0.022477, f[2,4,5] = 0.26303, f[2,4,6] = 0.048411, f[3,4,5] = -0.19690, \\[6pt]&
f[3,4,6] = -0.036239, f[4,2,5] = -0.01635, f[4,2,6] = -0.0030093, f[4,3,5] = -0.040938, \\[6pt]&
f[4,3,6] = -0.0075347, f[5,2,4] = 0.0044539, f[5,3,4] = 0.011152, f[6,2,4] = 0.0025194, \\[6pt]&
f[6,3,4] = 0.0063081,
\end{aligned}
\end{equation*}
\begin{equation*}
{\cal R}_4 = 0.0025394 \,, \quad  {\cal R}_6 = -0.069048 \,, \quad \eta_V= -3.6764 \,,
\end{equation*}
\begin{equation*}
\text{masses}^2 = (0.23781, 0.043113, 0.0095625, 0.0042868, -0.002334) \,,
\end{equation*}
\begin{equation*}
\vec{v} = (-0.49298, -0.86049, -0.048158, -0.11911, -0.0044811) \,.
\end{equation*}

\subsection*{$\boldsymbol{m_{46}^+ 2}$}

\begin{equation*}
\begin{aligned}
& I=1\!: 4\, ,\ I=2\!: 123\, ,\ I=3\!: 156,\\[6pt]&
T_{10}[1] = 0.021786, T_{10}[2] = 0.21625, T_{10}[3] = -0.0050741, F_{2}[1,5] = 0.05579, \\[6pt]&
F_{2}[2,5] = -0.053314, F_{2}[2,6] = 0.036657, F_{2}[3,5] = -0.13111, F_{2}[3,6] = 0.094822, \\[6pt]&
F_{4}[1,2,4,6] = 0.0014442, F_{4}[1,3,4,5] = 0.004467, F_{4}[1,3,4,6] = 0.00090447, H[1,2,5] = 0.045867, \\[6pt]&
H[1,2,6] = -0.0071595, H[1,3,5] = -0.020177, H[2,3,5] = -0.015322, H[2,3,6] = -0.0021738, \\[6pt]&
f[1,4,5] = 0.022176, f[1,4,6] = 0.036871, f[2,4,5] = -0.031927, f[2,4,6] = -0.053084, \\[6pt]&
f[3,4,5] = -0.079393, f[3,4,6] = -0.132, f[4,2,5] = -0.013161, f[4,2,6] = -0.021882, \\[6pt]&
f[6,2,4] = 0.0076445,
\end{aligned}
\end{equation*}
\begin{equation*}
{\cal R}_4 = 0.0010218 \,, \quad  {\cal R}_6 = -0.016884 \,, \quad \eta_V= -3.7145 \,,
\end{equation*}
\begin{equation*}
\text{masses}^2 = (0.05838, 0.0064771, 0.0036211, 0.0016224, -0.0009489) \,,
\end{equation*}
\begin{equation*}
\vec{v} = (-0.49448, -0.85989, -0.046114, -0.1178, -0.008685) \,.
\end{equation*}

\subsection*{$\boldsymbol{m_{46}^+ 3}$}

\begin{equation*}
\begin{aligned}
& I=1\!: 4\, ,\ I=2\!: 123\, ,\ I=3\!: 156,\\[6pt]&
T_{10}[1] = 0.52281, T_{10}[2] = 0.065807, T_{10}[3] = -0.029745, F_{2}[1,5] = -0.078038, \\[6pt]&
F_{2}[1,6] = -0.033188, F_{2}[2,5] = -0.092832, F_{2}[2,6] = 0.25624, F_{2}[3,5] = 0.012313, \\[6pt]&
F_{2}[3,6] = 0.093103, F_{4}[1,2,4,5] = 0.03363, F_{4}[1,2,4,6] = 0.052371, F_{4}[1,3,4,5] = -0.046776, \\[6pt]&
F_{4}[1,3,4,6] = -0.10158, H[1,2,5] = 0.097704, H[1,2,6] = -0.010041, H[1,3,5] = -0.23018, \\[6pt]&
H[1,3,6] = -0.12995, H[2,3,5] = -0.036589, H[2,3,6] = 0.052514, f[1,4,5] = -0.0063799, \\[6pt]&
f[1,4,6] = 0.0030219, f[2,4,5] = -0.043334, f[2,4,6] = 0.020526, f[3,4,5] = -0.0065711, \\[6pt]&
f[3,4,6] = 0.0031125, f[4,2,5] = -0.11424, f[4,2,6] = 0.054112, f[4,3,5] = -0.070979, \\[6pt]&
f[4,3,6] = 0.03362, f[5,2,4] = 0.018851, f[5,3,4] = 0.011712, f[6,2,4] = -0.044327, \\[6pt]&
f[6,3,4] = -0.02754,
\end{aligned}
\end{equation*}
\begin{equation*}
{\cal R}_4 = 0.0030452 \,, \quad  {\cal R}_6 = -0.016095 \,, \quad \eta_V= -2.2769 \,,
\end{equation*}
\begin{equation*}
\text{masses}^2 = (0.16326, 0.049326, 0.014903, 0.0022091, -0.0017334) \,,
\end{equation*}
\begin{equation*}
\vec{v} = (-0.5219, -0.81615, -0.17652, -0.16336, -0.06057) \,.
\end{equation*}

\subsection*{$\boldsymbol{m_{46}^+ 4}$}

\begin{equation*}
\begin{aligned}
& I=1\!: 4\, ,\ I=2\!: 123\, ,\ I=3\!: 156\, ,\ I=4\!: 256\, ,\ I=5\!: 356,\\[6pt]&
T_{10}[1] = 10, T_{10}[2] = 5.5434, T_{10}[3] = -0.28263, T_{10}[4] = -0.34668, T_{10}[5] = -3.0483, \\[6pt]&
F_{2}[1,5] = -0.0046132, F_{2}[1,6] = -0.86711, F_{2}[2,5] = -0.0071933, F_{2}[2,6] = 0.62328, \\[6pt]&
F_{2}[3,5] = 0.28752, F_{2}[3,6] = -0.56401, F_{4}[1,2,4,5] = 0.069428, F_{4}[1,2,4,6] = -0.073577, \\[6pt]&
F_{4}[1,3,4,5] = 0.41804, F_{4}[1,3,4,6] = -0.47935, F_{4}[2,3,4,5] = -0.23603, F_{4}[2,3,4,6] = 0.20559, \\[6pt]&
H[1,2,5] = -0.42143, H[1,2,6] = 0.16444, H[1,3,5] = -0.36009, H[1,3,6] = 0.40252, \\[6pt]&
H[2,3,5] = -0.68041, H[2,3,6] = 0.45725, f[1,4,5] = 0.47613, f[1,4,6] = 0.26763, \\[6pt]&
f[2,4,5] = -0.343, f[2,4,6] = -0.1928, f[3,4,5] = 0.22756, f[3,4,6] = 0.12791, \\[6pt]&
f[4,1,5] = -0.5254, f[4,1,6] = -0.29533, f[4,2,5] = -1.1177, f[4,2,6] = -0.62827, \\[6pt]&
f[4,3,5] = 0.17583, f[4,3,6] = 0.098834, f[5,1,4] = 0.35275, f[5,2,4] = 0.75044, \\[6pt]&
f[5,3,4] = -0.11805, f[6,1,4] = 0.17608, f[6,2,4] = 0.37458, f[6,3,4] = -0.058926,
\end{aligned}
\end{equation*}
\begin{equation*}
{\cal R}_4 = 0.21259 \,, \quad  {\cal R}_6 = -0.76168 \,, \quad \eta_V= -2.8266 \,,
\end{equation*}
\begin{equation*}
\text{masses}^2 = (5.1417, 2.5653, 1.2597, 0.92638, 0.34999, -0.15023) \,,
\end{equation*}
\begin{equation*}
\vec{v} = (0.476, 0.87352, -0.049169, 0.04948, -0.037752, -0.064036) \,.
\end{equation*}

\subsection*{$\boldsymbol{m_{46}^+ 5}$}

\begin{equation*}
\begin{aligned}
& I=1\!: 4\, ,\ I=2\!: 123\, ,\ I=3\!: 156\, ,\ I=4\!: 256\, ,\ I=5\!: 356,\\[6pt]&
T_{10}[1] = 10, T_{10}[2] = 0.089559, T_{10}[3] = -0.0014549, T_{10}[4] = -0.12442, T_{10}[5] = -0.020044, \\[6pt]&
F_{2}[1,5] = -0.054582, F_{2}[1,6] = -0.87864, F_{2}[2,5] = 0.18170, F_{2}[2,6] = -0.0428, \\[6pt]&
F_{2}[3,5] = -1.0043, F_{2}[3,6] = 0.31220, F_{4}[1,2,4,6] = -0.038841, F_{4}[1,3,4,5] = 0.0031155, \\[6pt]&
F_{4}[1,3,4,6] = -0.0062571, F_{4}[2,3,4,5] = 0.26643, H[1,2,5] = 0.36510, H[1,2,6] = 0.99268, \\[6pt]&
H[1,3,5] = 0.052376, H[1,3,6] = 0.17775, H[2,3,5] = -0.87116, H[2,3,6] = 0.10088, \\[6pt]&
f[1,4,5] = -0.0024863, f[1,4,6] = 0.032591, f[2,4,5] = -0.00018986, f[2,4,6] = 0.0024887, \\[6pt]&
f[3,4,5] = 0.0012693, f[3,4,6] = -0.016639, f[4,1,5] = -0.023942, f[4,1,6] = 0.31384, \\[6pt]&
f[4,2,5] = 0.00036106, f[4,2,6] = -0.0047329, f[4,3,5] = -0.00050343, f[4,3,6] = 0.0065991, \\[6pt]&
f[5,1,4] = -0.013785, f[5,2,4] = 0.00020788, f[5,3,4] = -0.00028985, f[6,1,4] = -0.095571, \\[6pt]&
f[6,2,4] = 0.0014413, f[6,3,4] = -0.0020096,
\end{aligned}
\end{equation*}
\begin{equation*}
{\cal R}_4 = 0.0035366 \,, \quad  {\cal R}_6 = -0.038434 \,, \quad \eta_V= -0.36462 \,,
\end{equation*}
\begin{equation*}
\text{masses}^2 = (3.9077, 1.8356, 0.20839, 0.07304, 0.00126, -0.00032237) \,,
\end{equation*}
\begin{equation*}
\vec{v} = (-0.51698, -0.8149, -0.17672, -0.19284, 0.00063048, -0.015972) \,.
\end{equation*}

\subsection*{$\boldsymbol{m_{46}^+ 6}$}

\begin{equation*}
\begin{aligned}
& I=1\!: 4\, ,\ I=2\!: 123\, ,\ I=3\!: 156\, ,\ I=4\!: 256\, ,\ I=5\!: 356,\\[6pt]&
T_{10}[1] = 10, T_{10}[2] = -0.056703, T_{10}[3] = -0.000039105, T_{10}[4] = -0.17140, T_{10}[5] = -0.89612, \\[6pt]&
F_{2}[1,5] = 0.17355, F_{2}[1,6] = -0.16826, F_{2}[2,5] = 0.78091, F_{2}[2,6] = 0.43765, \\[6pt]&
F_{2}[3,5] = -0.093708, F_{2}[3,6] = -0.016276, F_{4}[1,2,4,5] = 0.0097804, F_{4}[1,2,4,6] = -0.0010561, \\[6pt]&
F_{4}[1,3,4,5] = 0.051354, F_{4}[1,3,4,6] = -0.0055164, F_{4}[2,3,4,5] = 0.97045, F_{4}[2,3,4,6] = 0.021534, \\[6pt]&
H[1,2,5] = -0.038322, H[1,2,6] = 0.080497, H[1,3,5] = -0.66493, H[1,3,6] = 0.67871, \\[6pt]&
H[2,3,5] = -0.17221, H[2,3,6] = -0.27621, f[1,4,6] = 0.0028299, f[2,4,6] = -f[6,2,4], \\[6pt]&
f[3,4,6] = 0.00062974, f[4,1,6] = 1.1245, f[4,2,6] = -0.050407, f[4,3,6] = 0.0095924, \\[6pt]&
f[5,1,4] = 0.29524, f[5,2,4] = -0.013235, f[5,3,4] = 0.0025187, f[6,1,4] = -0.24375, \\[6pt]&
f[6,2,4] = 0.010927, f[6,3,4] = -0.0020793,
\end{aligned}
\end{equation*}
\begin{equation*}
{\cal R}_4 = 0.025555 \,, \quad  {\cal R}_6 = -0.43692 \,, \quad \eta_V= -3.0124 \,,
\end{equation*}
\begin{equation*}
\text{masses}^2 = (3.9604, 2.0538, 0.31293, 0.080416, 0.013095, -0.019246) \,,
\end{equation*}
\begin{equation*}
\vec{v} = (0.39122, 0.81467, -0.25082, -0.27852, -0.1723, -0.11441) \,.
\end{equation*}

\subsection*{$\boldsymbol{m_{46}^+ 7}$}

\begin{equation*}
\begin{aligned}
& I=1\!: 4\, ,\ I=2\!: 123\, ,\ I=3\!: 156\, ,\ I=4\!: 256\, ,\ I=5\!: 356,\\[6pt]&
T_{10}[1] = 10, T_{10}[2] = 1.231, T_{10}[4] = -0.41407, T_{10}[5] = -0.22002, \\[6pt]&
F_{2}[1,5] = -0.86747, F_{2}[1,6] = 0.80575, F_{2}[2,5] = -0.33502, F_{2}[2,6] = -0.080621, \\[6pt]&
F_{2}[3,5] = 0.29941, F_{2}[3,6] = -0.12093, F_{4}[1,2,4,5] = -0.0059089, F_{4}[1,2,4,6] = 0.0071751, \\[6pt]&
F_{4}[1,3,4,5] = -0.0031397, F_{4}[1,3,4,6] = 0.0038126, F_{4}[2,3,4,5] = -0.63918, \\[6pt]&
F_{4}[2,3,4,6] = 0.0013336, H[1,2,5] = -0.16391, H[1,2,6] = -0.28872, H[1,3,6] = -0.31351, \\[6pt]&
H[2,3,5] = 0.88184, H[2,3,6] = 0.72621, f[1,4,6] = -0.19050, f[3,4,6] = 0.035408, \\[6pt]&
f[4,1,6] = -0.70120, f[5,1,4] = -0.13255, f[6,1,4] = 0.16096,
\end{aligned}
\end{equation*}
\begin{equation*}
{\cal R}_4 = 0.051861 \,, \quad  {\cal R}_6 = -0.33773 \,, \quad \eta_V= -2.0672 \,,
\end{equation*}
\begin{equation*}
\text{masses}^2 = (4.2779, 1.3549, 0.2529, 0.2332, 0.03157, -0.026802) \,,
\end{equation*}
\begin{equation*}
\vec{v} = (0.52897, 0.82878, 0.12912, 0.1256, -0.023729, 0.017464) \,.
\end{equation*}

\subsection*{$\boldsymbol{m_{46}^+ 8}$}

\begin{equation*}
\begin{aligned}
& I=1\!: 4\, ,\ I=2\!: 123\, ,\ I=3\!: 156\, ,\ I=4\!: 256\, ,\ I=5\!: 356,\\[6pt]&
T_{10}[1] = 10, T_{10}[2] = 1.1329, T_{10}[3] = -0.047213, T_{10}[4] = -0.36849, T_{10}[5] = -0.34388, \\[6pt]&
F_{2}[1,5] = 1.16, F_{2}[1,6] = 0.35495, F_{2}[2,5] = 0.058393, F_{2}[2,6] = -0.10173, \\[6pt]&
F_{2}[3,5] = -0.20905, F_{2}[3,6] = -0.017627, F_{4}[1,3,4,5] = 0.047094, F_{4}[1,3,4,6] = 0.072761, \\[6pt]&
F_{4}[2,3,4,5] = 0.36756, F_{4}[2,3,4,6] = 0.56788, H[1,2,6] = 0.20821, H[1,3,5] = 0.081993, \\[6pt]&
H[1,3,6] = 0.047365, H[2,3,5] = 0.18002, H[2,3,6] = -1.167, f[1,4,5] = -0.14323, \\[6pt]&
f[1,4,6] = 0.092704, f[2,4,5] = -0.013694, f[2,4,6] = 0.0088635, f[3,4,5] = 0.02839, \\[6pt]&
f[3,4,6] = -0.018375, f[4,1,5] = -0.64444, f[4,1,6] = 0.41710, f[4,2,5] = 0.082569, \\[6pt]&
f[4,2,6] = -0.053442, f[5,1,4] = 0.24838, f[5,2,4] = -0.031823, f[6,1,4] = 0.040763, \\[6pt]&
f[6,2,4] = -0.0052227,
\end{aligned}
\end{equation*}
\begin{equation*}
{\cal R}_4 = 0.058901 \,, \quad  {\cal R}_6 = -0.36229 \,, \quad \eta_V= -2.3554 \,,
\end{equation*}
\begin{equation*}
\text{masses}^2 = (4.5142, 1.4638, 0.12601, 0.072419, 0.019515, -0.034684) \,,
\end{equation*}
\begin{equation*}
\vec{v} = (-0.59793, -0.79614, -0.032498, -0.0074837, 0.076122, -0.04174) \,.
\end{equation*}

\subsection*{$\boldsymbol{m_{46}^+ 9}$}

\begin{equation*}
\begin{aligned}
& I=1\!: 4\, ,\ I=2\!: 123\, ,\ I=3\!: 156,\\[6pt]&
T_{10}[1] = 2.3582, T_{10}[2] = 0.48564, T_{10}[3] = -0.14556, F_{2}[1,5] = -0.20255, \\[6pt]&
F_{2}[1,6] = -0.039083, F_{2}[2,5] = -0.0014989, F_{2}[2,6] = 0.0077682, F_{2}[3,5] = 0.0015074, \\[6pt]&
F_{2}[3,6] = -0.62275, F_{4}[1,2,4,5] = -0.16121, F_{4}[1,2,4,6] = -0.19422, F_{4}[1,3,4,5] = -0.05474, \\[6pt]&
F_{4}[1,3,4,6] = -0.059983, H[1,2,5] = -0.58038, H[1,2,6] = -0.11199, H[2,3,5] = -0.071827, \\[6pt]&
H[2,3,6] = 0.13254, f[1,4,5] = -0.014297, f[1,4,6] = 0.011866, f[3,4,5] = 0.11639, \\[6pt]&
f[3,4,6] = -0.096608, f[4,2,5] = -0.073867, f[4,2,6] = 0.061309, f[4,3,5] = 0.22131, \\[6pt]&
f[4,3,6] = -0.18369, f[5,2,4] = 0.0066995, f[5,3,4] = -0.020072, f[6,2,4] = -0.03472, \\[6pt]&
f[6,3,4] = 0.10402,
\end{aligned}
\end{equation*}
\begin{equation*}
{\cal R}_4 = 0.01948 \,, \quad  {\cal R}_6 = -0.093534 \,, \quad \eta_V= -2.6418 \,,
\end{equation*}
\begin{equation*}
\text{masses}^2 = (0.73377, 0.22015, 0.084047, 0.020082, -0.012866) \,,
\end{equation*}
\begin{equation*}
\vec{v} = (0.5243, 0.81491, 0.16459, 0.17724, 0.050354) \,.
\end{equation*}

\subsection*{$\boldsymbol{m_{46}^+ 10}$}

\begin{equation*}
\begin{aligned}
& I=1\!: 4\, ,\ I=2\!: 123\, ,\ I=3\!: 156,\\[6pt]&
T_{10}[1] = 1.0414, T_{10}[2] = -0.024954, T_{10}[3] = -0.18988, F_{2}[1,5] = -0.054017, \\[6pt]&
F_{2}[1,6] = 0.049286, F_{2}[2,5] = 0.11254, F_{2}[2,6] = -0.063389, F_{2}[3,5] = -0.20671, \\[6pt]&
F_{2}[3,6] = 0.064899, F_{4}[1,2,4,5] = 0.16296, F_{4}[1,2,4,6] = 0.035464, F_{4}[1,3,4,5] = -0.27918, \\[6pt]&
F_{4}[1,3,4,6] = -0.028805, H[1,2,5] = 0.20416, H[1,2,6] = 0.14499, H[1,3,5] = 0.18054, \\[6pt]&
H[1,3,6] = 0.0063486, H[2,3,5] = 0.049823, H[2,3,6] = 0.028726, f[1,4,5] = 0.0002223, \\[6pt]&
f[1,4,6] = 0.0026105, f[2,4,5] = -0.00084459, f[2,4,6] = -0.0099182, f[3,4,5] = 0.00090002, \\[6pt]&
f[3,4,6] = 0.010569, f[4,2,5] = 0.035286, f[4,2,6] = 0.41437, f[4,3,5] = 0.018223, \\[6pt]&
f[4,3,6] = 0.21399, f[5,2,4] = -0.17188, f[5,3,4] = -0.088762, f[6,2,4] = -0.18838, \\[6pt]&
f[6,3,4] = -0.097283,
\end{aligned}
\end{equation*}
\begin{equation*}
{\cal R}_4 = 0.0020327 \,, \quad  {\cal R}_6 = -0.041506 \,, \quad \eta_V= -1.2539 \,,
\end{equation*}
\begin{equation*}
\text{masses}^2 = (0.47078, 0.1756, 0.042927, 0.0029872, -0.0006372) \,,
\end{equation*}
\begin{equation*}
\vec{v} = (-0.31927, -0.90628, -0.14898, 0.19575, -0.12736) \,.
\end{equation*}

\subsection*{$\boldsymbol{m_{55}^+ 1}$}

\begin{equation*}
\begin{aligned}
& I=1\!: 12\, ,\ I=2\!: 34\, ,\ I=3\!: 56\, ,\ I=4\!: 2456\, ,\ I=5\!: 2356\, ,\ I=6\!: 1456\, ,\ I=7\!: 1356\,,\\[6pt]&
T_{10}[1]= 1, T_{10}[2]= 0.40965,T_{10}[3]= -3.1313\cdot 10^{-6},T_{10}[4]= -0.053217,T_{10}[5]= -0.0092735, \\[6pt]&
T_{10}[6]= -0.00016486,T_{10}[7]= -0.043252,F_1[5]= -0.22187,F_1[6]= -0.078553,\\[6pt]&
F_3[1,3,5]= -0.15921,F_3[1,3,6]= -0.15796,F_3[1,4,5]= -0.20668,F_3[1,4,6]= -0.063676,\\[6pt]&
F_3[2,3,5]= 0.029101,F_3[2,3,6]= 0.0063373,F_3[2,4,5]= 0.15168,F_3[2,4,6]= 0.18072,\\[6pt]&
H[1,2,5]= 0.1014,H[1,2,6]= -0.047922,H[3,4,5]= 0.16404,H[3,4,6]= -0.12833,\\[6pt]&
f[1,3,5]= -0.13246,f[1,3,6]= 0.2164,f[1,4,5]= 0.14238,f[1,4,6]= -0.094809,\\[6pt]&
f[2,3,5]= 0.14693,f[2,3,6]= -0.069881,f[2,4,5]= -0.039316,f[2,4,6]= -0.097488,\\[6pt]&
f[3,1,5]= -0.068956,f[3,1,6]= 0.12569,f[3,2,5]= -0.093779,f[3,2,6]= 0.056352,\\[6pt]&
f[4,1,5]= -0.10651,f[4,1,6]= 0.08305,f[4,2,5]= -0.042607,f[4,2,6]= -0.061736,\\[6pt]&
f[5,1,3]= -0.027059,f[5,1,4]= 0.008422,f[5,2,3]= -0.0048102,f[5,2,4]= -0.017341,\\[6pt]&
f[6,1,3]= -0.0082548,f[6,1,4]= -0.0090315,f[6,2,3]= 0.013849,f[6,2,4]= -0.019847,
\end{aligned}
\end{equation*}
\begin{equation*}
{\cal R}_4 = 0.012683  \,, \quad  {\cal R}_6 =-0.095557 \,, \quad \eta_V= -2.5435 \,,
\end{equation*}
\begin{equation*}
\text{masses}^2 = (0.25785,0.19856,0.13284,0.061518,0.054618,-0.0080648) \,,
\end{equation*}
\begin{equation*}
 \vec{v} = (0.47634,0.83174,0.19612,0.20396,0.0040682,0.034975) \,.
\end{equation*}

\subsection*{$\boldsymbol{m_{55}^+ 2}$}

\begin{equation*}
	\begin{aligned}
		& I=1\!: 12\, ,\ I=2\!: 34\, ,\ I=4\!: 2456\, ,\ I=6\!: 1456\, ,\ I=7\!: 1356\,,\\[6pt]&
T_{10}[1]= 1,T_{10}[2]= 0.52948,
T_{10}[4]= -0.10152,T_{10}[6]= -0.046308,T_{10}[7]= -0.037988,\\[6pt]&
F_1[5]= 0.1755,F_1[6]= 0.091816,
F_3[1,3,5]= -0.15299,F_3[1,3,6]= -0.057731,\\[6pt]&
F_3[1,4,5]= 0.20799,F_3[1,4,6]= -0.046086,
F_3[2,3,5]= -0.19793,F_3[2,3,6]= -0.15965,\\[6pt]&
F_3[2,4,5]= 0.16758,F_3[2,4,6]= -0.040334,
H[1,2,5]= -0.032156,H[1,2,6]= 0.1053\\[6pt]&
,H[3,4,5]= -0.051015,H[3,4,6]= 0.14757,
f[1,3,5]= -0.031329,f[1,3,6]= -0.17231,\\[6pt]&
f[1,4,5]= 0.085189,f[1,4,6]= -0.10795,
f[2,3,5]= -0.00037676,f[2,3,6]= -0.12381,\\[6pt]&
f[2,4,5]= 0.18707,f[2,4,6]= -0.1351,
f[3,1,5]= -0.034965,f[3,1,6]= -0.19981,\\[6pt]&
f[3,2,5]= -0.0015592,f[3,2,6]= 0.1123,
f[4,1,5]= -0.046789,f[4,1,6]= 0.05653,\\[6pt]&
f[4,2,5]= 0.079254,f[4,2,6]= -0.05598,
f[5,1,3]= 0.031953,f[5,1,4]= -0.018806,\\[6pt]&
f[5,2,3]= -0.011485,f[5,2,4]= 0.041502,
f[6,1,3]= 0.077132,f[6,1,4]= 0.035947,\\[6pt]&
f[6,2,3]= -0.04109,f[6,2,4]= -0.027612,
	\end{aligned}
\end{equation*}
\begin{equation*}
	{\cal R}_4 = 0.025958  \,, \quad  {\cal R}_6 =-0.12355 \,, \quad \eta_V= -2.6059 \,,
\end{equation*}
\begin{equation*}
\text{masses}^2 = (0.26217,0.16278,0.12817,0.066834,0.06477,-0.016911) \,,
\end{equation*}
\begin{equation*}
\vec{v} = (0.45995,0.86092,0.1783,0.12363,0.011867,0.0073728) \,.
\end{equation*}

\subsection*{$\boldsymbol{m_{55}^+ 3}$}

\begin{equation*}
	\begin{aligned}
		& I=1\!: 12\, ,\ I=2\!: 34\, ,\ I=4\!: 2456\, ,\ I=6\!: 1456\, ,\ I=7\!: 1356\,,\\[6pt]&
		T_{10}[1]= 1,T_{10}[2]= 0.39649,T_{10}[4]= -0.13597,T_{10}[6]= -0.0044617,T_{10}[7]= -0.0016095,\\[6pt]&
		F_1[5]= 0.059917,F_1[6]= 0.02568,F_3[1,3,5]= 0.00041564,F_3[1,3,6]= 0.00015725,\\[6pt]&
		F_3[1,4,5]= -0.037163,F_3[1,4,6]= 0.23926,F_3[2,3,5]= 0.10469,F_3[2,3,6]= -0.075765,\\[6pt]&
		F_3[2,4,5]= 0.25509,F_3[2,4,6]= -0.22331,H[1,2,5]= 0.018151,H[1,2,6]= 0.012143,\\[6pt]&
		H[3,4,5]= -0.024397,H[3,4,6]= 0.02125,f[1,3,5]= 0.34509,f[1,3,6]= 0.21035,\\[6pt]&
		f[1,4,5]= -0.0017562,f[1,4,6]= -0.00080959,f[2,3,6]= -0.27667,f[2,4,5]= -0.053521,\\[6pt]&
		f[2,4,6]= 0.075134,f[3,1,5]= 0.055747,f[3,1,6]= 0.033981,f[3,2,6]= -0.00027177,\\[6pt]&
		f[4,1,5]= 0.15453,f[4,1,6]= 0.071239,f[4,2,5]= 0.028637,f[4,2,6]= -0.040202,\\[6pt]&
		f[5,1,3]= 0.20748,f[5,1,4]= 0.0034883,f[5,2,3]= -0.0293,f[5,2,4]= 0.011216,\\[6pt]&
		f[6,1,3]= 0.17777,f[6,1,4]= -0.0081389,f[6,2,3]= 0.046644,f[6,2,4]= -0.018335,
	\end{aligned}
\end{equation*}
\begin{equation*}
	{\cal R}_4 = 0.020481  \,, \quad  {\cal R}_6 =-0.12722 \,, \quad \eta_V= -2.7126 \,,
\end{equation*}
\begin{equation*}
\text{masses}^2 = (0.43621,0.3025,0.07476,0.029728,0.017301,-0.01389) \,,
\end{equation*}
\begin{equation*}
\vec{v} = (0.55604,0.81549,0.098113,0.096111,0.07969,0.024069) \,.
\end{equation*}

\subsection*{$\boldsymbol{m_{55}^+ 4}$}

\begin{equation*}
	\begin{aligned}
		& I=1\!: 12\, ,\ I=2\!: 34\, ,\ I=3\!: 56\, ,\ I=4\!: 2456\, ,\ I=5\!: 2356\, ,\ I=6\!: 1456\, ,\ I=7\!: 1356\,,\\[6pt]&
		T_{10}[1]= 1,T_{10}[2]= 0.79176,T_{10}[3]= -0.026181,T_{10}[4]= -0.11482,T_{10}[5]= -0.058768,\\[6pt]&
		T_{10}[6]= -0.07157,T_{10}[7]= -0.036564,F_1[5]= 0.049133,F_1[6]= -0.3157,\\[6pt]&
		F_3[1,3,5]= 0.0023854,F_3[1,3,6]= 0.13983,F_3[1,4,5]= -0.0018503,F_3[1,4,6]= -0.10956,\\[6pt]&
		F_3[2,3,5]= 0.0093444,F_3[2,3,6]= 0.30172,F_3[2,4,5]= 0.0078262,F_3[2,4,6]= -0.024632,\\[6pt]&
		H[1,2,5]= 0.23093,H[1,2,6]= -0.053949,H[3,4,5]= 0.21663,H[3,4,6]= 0.022755,\\[6pt]&
		f[1,3,5]= -0.058367,f[1,3,6]= 0.011708,f[1,4,5]= -0.13198,f[1,4,6]= -0.065907,\\[6pt]&
		f[2,3,5]= -0.042264,f[2,3,6]= -0.037247,f[2,4,5]= -0.23808,f[2,4,6]= -0.077091,\\[6pt]&
		f[3,1,5]= -0.15019,f[3,1,6]= 0.1257,f[3,2,5]= 0.037172,f[3,2,6]= -0.099223,\\[6pt]&
		f[4,1,5]= -0.020208,f[4,1,6]= -0.052354,f[4,2,5]= -0.097708,f[4,2,6]= -0.0082088,\\[6pt]&
		f[5,1,3]= 0.015919,f[5,1,4]= -0.0081269,f[5,2,3]= -0.0099126,f[5,2,4]= 0.005096,\\[6pt]&
		f[6,1,3]= -0.042985,f[6,1,4]= 0.022004,f[6,2,3]= 0.026795,f[6,2,4]= -0.013684,
	\end{aligned}
\end{equation*}
\begin{equation*}
	{\cal R}_4 = 0.033613  \,, \quad  {\cal R}_6 =-0.11129 \,, \quad \eta_V= -3.3574 \,,
\end{equation*}
\begin{equation*}
\text{masses}^2 = (0.32162,0.2238,0.097703,0.080953,0.04155,-0.028213) \,,
\end{equation*}
\begin{equation*}
\vec{v} = (0.42536,0.89622,0.095324,0.069018,-0.01847,-0.040805) \,.
\end{equation*}

\subsection*{$\boldsymbol{m_{5577}^+ 1}$}

\begin{equation*}
	\begin{aligned}
		& I=1\!: 12\, ,\ I=2\!: 34\, ,\ I=3\!: 1356\, , I=4\!: 2456\,,\\[6pt]&
		T_{10}[1]= 1,T_{10}[2]= -0.07936,T_{10}[3]= 0.40065,T_{10}[4]= -0.40663,F_1[5]= 0.26286,\\[6pt]&
		F_1[6]= -0.053736,F_3[1,4,5]= -0.25386,F_3[2,3,5]= 0.044692,F_3[1,4,6]= 0.091146,\\[6pt]&
		F_3[2,3,6]= -0.0632,H[1,2,5]= -0.016356,H[3,4,5]= 0.25937,H[1,2,6]= -0.041749,\\[6pt]&
		H[3,4,6]= 0.038482,f[5,1,3]= 0.25208,f[5,2,4]= -0.17534,f[6,1,3]= 0.28717,\\[6pt]&
		f[6,2,4]= 0.074302,f[1,3,5]= 0.2044,f[1,3,6]= 0.48235,f[2,4,5]= 0.063944,\\[6pt]&
		f[2,4,6]= -0.05613,f[3,1,5]= -0.0062549,f[3,1,6]= -0.01476,f[4,2,5]= 0.0024215,\\[6pt]&
		f[4,2,6]= -0.0021256,
	\end{aligned}
\end{equation*}
\begin{equation*}
	{\cal R}_4 = 0.0019652  \,, \quad  {\cal R}_6 =-0.042929 \,, \quad \eta_V= -4.7535 \,,
\end{equation*}
\begin{equation*}
\text{masses}^2 = (0.4787,0.24925,0.029271,0.014598,-0.0023354) \,,
\end{equation*}
\begin{equation*}
\vec{v} = (0.12012,-0.95576,-0.028666,0.25498,-0.079111) \,.
\end{equation*}

\subsection*{$\boldsymbol{m_{5577}^+ 2}$}

\begin{equation*}
	\begin{aligned}
		& I=1\!: 12\, ,\ I=2\!: 34\, ,\ I=3\!: 1356\, , I=4\!: 2456\,,\\[6pt]&			
T_{10}[1]= 0.39149,T_{10}[2]= -0.037084,T_{10}[3]= -0.071447,T_{10}[4]= 0.65108,F_1[5]= -0.084985,\\[6pt]&
F_1[6]= -0.31118,F_3[1,4,5]= -0.0059985,F_3[2,3,5]= -0.01598,F_3[1,4,6]= -0.093141,\\[6pt]&
F_3[2,3,6]= -0.032537,H[1,2,5]= -0.034905,H[3,4,5]= 0.11369,H[1,2,6]= 0.0038944,\\[6pt]&
H[3,4,6]= -0.0010591,f[5,1,3]= 0.073946,f[5,2,4]= -0.0016865,f[6,1,3]= 0.24134,\\[6pt]&
f[6,2,4]= -0.02824,f[1,3,5]= -0.25493,f[1,3,6]= 0.015225,f[2,4,5]= 0.15877,\\[6pt]&
f[2,4,6]= -0.048646,f[3,1,5]= 0.18839,f[3,1,6]= -0.011251,f[4,2,5]= 0.01366,\\[6pt]&
f[4,2,6]= -0.0041853,
	\end{aligned}
\end{equation*}
\begin{equation*}
	{\cal R}_4 = 0.017441  \,, \quad  {\cal R}_6 =-0.076121 \,, \quad \eta_V= -3.5034 \,,
\end{equation*}
\begin{equation*}
\text{masses}^2 = (0.26179,0.13316,0.014967,0.0080815, -0.015276) \,,
\end{equation*}
\begin{equation*}
\vec{v} = (0.37062,0.92139,-0.027258,-0.078704,-0.082072) \,.
\end{equation*}

\subsection*{$\boldsymbol{m_{5577}^+ 3}$}

\begin{equation*}
	\begin{aligned}
		& I=1\!: 12\, ,\ I=2\!: 34\, ,\ I=3\!: 1356\, , I=4\!: 2456\,,\\[6pt]&			
T_{10}[1]= 0.10391,T_{10}[2]= 0.16488,T_{10}[3]= -0.016864,T_{10}[4]= -0.016864,F_1[5]= -0.025133,\\[6pt]&
F_1[6]= -0.16238,F_3[1,4,5]= -0.021515,F_3[2,3,5]= -0.01636,F_3[1,4,6]= 0.074891,\\[6pt]&
F_3[2,3,6]= -0.051884,H[1,2,5]= 0.1121,H[3,4,5]= 0.076829,H[1,2,6]= -0.069412,\\[6pt]&
H[3,4,6]= 0.058085,
f[6,1,3]= -0.012982,f[6,2,4]= -0.012982,f[1,3,5]= 0.055126,\\[6pt]&
f[2,4,5]= 0.041919,f[3,1,5]= 0.050093,f[4,2,5]= 0.065875,
	\end{aligned}
\end{equation*}
\begin{equation*}
	{\cal R}_4 = 0.0045531  \,, \quad  {\cal R}_6 =-0.011514 \,, \quad \eta_V= -3.2722 \,,
\end{equation*}
\begin{equation*}
\text{masses}^2 = (0.070903,0.014266,0.0087496,0.0024451, -0.0037247) \,,
\end{equation*}
\begin{equation*}
\vec{v} = (0.4319,0.88855,0.1044,0.1142,0) \,.
\end{equation*}

\subsection*{$\boldsymbol{m_{5577}^+ 4}$}

\begin{equation*}
	\begin{aligned}
		& I=1\!: 12\, ,\ I=2\!: 34\, ,\ I=3\!: 1356\, , I=4\!: 2456\,,\\[6pt]&			
T_{10}[1]= 0.45199,T_{10}[2]= 0.29433,T_{10}[3]= -0.043543,T_{10}[4]= -0.043543,F_1[5]= 0.051509,\\[6pt]&
F_1[6]= 0.26368,F_3[1,4,5]= -0.030379,F_3[2,3,5]= -0.052113,F_3[1,4,6]= -0.072095,\\[6pt]&
F_3[2,3,6]= 0.14403,H[1,2,5]= -0.13471,H[3,4,5]= -0.17112,H[1,2,6]= -0.094273,\\[6pt]&
H[3,4,6]= 0.12252,f[6,1,3]= 0.020642,f[6,2,4]= 0.020642,f[1,3,5]= -0.074879,\\[6pt]&
f[2,4,5]= -0.12845,f[3,1,5]= -0.098833,f[4,2,5]= -0.057615,
	\end{aligned}
\end{equation*}
\begin{equation*}
	{\cal R}_4 = 0.011738  \,, \quad  {\cal R}_6 =-0.032824 \,, \quad \eta_V= -3.1779 \,,
\end{equation*}
\begin{equation*}
\text{masses}^2 = (0.18893,0.041491,0.024343,0.006295, -0.0093255) \,,
\end{equation*}
\begin{equation*}
\vec{v} = (0.43789,0.88332,0.12479,0.11147,0) \,.
\end{equation*}

\subsection*{$\boldsymbol{m_{5577}^+ 5}$}

\begin{equation*}
	\begin{aligned}
		& I=1\!: 12\, ,\ I=2\!: 34\, ,\ I=3\!: 1356\, , I=4\!: 2456\,,\\[6pt]&			
T_{10}[1]= -0.064412,T_{10}[2]= 0.4785,T_{10}[3]= -0.007451,T_{10}[4]= -0.007451,F_1[5]= 0.15359,\\[6pt]&
F_1[6]= 0.18186,F_3[1,4,5]= 0.0023579,F_3[2,3,5]= -0.0023524,F_3[1,4,6]= -0.067459,\\[6pt]&
F_3[2,3,6]= -0.063327,H[1,2,5]= -0.12592,H[3,4,5]= 0.11951,H[1,2,6]= 0.13358,\\[6pt]&
H[3,4,6]= -0.092897,f[6,1,3]= 0.0051214,f[6,2,4]= 0.0051214,f[1,3,5]= -0.19341,\\[6pt]&
f[2,4,5]= 0.19295,f[3,1,5]= 0.27746,f[4,2,5]= -0.27812,
	\end{aligned}
\end{equation*}
\begin{equation*}
	{\cal R}_4 = 0.0019152  \,, \quad  {\cal R}_6 =-0.0071851 \,, \quad \eta_V= -4.7957 \,,
\end{equation*}
\begin{equation*}
\text{masses}^2 = (0.26356,0.13546,0.0012618,0.00098383, -0.0022962) \,,
\end{equation*}
\begin{equation*}
\vec{v} = (0.35331,0.93501,-0.027321,0.013431,0) \,.
\end{equation*}

\subsection*{$\boldsymbol{m_{5577}^+ 6}$}

\begin{equation*}
	\begin{aligned}
		& I=1\!: 12\, ,\ I=2\!: 34\, ,\ I=3\!: 1356\, , I=4\!: 2456\,,\\[6pt]&			
T_{10}[1]= 0.67182,T_{10}[2]= -0.094833,T_{10}[3]= -0.009195,T_{10}[4]= T_{10}[3],F_1[5]= -0.18732,\\[6pt]&
F_1[6]= -0.21629,F_3[1,4,6]= -0.069422,
F_3[2,3,6]= F_3[1,4,6],H[1,2,5]= -0.14442,
\\[6pt]&
H[3,4,5]= 0.14945,H[1,2,6]= 0.11761,
H[3,4,6]= -0.15745,f[1,3,5]= -0.36121,\\[6pt]&
f[2,4,5]= -f[1,3,5],f[3,1,5]= 0.26982,
f[4,2,5]= -f[3,1,5],f[6,1,3]= -0.0053141,\\[6pt]&
f[6,2,4]= f[6,1,3],
	\end{aligned}
\end{equation*}
\begin{equation*}
	{\cal R}_4 = 0.0023552  \,, \quad  {\cal R}_6 =-0.0083805 \,, \quad \eta_V= -4.9129 \,,
\end{equation*}
\begin{equation*}
\text{masses}^2 = (0.45272,0.19968,0.0014455,0.0012059, -0.0028928) \,,
\end{equation*}
\begin{equation*}
\vec{v} = (0.35065,0.93594,0.0020433,-0.032402,0) \,.
\end{equation*}

\subsection*{$\boldsymbol{m_{5577}^+ 7}$}

\begin{equation*}
	\begin{aligned}
		& I=1\!: 12\, ,\ I=2\!: 34\, ,\ I=3\!: 1356\, , I=4\!: 2456\,,\\[6pt]&			
T_{10}[1]= 1,T_{10}[2]= 0.96829,T_{10}[4]= -0.28566,T_{10}[3]= -0.13815,F_1[5]= -0.30382,\\[6pt]&
F_1[6]= -0.33842,F_3[1,4,5]= 0.048987,F_3[1,4,6]= 0.05421,F_3[2,3,5]= -0.12112,\\[6pt]&
F_3[2,3,6]= -0.10184,H[1,2,5]= -0.024807,H[1,2,6]= -0.3069,H[3,4,5]= 0.29781,\\[6pt]&
H[3,4,6]= -0.1961,f[1,3,5]= 0.049702,f[1,3,6]= -0.030596,f[2,4,5]= -0.020729,\\[6pt]&
f[2,4,6]= -0.034844,f[3,1,5]= 0.4302,f[3,1,6]= -0.26482,f[4,2,5]= -0.022984,\\[6pt]&
f[4,2,6]= -0.038634,f[5,1,3]= -0.34838,f[5,2,4]= -0.020232,f[6,1,3]= 0.20726,\\[6pt]&
f[6,2,4]= -0.032866,
	\end{aligned}
\end{equation*}
\begin{equation*}
	{\cal R}_4 = 0.037858  \,, \quad  {\cal R}_6 =-0.064418 \,, \quad \eta_V= -3.421 \,,
\end{equation*}
\begin{equation*}
\text{masses}^2 = (0.62528,0.47689,0.057455,0.026924, -0.032378) \,,
\end{equation*}
\begin{equation*}
\vec{v} = (0.39299,0.91772,0.03649,0.021945,-0.039089) \,.
\end{equation*}

\subsection*{$\boldsymbol{m_{5577}^+ 8}$}

\begin{equation*}
	\begin{aligned}
		& I=1\!: 12\, ,\ I=2\!: 34\, ,\ I=3\!: 1356\, , I=4\!: 2456\,,\\[6pt]&			
T_{10}[1]= 1,f[2,4,5]= 0,T_{10}[2]= 0.38463,T_{10}[4]= -0.075911,T_{10}[3]= -0.070294,\\[6pt]&
F_1[5]= -0.40134,F_1[6]= -0.1518,F_3[1,4,5]= 0.069139,F_3[1,4,6]= 0.025857,\\[6pt]&
F_3[2,3,5]= -0.050848,F_3[2,3,6]= -0.025092,H[1,2,5]= -0.10906,H[1,2,6]= -0.1663,\\[6pt]&
H[3,4,5]= 0.16742,H[3,4,6]= -0.33927,f[1,3,5]= 0.030643,f[1,3,6]= -0.062096,\\[6pt]&
f[2,4,6]= 0,f[3,1,5]= 0.10625,f[3,1,6]= -0.2153,f[4,2,5]= -0.027693,\\[6pt]&
f[4,2,6]= -0.014767,f[5,1,3]= -0.081342,f[5,2,4]= -0.01845,f[6,1,3]= 0.15255,\\[6pt]&
f[6,2,4]= -0.0091045,
	\end{aligned}
\end{equation*}
\begin{equation*}
	{\cal R}_4 = 0.019711  \,, \quad  {\cal R}_6 =-0.034616 \,, \quad \eta_V= -3.5611 \,,
\end{equation*}
\begin{equation*}
\text{masses}^2 = (0.509,0.1253,0.029292,0.013056, -0.017548) \,,
\end{equation*}
\begin{equation*}
\vec{v} = (0.39247,0.91709,0.056825,0.033463,-0.023711) \,.
\end{equation*}

\subsection*{$\boldsymbol{m_{5577}^+ 9}$}

\begin{equation*}
	\begin{aligned}
		& I=1\!: 12\, ,\ I=2\!: 34\, ,\ I=3\!: 1356\, , I=4\!: 2456\,,\\[6pt]&			
T_{10}[1]= 1,T_{10}[2]= 0.35086,T_{10}[4]= -0.000037783,T_{10}[3]= -0.090423,F_1[5]= 0.17143,\\[6pt]&
F_1[6]= 0.17787,F_3[1,4,5]= 0.35651,F_3[1,4,6]= 0.11205,F_3[2,3,5]= -0.025709,\\[6pt]&
F_3[2,3,6]= -0.015791,H[1,2,5]= -0.00038072,H[1,2,6]= 0.16276,H[3,4,5]= -0.095935,\\[6pt]&
H[3,4,6]= 0.15619,f[1,3,5]= 0.17737,f[1,3,6]= -0.28878,f[3,1,5]= 0.027193,\\[6pt]&
f[3,1,6]= -0.044273,f[4,2,5]= -0.11253,f[4,2,6]= -0.11655,f[5,1,3]= -0.014749,\\[6pt]&
f[5,2,4]= 0.04027,f[6,1,3]= 0.014241,f[6,2,4]= 0.024734,
	\end{aligned}
\end{equation*}
\begin{equation*}
	{\cal R}_4 = 0.012244  \,, \quad  {\cal R}_6 =-0.089117 \,, \quad \eta_V= -2.9333 \,,
\end{equation*}
\begin{equation*}
\text{masses}^2 = (0.19582,0.089391,0.046288,0.0069341, -0.0089789) \,,
\end{equation*}
\begin{equation*}
\vec{v} = (0.45671,0.85731,0.20184,0.1253,-0.00083919) \,.
\end{equation*}

\subsection*{$\boldsymbol{m_{5577}^+ 10}$}

\begin{equation*}
	\begin{aligned}
		& I=1\!: 12\, ,\ I=2\!: 34\, ,\ I=3\!: 1356\, , I=4\!: 2456\,,\\[6pt]&			
T_{10}[1]= 1,T_{10}[2]= 0.22554,T_{10}[4]= -0.0017644,T_{10}[3]= -0.05895,F_1[5]= 0.018004,\\[6pt]&
F_1[6]= 0.22278,F_3[1,4,5]= 0.13535,F_3[1,4,6]= -0.34611,F_3[2,3,5]= -0.003407,\\[6pt]&
F_3[2,3,6]= 0.01705,H[1,2,5]= -0.093626,H[1,2,6]= -0.086751,H[3,4,5]= -0.17802,\\[6pt]&
H[3,4,6]= -0.035572,f[1,3,5]= -0.34208,f[1,3,6]= -0.068354,f[3,1,5]= -0.033679,\\[6pt]&
f[3,1,6]= -0.0067298,f[4,2,5]= 0,f[4,2,6]= -0.13075,f[5,1,3]= 0.01225,\\[6pt]&
f[5,2,4]= -0.0067179,f[6,2,4]= 0.033619,
	\end{aligned}
\end{equation*}
\begin{equation*}
	{\cal R}_4 = 0.008297  \,, \quad  {\cal R}_6 =-0.082007 \,, \quad \eta_V= -2.9003 \,,
\end{equation*}
\begin{equation*}
\text{masses}^2 = (0.18601,0.074824,0.031424,0.0045174, -0.0060158) \,,
\end{equation*}
\begin{equation*}
\vec{v} = (0.4472,0.85893,0.21089,0.1333,-0.0034951) \,.
\end{equation*}

\subsection*{$\boldsymbol{m_{5577}^+ 11}$}

\begin{equation*}
	\begin{aligned}
		& I=1\!: 12\, ,\ I=2\!: 34\, ,\ I=3\!: 1356\, , I=4\!: 2456\,,\\[6pt]&			
T_{10}[1]= 1,T_{10}[2]= 0.22558,T_{10}[4]= -0.03301,T_{10}[3]= -0.047252,F_1[5]= 0.07129,\\[6pt]&
F_1[6]= 0.40424,F_3[1,4,5]= -0.019431,F_3[1,4,6]= -0.10771,F_3[2,3,5]= 0.0026386,\\[6pt]&
F_3[2,3,6]= 0.022276,H[1,2,5]= -0.060697,H[1,2,6]= 0.13547,H[3,4,5]= -0.37583,\\[6pt]&
H[3,4,6]= 0.044517,f[1,3,5]= -0.10513,f[1,3,6]= 0.012452,f[3,1,5]= -0.11979,\\[6pt]&
f[3,1,6]= 0.014189,f[4,2,6]= 0.035884,f[5,1,3]= 0.057879,f[5,2,4]= 0.0016953,\\[6pt]&
f[6,2,4]= 0.014312,
	\end{aligned}
\end{equation*}
\begin{equation*}
	{\cal R}_4 = 0.013256  \,, \quad  {\cal R}_6 =-0.027736 \,, \quad \eta_V= -3.4806 \,,
\end{equation*}
\begin{equation*}
\text{masses}^2 = (0.47349,0.044026,0.019742,0.0073066, -0.011535) \,,
\end{equation*}
\begin{equation*}
\vec{v} = (0.40291,0.90746,0.098306,0.066818,-0.007421) \,.
\end{equation*}

\subsection*{$\boldsymbol{m_{5577}^+ 12}$}

\begin{equation*}
	\begin{aligned}
		& I=1\!: 12\, ,\ I=2\!: 34\, ,\ I=3\!: 1356\, , I=4\!: 2456\,,\\[6pt]&			
T_{10}[1]= 1,T_{10}[2]= -0.020994,T_{10}[3]= -0.44158,T_{10}[4]= 0.35501,F_1[5]= -0.28314,\\[6pt]&
F_1[6]= 0.020063,F_3[1,4,5]= -0.096704,F_3[1,4,6]= 0.01269,F_3[2,3,5]= -0.14117,\\[6pt]&
F_3[2,3,6]= 0.16128,H[1,2,5]= 0.0073583,H[1,2,6]= 0.0098314,H[3,4,5]= 0.16407,\\[6pt]&
H[3,4,6]= -0.21909,f[1,3,5]= -0.083967,f[1,3,6]= 0.062978,f[2,4,5]= -0.28112,\\[6pt]&
f[2,4,6]= -0.40774,f[3,1,5]= 0.0024625,f[3,1,6]= -0.001847,f[5,1,3]= 0.14943,\\[6pt]&
f[5,2,4]= -0.21529,f[6,1,3]= -0.10302,f[6,2,4]= -0.28704,f[4,2,5]= -0.0028201,\\[6pt]&
f[4,2,6]= -0.0040902,
	\end{aligned}
\end{equation*}
\begin{equation*}
	{\cal R}_4 = 0.016322  \,, \quad  {\cal R}_6 =-0.05496 \,, \quad \eta_V= -2.8966 \,,
\end{equation*}
\begin{equation*}
\text{masses}^2 = (0.44749,0.25896,0.026744,0.010937, -0.011819) \,,
\end{equation*}
\begin{equation*}
\vec{v} = (0.19862,0.96945,0.0087219,-0.14259,-0.01742) \,.
\end{equation*}

\subsection*{$\boldsymbol{m_{5577}^{*\, +} 1}$}

\begin{equation*}
	\begin{aligned}
		& I=1\!: 12\, ,\ I=2\!: 34\, ,\ I=3\!: 1456\, , I=4\!: 2356\,,\\[6pt]&			
T_{10}[1]= 1,T_{10}[2]= -0.14921,T_{10}[3]= -0.010785,T_{10}[4]= T_{10}[3],,F_1[5]= 0.23733,\\[6pt]&
F_1[6]= 0.26419,F_3[1,3,6]= -0.071529,F_3[2,4,6]= -F_3[1,3,6],H[1,2,5]= 0.17952,\\[6pt]&
H[1,2,6]= -0.15497,H[3,4,5]= -0.18274,H[3,4,6]= 0.19088,f[1,4,5]= 0.5109,\\[6pt]&
f[2,3,5]= f[1,4,5],f[3,2,5]= -0.41478,f[4,1,5]= f[3,2,5],f[6,1,4]= -0.0051027,\\[6pt]&
f[6,2,3]= f[6,1,4],
	\end{aligned}
\end{equation*}
\begin{equation*}
	{\cal R}_4 = 0.0027482  \,, \quad  {\cal R}_6 =-0.0092648 \,, \quad \eta_V= -5.0483 \,,
\end{equation*}
\begin{equation*}
\text{masses}^2 = (0.93309,0.31193,0.00155,0.0014002, -0.0034685) \,,
\end{equation*}
\begin{equation*}
\vec{v} = (0.34756,0.93684,-0.011322,-0.037404,0) \,.
\end{equation*}

\subsection{Minkowski solutions}\label{ap:Mink}

\subsection*{$\boldsymbol{s_{555}^0 1}$}

\begin{equation*}
	\begin{aligned}
		& I=1\!: 12\, ,\ I=2\!: 34\, ,\ I=3\!: 56\,,\\[6pt]&			
T_{10}[1]= 0.0053035,T_{10}[2]= -0.036698,T_{10}[3]= 0.23829,F_3[1,3,5]= 0.079466,\\[6pt]&
F_3[1,3,6]= -0.10085,F_3[1,4,5]= 0.083997,F_3[1,4,6]= -0.029797,F_3[2,3,5]= 0.052531,\\[6pt]&
F_3[2,3,6]= 0.047731,F_3[2,4,5]= -0.069737,F_3[2,4,6]= -0.01231,f[1,3,5]= 0.003039,\\[6pt]&
f[1,3,6]= 0.0013338,f[1,4,5]= 0.0048401,f[1,4,6]= 0.0089686,f[2,3,5]= -0.0020841,\\[6pt]&
f[2,3,6]= 0.0024116,f[2,4,5]= 0.00016488,f[2,4,6]= -0.0057166,f[3,1,5]= 0.030623,\\[6pt]&
f[3,1,6]= 0.032105,f[3,2,5]= 0.020769,f[3,2,6]= 0.0031838,f[4,1,5]= 0.023694,\\[6pt]&
f[4,1,6]= 0.032186,f[4,2,5]= 0.0093169,f[4,2,6]= 0.020989,f[5,1,3]= -0.073303,\\[6pt]&
f[5,1,4]= -0.091413,f[5,2,3]= -0.11303,f[5,2,4]= 0.045887,f[6,1,3]= -0.0028902,\\[6pt]&
f[6,1,4]= -0.07716,f[6,2,3]= 0.060108,f[6,2,4]= -0.11056,
	\end{aligned}
\end{equation*}
\begin{equation*}
	{\cal R}_4 = 0  \,, \quad  {\cal R}_6 =-0.017241 \,,
\end{equation*}
\begin{equation*}
\text{masses}^2 = (0.052928,0.0021215,0.00005291,0) \,.
\end{equation*}

\subsection*{$\boldsymbol{s_{555}^0 2}$}

\begin{equation*}
	\begin{aligned}
		& I=1\!: 12\, ,\ I=2\!: 34\, ,\ I=3\!: 56\,,\\[6pt]&			
T_{10}[1]= 1,T_{10}[2]= -0.01029,T_{10}[3]= 0.40819,F_3[1,3,6]= 0.25963,F_3[1,4,6]= 0.1769,\\[6pt]&
F_3[2,3,6]= 0.31373,F_3[2,4,6]= -0.18935,f[1,3,5]= 0.25151,f[1,4,5]= 0.19442,\\[6pt]&
f[2,3,5]= 0.15194,f[2,4,5]= -0.64202,f[3,1,5]= 0.13744,f[3,2,5]= -0.076135,\\[6pt]&
f[4,1,5]= -0.19357,
f[4,2,5]= 0.55299,f[6,1,3]= 0.084844,f[6,1,4]= -0.071486,\\[6pt]&
f[6,2,3]= -0.0021672,f[6,2,4]= 0.23605,
	\end{aligned}
\end{equation*}
\begin{equation*}
	{\cal R}_4 = 0  \,, \quad  {\cal R}_6 =-0.11649 \,,
\end{equation*}
\begin{equation*}
\text{masses}^2 = (0.83127,0.07301,0.068032,0) \,.
\end{equation*}

\subsection*{$\boldsymbol{s_{555}^0 3}$}

\begin{equation*}
	\begin{aligned}
		& I=1\!: 12\, ,\ I=2\!: 34\, ,\ I=3\!: 56\,,\\[6pt]&			
T_{10}[1]= 1,T_{10}[2]= 0.45016,T_{10}[3]= 0.2758,F_3[1,3,6]= -0.15407,F_3[1,4,6]= 0.44154,\\[6pt]&
F_3[2,3,6]= 0.21176,F_3[2,4,6]= 0.15532,f[1,3,5]= 0.40512,f[1,4,5]= 0.010458,\\[6pt]&
f[2,3,5]= 0.12269,f[2,4,5]= 0.15527,f[3,1,5]= -0.0096352,f[3,2,5]= -0.24852,\\[6pt]&
f[4,1,5]= -0.22339,f[6,1,3]= -0.1286,f[6,2,3]= -0.10094,f[6,2,4]= -0.13871,
	\end{aligned}
\end{equation*}
\begin{equation*}
	{\cal R}_4 = 0 \,, \quad  {\cal R}_6 =-0.14383 \,,
\end{equation*}
\begin{equation*}
\text{masses}^2 = (0.2163,0.098852,0.045967,0) \,.
\end{equation*}

\subsection*{$\boldsymbol{s_{555}^0 4}$}

\begin{equation*}
	\begin{aligned}
		& I=1\!: 12\, ,\ I=2\!: 34\, ,\ I=3\!: 56\,,\\[6pt]&			
T_{10}[1]= 1,T_{10}[2]= -0.11111,T_{10}[3]= 0.46692,F_3[1,3,6]= -0.11301,F_3[1,4,6]= 0.34733,\\[6pt]&
F_3[2,3,6]= 0.30083,F_3[2,4,6]= -0.045329,f[1,3,5]= 0.2766,f[1,4,5]= -0.17052,\\[6pt]&
f[2,3,5]= -0.27582,f[2,4,5]= -0.25716,f[3,2,5]= 0.21512,f[4,1,5]= 0.12779,\\[6pt]&
f[6,1,3]= -0.12305,f[6,2,3]= -0.14059,f[6,2,4]= -0.20715,
	\end{aligned}
\end{equation*}
\begin{equation*}
	{\cal R}_4 = 0  \,, \quad  {\cal R}_6 =-0.11298 \,,
\end{equation*}
\begin{equation*}
\text{masses}^2 = (0.27831,0.077819,0.032095,0) \,.
\end{equation*}

\subsection*{$\boldsymbol{m_{46}^0 1}$}

\begin{equation*}
\begin{aligned}
& I=1\!: 4\, ,\ I=2\!: 123\, ,\ I=3\!: 156\,,\\[6pt]&
T_{10}[1]=2.1676,T_{10}[2]=0.0094995,T_{10}[3]=-1,F_{2}[2,5]=0.019929,F_{2}[2,6]=0.33076,\\[6pt]&
F_{2}[3,5]=0.010943,F_{2}[3,6]=0.099361,F_{4}[1,2,4,5]=-0.13174,F_{4}[1,2,4,6]=0.38654,\\[6pt]&
F_{4}[1,3,4,5]=-0.049892,F_{4}[1,3,4,6]=0.054594,H[1,2,5]=0.063142,H[1,2,6]=0.09965, \\[6pt]&
H[1,3,5]=-0.30783,H[1,3,6]=-0.39066, f[2,4,5]=-0.0037153,f[2,4,6]=0.0017274,\\[6pt]&
f[3,4,5]=-0.00089793,f[3,4,6]=0.00041749,f[4,2,5]=0.087441,f[4,2,6]=-0.040656,\\[6pt]&
f[4,3,5]=-1.0193,f[4,3,6]=0.47393,f[5,2,4]=-0.073129,f[5,3,4]=0.85247,\\[6pt]&
f[6,2,4]=0.040618,f[6,3,4]=-0.47349,
\end{aligned}
\end{equation*}
\begin{equation*}
{\cal R}_4 = 0 \,, \quad  {\cal R}_6 = -0.015368 \,,
\end{equation*}
\begin{equation*}
\text{masses}^2 = (3.3631, 0.45394, 0.067729, 9.1638 \cdot 10^{-6}, 0) \,.
\end{equation*}

\subsection*{$\boldsymbol{m_{46}^0 2}$}

\begin{equation*}
\begin{aligned}
& I=1\!: 4\, ,\ I=2\!: 123\, ,\ I=3\!: 156\,,\\[6pt]&
T_{10}[1]=0.13033,T_{10}[2]=0.59346,T_{10}[3]=-0.25891,F_{2}[2,5]=-0.11038,F_{2}[2,6]=-0.23729,\\[6pt]&
F_{2}[3,5]=-0.0057951,F_{2}[3,6]=-0.073035,H[1,2,5]=-0.11108,H[1,2,6]=0.071214,\\[6pt]&
H[1,3,5]=0.033558,H[1,3,6]=-0.086773,f[2,4,5]=0.26228,f[2,4,6]=-0.40397,\\[6pt]&
f[3,4,5]=-0.30188,f[5,2,4]=0.18217,f[5,3,4]=-0.20391,f[6,2,4]=-0.31933,\\[6pt]&
f[6,3,4]=0.15688,
\end{aligned}
\end{equation*}
\begin{equation*}
{\cal R}_4 = 0 \,, \quad  {\cal R}_6 = -0.023897 \,,
\end{equation*}
\begin{equation*}
\text{masses}^2 = (0.52608, 0.077079, 0.021226, 0, 0) \,.
\end{equation*}

\subsection*{$\boldsymbol{m_{466}^0 1}$}

\begin{equation*}
\begin{aligned}
& I=1\!: 4\, ,\ I=2\!: 123\, ,\ I=3\!: 156,\\[6pt]&
T_{10}[1] = 0.15998, T_{10}[2] = -0.17570, T_{10}[3] = 0.54357, F_{2}[2,5] = 0.23575, \\[6pt]&
F_{2}[2,6] = -0.058553, F_{2}[3,5] = -0.07524, F_{2}[3,6] = 0.14100, H[1,2,5] = 0.076732, \\[6pt]&
H[1,2,6] = 0.13386, H[1,3,5] = -0.064786, H[1,3,6] = 0.063202, f[2,4,5] = -0.05151, \\[6pt]&
f[2,4,6] = 0.01468, f[3,4,5] = 0.24009, f[3,4,6] = -0.11041, f[5,2,4] = 0.098223, \\[6pt]&
f[5,3,4] = 0.33923, f[6,2,4] = 0.14162, f[6,3,4] = -0.17514,
\end{aligned}
\end{equation*}
\begin{equation*}
{\cal R}_4 = 0 \,, \quad  {\cal R}_6 = -0.026276 \,,
\end{equation*}
\begin{equation*}
\text{masses}^2 = (0.26972, 0.074729, 0.020261, 0, 0) \,.
\end{equation*}

\subsection*{$\boldsymbol{m_{466}^0 2}$}

\begin{equation*}
\begin{aligned}
& I=1\!: 4\, ,\ I=2\!: 123\, ,\ I=3\!: 156,\\[6pt]&
T_{10}[1] = 0.00089705, T_{10}[2] = -0.12343, T_{10}[3] = 0.67703, F_{2}[2,6] = -0.073204, \\[6pt]&
F_{2}[3,5] = 0.095626, F_{2}[3,6] = -0.25448, H[1,2,5] = -0.0019212, H[1,2,6] = 0.0099455, \\[6pt]&
H[1,3,5] = 0.0087639, H[1,3,6] = 0.000016316, f[2,4,5] = -0.11340, f[2,4,6] = 0.12143, \\[6pt]&
f[3,4,5] = 0.001685, f[3,4,6] = -0.10207, f[5,2,4] = -0.09093, f[5,3,4] = 0.17684, \\[6pt]&
f[6,2,4] = 0.33283, f[6,3,4] = -0.04542,
\end{aligned}
\end{equation*}
\begin{equation*}
{\cal R}_4 = 0 \,, \quad  {\cal R}_6 = -0.039542 \,,
\end{equation*}
\begin{equation*}
\text{masses}^2 = (0.23513, 0.03448, 0.00023868, 0, 0) \,.
\end{equation*}

\subsection*{$\boldsymbol{m_{466}^0 3}$}

\begin{equation*}
\begin{aligned}
& I=1\!: 4\, ,\ I=2\!: 123\, ,\ I=3\!: 156,\\[6pt]&
T_{10}[1] = 0.038745, T_{10}[2] = -0.0089326, T_{10}[3] = 0.015046, F_{2}[2,6] = 0.090971, \\[6pt]&
F_{2}[3,5] = 0.016576, F_{2}[3,6] = 0.0084743, H[1,2,6] = 0.0093718, H[1,3,5] = -0.086888, \\[6pt]&
H[1,3,6] = -0.010563, f[2,4,5] = 0.0067077, f[2,4,6] = 0.029515, f[3,4,5] = 0.10123, \\[6pt]&
f[3,4,6] = 0.011584, f[5,2,4] = -0.0087884, f[5,3,4] = 0.10755, f[6,2,4] = 0.048721, \\[6pt]&
f[6,3,4] = 0.026565,
\end{aligned}
\end{equation*}
\begin{equation*}
{\cal R}_4 = 0 \,, \quad  {\cal R}_6 = -0.00043667 \,,
\end{equation*}
\begin{equation*}
\text{masses}^2 = (0.026127, 0.015642, 0.00062489, 0, 0) \,.
\end{equation*}

\subsection*{$\boldsymbol{m_{466}^0 4}$}

\begin{equation*}
\begin{aligned}
& I=1\!: 4\, ,\ I=2\!: 123\, ,\ I=3\!: 156,\\[6pt]&
T_{10}[1] = 0.036472, T_{10}[2] = 0.56902, T_{10}[3] = -0.054643, F_{2}[2,5] = 0.18885, \\[6pt]&
F_{2}[2,6] = 0.11872, F_{2}[3,5] = -0.16477, F_{2}[3,6] = -0.062198, H[1,2,5] = 0.049683, \\[6pt]&
H[1,2,6] = -0.011537, H[1,3,5] = 0.0051745, H[1,3,6] = 0.068309, f[2,4,5] = -0.052342, \\[6pt]&
f[2,4,6] = 0.22423, f[3,4,5] = 0.14330, f[3,4,6] = -0.14454, f[6,2,4] = 0.012182, \\[6pt]&
f[6,3,4] = -0.072132,
\end{aligned}
\end{equation*}
\begin{equation*}
{\cal R}_4 = 0 \,, \quad  {\cal R}_6 = -0.036741  \,,
\end{equation*}
\begin{equation*}
\text{masses}^2 = (0.17069, 0.012707, 0.0044701, 0, 0) \,.
\end{equation*}

\subsection*{$\boldsymbol{m_{466}^0 5}$}

\begin{equation*}
\begin{aligned}
& I=1\!: 4\, ,\ I=2\!: 123\, ,\ I=3\!: 156,\\[6pt]&
T_{10}[1] = 0.65414, T_{10}[2] = 0.55886, T_{10}[3] = -0.070139, F_{2}[2,5] = -0.31326, \\[6pt]&
F_{2}[2,6] = -0.19336, F_{2}[3,5] = 0.20421, F_{2}[3,6] = -0.15304, H[1,2,5] = -0.20324, \\[6pt]&
H[1,2,6] = -0.014823, H[1,3,5] = 0.016713, H[1,3,6] = -0.29836, f[2,4,5] = 0.043513, \\[6pt]&
f[2,4,6] = -0.23363, f[3,4,5] = -0.16787, f[3,4,6] = 0.11717, f[5,3,4] = -0.028836, \\[6pt]&
f[6,2,4] = -0.015351, f[6,3,4] = 0.086384,
\end{aligned}
\end{equation*}
\begin{equation*}
{\cal R}_4 = 0 \,, \quad  {\cal R}_6 = -0.034908 \,,
\end{equation*}
\begin{equation*}
\text{masses}^2 = (0.32049, 0.11059, 0.0073101, 0, 0) \,.
\end{equation*}

\subsection*{$\boldsymbol{m_{466}^0 6}$}

\begin{equation*}
\begin{aligned}
& I=1\!: 4\, ,\ I=2\!: 123\, ,\ I=3\!: 156,\\[6pt]&
T_{10}[1] = 0.1391, T_{10}[2] = 0.21921, T_{10}[3] = 0.6068, F_{2}[2,5] = 0.17163, \\[6pt]&
F_{2}[2,6] = 0.030109, F_{2}[3,5] = 0.032577, F_{2}[3,6] = -0.33822, H[1,2,5] = -0.083254, \\[6pt]&
H[1,2,6] = 0.098809, H[1,3,5] = 0.099749, H[1,3,6] = 0.034276, f[2,4,6] = -0.048187, \\[6pt]&
f[3,4,5] = 0.11704, f[5,2,4] = -0.037857, f[5,3,4] = -0.023053, f[6,2,4] = 0.26141, \\[6pt]&
f[6,3,4] = 0.033041,
\end{aligned}
\end{equation*}
\begin{equation*}
{\cal R}_4 = 0 \,, \quad  {\cal R}_6 = -0.059001 \,,
\end{equation*}
\begin{equation*}
\text{masses}^2 = (0.21201, 0.035651, 0.013395, 0, 0) \,.
\end{equation*}

\subsection{Anti-de Sitter solutions}\label{ap:AdS}

\subsection*{$\boldsymbol{s_{55}^- 1}$}

\begin{equation*}
	\begin{aligned}
		& I=1\!: 12\, ,\ I=2\!: 34\,,\\[6pt]&			
T_{10}[1]= 0.65385,T_{10}[2]= 0.067793,F_1[5]= 0.011227,F_1[6]= -0.070069,F_3[1,3,5]= 0.16616,\\[6pt]&
F_3[1,3,6]= 0.17837,F_3[1,4,5]= 0.11969,F_3[1,4,6]= -0.2383,F_3[2,3,5]= -0.075801,\\[6pt]&
F_3[2,3,6]= -0.077512,F_3[2,4,5]= -0.0060063,F_3[2,4,6]= -0.079968,H[1,2,5]= 0.064459,\\[6pt]&
H[1,2,6]= 0.0056405,H[3,4,5]= 0.069322,H[3,4,6]= -0.17184,F_5[1,2,3,4,5]= -0.008645,\\[6pt]&
F_5[1,2,3,4,6]= -0.0013851,f[1,3,5]= -0.27755,f[1,3,6]= -0.0073778,f[1,4,5]= -0.25336,\\[6pt]&
f[1,4,6]= -0.039444,f[2,3,5]= 0.054799,f[2,3,6]= 0.057591,f[2,4,5]= 0.059283,\\[6pt]&
f[2,4,6]= 0.0022988,f[3,1,5]= 0.0028896,f[3,1,6]= -0.0044766,f[3,2,5]= 0.0019835,\\[6pt]&
f[3,2,6]= -0.020978,f[4,1,5]= 0.016741,f[4,1,6]= 0.0075944,f[4,2,5]= -0.014713,\\[6pt]&
f[4,2,6]= 0.020748,f[5,1,3]= -0.14682,f[5,1,4]= -0.13679,f[5,2,3]= 0.093115,\\[6pt]&
f[5,2,4]= 0.086751,f[6,1,3]= -0.023524,f[6,1,4]= -0.021917,f[6,2,3]= 0.014919,\\[6pt]&
f[6,2,4]= 0.0139,
	\end{aligned}
\end{equation*}
\begin{equation*}
	{\cal R}_4 = -0.033561  \,, \quad  {\cal R}_6 =-0.0073162 \,, \quad \eta_V= 0.7785 \,,
\end{equation*}
\begin{equation*}
\text{masses}^2 = (0.19854,0.060726,0.04147,-0.0065318) \,.
\end{equation*}

\subsection*{$\boldsymbol{s_{55}^- 2}$}

\begin{equation*}
	\begin{aligned}
		& I=1\!: 12\, ,\ I=2\!: 34\,,\\[6pt]&			
T_{10}[1]= 0.28653,T_{10}[2]= 0.19465,F_3[1,3,5]= 0.030441,F_3[1,3,6]= -0.18245,\\[6pt]&
F_3[1,4,5]= 0.1287,F_3[1,4,6]= -0.047985,F_3[2,3,5]= 0.09543,F_3[2,3,6]= -0.049685,\\[6pt]&
F_3[2,4,5]= 0.16944,F_3[2,4,6]= 0.045125,H[1,2,5]= 0.013663,H[1,2,6]= -0.086069,\\[6pt]&
H[3,4,5]= -0.087243,H[3,4,6]= 0.0012479,f[1,3,6]= -0.15281,f[1,4,6]= 0.13141,\\[6pt]&
f[2,3,6]= -0.095775,f[2,4,6]= 0.082363,f[3,1,5]= 0.067638,f[3,1,6]= 0.011932,\\[6pt]&
f[3,2,5]= -0.10792,f[3,2,6]= -0.019038,f[4,1,5]= 0.078652,f[4,1,6]= 0.013875,\\[6pt]&
f[4,2,5]= -0.12549,f[4,2,6]= -0.022138,f[5,1,3]= -0.02027,f[5,1,4]= -0.061815,\\[6pt]&
f[5,2,3]= 0.057795,f[5,2,4]= 0.076739,f[6,1,3]= -0.09923,f[6,1,4]= 0.085335,\\[6pt]&
f[6,2,3]= 0.014035,f[6,2,4]= -0.012069,
	\end{aligned}
\end{equation*}
\begin{equation*}
	{\cal R}_4 = -0.015208  \,, \quad  {\cal R}_6 =-0.017287 \,, \quad \eta_V= -4 \,,
\end{equation*}
\begin{equation*}
\text{masses}^2 = (0.070021,0.044657,0.027383,0.015208)  \,.
\end{equation*}

\subsection*{$\boldsymbol{s_{55}^- 3}$}

\begin{equation*}
	\begin{aligned}
		& I=1\!: 12\, ,\ I=2\!: 34\,,\\[6pt]&			
T_{10}[1]= 0.39238,T_{10}[2]= 0.5567,F_3[1,3,5]= 0.091642,F_3[1,3,6]= 0.19088,\\[6pt]&
F_3[1,4,5]= -0.21033,F_3[1,4,6]= 0.23536,F_3[2,3,5]= -0.010403,F_3[2,3,6]= -0.0046634,\\[6pt]&
F_3[2,4,5]= -0.20495,F_3[2,4,6]= -0.091878,H[1,2,5]= 0.0040374,H[1,2,6]= -0.16052,\\[6pt]&
H[3,4,5]= -0.10526,H[3,4,6]= 0.00052599,f[1,3,6]= 0.31796,f[1,4,6]= -0.016139,\\[6pt]&
f[3,2,5]= 0.018778,f[3,2,6]= 0.00052518,f[4,2,5]= 0.36995,f[4,2,6]= 0.010347,\\[6pt]&
f[5,1,3]= -0.0057443,f[5,1,4]= 0.00029156,f[5,2,3]= 0.08266,f[5,2,4]= -0.2441,\\[6pt]&
f[6,1,3]= 0.20538,f[6,1,4]= -0.010425,f[6,2,3]= -0.092728,f[6,2,4]= 0.0047066,
	\end{aligned}
\end{equation*}
\begin{equation*}
	{\cal R}_4 = -0.036862  \,, \quad  {\cal R}_6 =-0.023797 \,, \quad \eta_V= -3.8495  \,,
\end{equation*}
\begin{equation*}
\text{masses}^2 = (0.20393,0.11596,0.074406,0.035475) \,.
\end{equation*}

\subsection*{$\boldsymbol{s_{55}^- 4}$}

\begin{equation*}
	\begin{aligned}
		& I=1\!: 12\, ,\ I=2\!: 34\,,\\[6pt]&			
T_{10}[1]= 0.14258,T_{10}[2]= 0.58135,F_3[1,3,5]= 0.075163,F_3[1,3,6]= -0.1116,\\[6pt]&
F_3[1,4,5]= -0.11823,F_3[1,4,6]= -0.2689,F_3[2,4,5]= -0.19084,F_3[2,4,6]= 0.065352,\\[6pt]&
H[1,2,5]= -0.0073236,H[1,2,6]= -0.15603,H[3,4,5]= 0.0029354,H[3,4,6]= 0.0040146,\\[6pt]&
f[1,3,6]= 0.20098,f[4,2,5]= -0.35082,f[4,2,6]= 0.021637,f[5,1,3]= 0.0088588,\\[6pt]&
f[5,2,3]= -0.11714,f[5,2,4]= 0.18033,f[6,1,3]= 0.14363,f[6,2,3]= -0.027522,
	\end{aligned}
\end{equation*}
\begin{equation*}
	{\cal R}_4 = -0.024424  \,, \quad  {\cal R}_6 =-0.023691 \,, \quad \eta_V=-2.4901  \,,
\end{equation*}
\begin{equation*}
\text{masses}^2 = (0.15904,0.067206,0.039032,0.015205 )\,.
\end{equation*}

\subsection*{$\boldsymbol{m_{46}^- 1}$}

\begin{equation*}
\begin{aligned}
& I=1\!: 4\, ,\ I=2\!: 123\, ,\ I=3\!: 156,\\[6pt]&
T_{10}[1] = 1.1971, T_{10}[2] = 0.072312, T_{10}[3] = -0.062975, F_{2}[1,5] = -0.041978, \\[6pt]&
F_{2}[1,6] = 0.14026, F_{2}[2,5] = 0.066399, F_{2}[2,6] = 0.21752, F_{2}[3,5] = 0.011368, \\[6pt]&
F_{2}[3,6] = -0.011895, F_{4}[1,2,4,5] = -0.11379, F_{4}[1,2,4,6] = -0.19903, F_{4}[1,3,4,5] = -0.28462, \\[6pt]&
F_{4}[1,3,4,6] = 0.28620, H[1,2,5] = -0.079033, H[1,2,6] = -0.10719, H[1,3,5] = -0.072608, \\[6pt]&
H[1,3,6] = 0.077682, H[2,3,5] = 0.21997, H[2,3,6] = -0.13693, f[1,4,5] = -0.042794, \\[6pt]&
f[1,4,6] = -0.025141, f[2,4,5] = -0.019349, f[2,4,6] = -0.011367, f[3,4,5] = -0.0097687, \\[6pt]&
f[3,4,6] = -0.005739, f[4,2,5] = 0.37334, f[4,2,6] = 0.21933, f[4,3,5] = 0.16584, \\[6pt]&
f[4,3,6] = 0.097433, f[5,2,4] = -0.17777, f[5,3,4] = -0.078969, f[6,2,4] = -0.053204, \\[6pt]&
f[6,3,4] = -0.023634,
\end{aligned}
\end{equation*}
\begin{equation*}
{\cal R}_4 = -0.048164 \,, \quad  {\cal R}_6 = -0.02412 \,, \quad \eta_V= 1.2531\,,
\end{equation*}
\begin{equation*}
\text{masses}^2 = (0.49918, 0.13392, 0.060085, 0.054407, -0.015089) \,.
\end{equation*}

\subsection*{$\boldsymbol{m_{46}^- 2}$}

\begin{equation*}
\begin{aligned}
& I=1\!: 4\, ,\ I=2\!: 123\, ,\ I=3\!: 156,\\[6pt]&
T_{10}[1] = 0.7165, T_{10}[2] = 0.0763, T_{10}[3] = -0.10534, F_{2}[1,6] = 0.24176, \\[6pt]&
F_{2}[2,5] = 0.037154, F_{2}[2,6] = -0.000011568, F_{2}[3,5] = -0.042859, F_{2}[3,6] = 0.0067886, \\[6pt]&
F_{4}[1,2,4,5] = -0.13513, F_{4}[1,2,4,6] = -0.17510, F_{4}[1,3,4,5] = -0.13777, F_{4}[1,3,4,6] = 0.16894, \\[6pt]&
H[1,2,5] = 0.0036697, H[1,2,6] = 0.004547, H[1,3,5] = 0.0038034, H[1,3,6] = -0.0037413, \\[6pt]&
H[2,3,5] = 0.19615, H[2,3,6] = 0.19189, f[1,4,5] = -0.045081, f[1,4,6] = 0.010927, \\[6pt]&
f[2,4,5] = -0.00053083, f[2,4,6] = 0.00012866, f[3,4,5] = -0.0010236, f[3,4,6] = 0.0002481, \\[6pt]&
f[4,2,5] = 0.27295, f[4,2,6] = -0.066157, f[4,3,5] = 0.28289, f[4,3,6] = -0.068567, \\[6pt]&
f[5,2,4] = -0.18495, f[5,3,4] = -0.19169,
\end{aligned}
\end{equation*}
\begin{equation*}
{\cal R}_4 = -0.019002 \,, \quad  {\cal R}_6 = -0.012892 \,, \quad \eta_V=  1.5483 \,,
\end{equation*}
\begin{equation*}
\text{masses}^2 = (0.38901, 0.18817, 0.031941, 0.013066, -0.0073556) \,.
\end{equation*}

\subsection*{$\boldsymbol{m_{46}^- 3}$}

\begin{equation*}
\begin{aligned}
& I=1\!: 4\, ,\ I=2\!: 123\, ,\ I=3\!: 156,\\[6pt]&
T_{10}[1] = 5.9022, T_{10}[2] = 0.67951, T_{10}[3] = -0.88782, F_{2}[1,5] = 0.32655, \\[6pt]&
F_{2}[1,6] = 0.61787, F_{2}[2,5] = -0.060571, F_{2}[2,6] = 0.032012, F_{2}[3,5] = 0.13077, \\[6pt]&
F_{2}[3,6] = -0.069114, F_{4}[1,2,4,5] = -0.30798, F_{4}[1,2,4,6] = 0.49291, F_{4}[1,3,4,5] = -0.50027, \\[6pt]&
F_{4}[1,3,4,6] = -0.44836, H[2,3,5] = 0.75352, H[2,3,6] = 0.23160, f[1,4,5] = -0.10747, \\[6pt]&
f[1,4,6] = 0.093925, f[4,2,5] = -0.79910, f[4,2,6] = 0.69839, f[4,3,5] = -0.37013, \\[6pt]&
f[4,3,6] = 0.32348, f[5,2,4] = 0.62420, f[5,3,4] = 0.28912, f[6,2,4] = -0.32990, \\[6pt]&
f[6,3,4] = -0.15280,
\end{aligned}
\end{equation*}
\begin{equation*}
{\cal R}_4 = -0.1534 \,, \quad  {\cal R}_6 = -0.11122 \,, \quad \eta_V= 1.5537 \,,
\end{equation*}
\begin{equation*}
\text{masses}^2 = (3.2576, 1.5711, 0.26172, 0.109, -0.059584) \,.
\end{equation*}

\subsection*{$\boldsymbol{m_{46}^- 4}$}

\begin{equation*}
\begin{aligned}
& I=1\!: 4\, ,\ I=2\!: 123\, ,\ I=3\!: 156,\\[6pt]&
T_{10}[1] = 0.072202, T_{10}[2] = 1.0613, T_{10}[3] = -0.090953, F_{2}[1,6] = 0.35610, \\[6pt]&
F_{2}[3,5] = 0.075768, F_{4}[1,2,4,5] = -0.20253, H[1,3,5] = -0.046817, H[1,3,6] = 0.016544, \\[6pt]&
H[2,3,5] = 0.040552, H[2,3,6] = 0.0087218, f[1,4,5] = -0.42575, f[1,4,6] = -0.059404, \\[6pt]&
f[2,4,5] = 0.11424, f[2,4,6] = -0.21770, f[5,1,4] = -0.19798, f[5,2,4] = 0.17149, \\[6pt]&
f[6,3,4] = -0.042125,
\end{aligned}
\end{equation*}
\begin{equation*}
{\cal R}_4 = -0.020509 \,, \quad  {\cal R}_6 = -0.053926 \,, \quad \eta_V= 1.3004 \,,
\end{equation*}
\begin{equation*}
\text{masses}^2 = (0.4783, 0.10213, 0.034824, 0.030265, -0.0066676) \,.
\end{equation*}

\subsection*{$\boldsymbol{m_{46}^- 5}$}

\begin{equation*}
\begin{aligned}
& I=1\!: 4\, ,\ I=2\!: 123\, ,\ I=3\!: 156,\\[6pt]&
T_{10}[1] = 0.05707, T_{10}[2] = 1.3103, T_{10}[3] = -0.091524, F_{2}[1,5] = -0.28057, \\[6pt]&
F_{2}[1,6] = -0.092774, F_{2}[2,5] = 0.26093, F_{2}[2,6] = 0.086279, F_{2}[3,5] = 0.019, \\[6pt]&
F_{2}[3,6] = -0.057461, F_{4}[1,2,4,5] = -0.061201, F_{4}[1,2,4,6] = 0.18509, H[1,3,6] = 0.043744, \\[6pt]&
f[1,4,6] = -0.40037, f[2,4,5] = -0.25405, f[2,4,6] = 0.20285, f[5,1,4] = 0.07293, \\[6pt]&
f[5,3,4] = -0.04517, f[6,1,4] = -0.22056, f[6,3,4] = -0.014936,
\end{aligned}
\end{equation*}
\begin{equation*}
{\cal R}_4 = -0.019001 \,, \quad  {\cal R}_6 = -0.072802 \,, \quad \eta_V=  1.2548 \,,
\end{equation*}
\begin{equation*}
\text{masses}^2 = (0.4181, 0.16632, 0.043898, 0.028584, -0.0059604) \,.
\end{equation*}

\end{appendix}

\newpage

\providecommand{\href}[2]{#2}\begingroup\raggedright
\endgroup

\end{document}